# The impact and recovery of asteroid 2018 LA

Peter Jenniskens[1,2*], Mohutsiwa Gabadirwe[3], Qing-Zhu Yin[4], Alexander Proyer[5], Oliver Moses[6], Tomas Kohout[7,8], Fulvio Franchi[5], Roger L. Gibson[9], Richard Kowalski[10], Eric J. Christensen[10], Alex R. Gibbs[10], Aren Heinze[11], Larry Denneau[11], Davide Farnocchia[12], Paul W. Chodas[12], William Gray[13], Marco Micheli[14], Nick Moskovitz[15], Christopher A. Onken[16], Christian Wolf[16], Hadrien A. R. Devillepoix[17], Quanzhi Ye[18,19], Darrel K. Robertson[20], Peter Brown[21], Esko Lyytinen[8], Jarmo Moilanen[8], Jim Albers[1], Tim Cooper[22], Jelle Assink[23], Läslo Evers[23,24], Panu Lahtinen[8], Lesedi Seitshiro[5], Matthias Laubenstein[25], Nggie Wantlo[3], Phemo Moleje[3], Joseph Maritinkole[3], Heikki Suhonen[26], Michael E. Zolensky[27], Lewis Ashwal[9], Takahiro Hiroi[28], Derek W. Sears[29], Alexander Sehlke[29], Alessandro Maturilli[30], Matthew E. Sanborn[4], Magdalena H. Huyskens[4], Supratim Dey[4], Karen Ziegler[31], Henner Busemann[32], My E. I. Riebe[32], Matthias M. M. Meier[33], Kees C. Welten[34], Marc W. Caffee[35], Qin Zhou[36], Qiu-Li Li[36], Xian-Hua Li[37], Yu Liu[37], Guo-Qiang Tang[37], Hannah L. McLain[38,39], Jason P. Dworkin[39], Daniel P. Glavin[39], Philippe Schmitt-Kopplin[40,41], Hassan Sabbah[42], Christine Joblin[43], Mikael Granvik[26,44], Babutsi Mosarwa[43], and Koketso Botepe[3].

[1] SETI Institute, 189 Bernardo Avenue, Mountain View, CA 94043, USA.

[2] NASA Ames Research Center, Moffett Field, CA 94035, USA.

[3] Botswana Geoscience Institute, Plot 11566, Khama 1 Avenue, Private Bag 0014, Lobatse, Botswana.

[4] Department of Earth and Planetary Sciences, University of California Davis, One Shields Avenue, CA 95616, USA.

[5] Botswana International University of Science and Technology, Private Bag 16, Palapye, Botswana.

[6] University of Botswana, Okavango Research Institute, Private Bag 285, Maun, Botswana.

[7] Department of Geosciences and Geography, University of Helsinki, P. O. Box 64, FI-00014 Helsinki, Finland.

[8] Ursa Finnish Fireball Network, Kopernikuksentie 1, FI-00130 Helsinki, Finland.

[9] School of Geosciences, University of the Witwatersrand, P.O. WITS, Johannesburg 2050, South Africa.

[10] Catalina Sky Survey, Lunar & Planetary Laboratory, The University of Arizona, 1629 E University Blvd., Tucson, AZ 85721, USA.

[11] ATLAS, Institute for Astronomy, 2680 Woodlawn Drive, Honolulu, HI 96822-1839, USA.

[12] Jet Propulsion Laboratory, California Institute of Technology, 4800 Oak Grove Drive, Pasadena, CA 91109, USA.

[13] Project Pluto, 168 Ridge Road, Bowdoinham, ME 04008, USA.

[14] ESA NEO Coordination Centre, Largo Galileo Galilei 1, I-00044, Frascati, Italy.

[15] Lowell Observatory, 1400 W. Mars Hill Rd., Flagstaff, AZ 86001, USA.

[16] Research School of Astronomy and Astrophysics, The Australian National University, Canberra ACT 2611, Australia.

[17] School of Earth and Planetary Sciences, Curtin University, Perth WA 6845, Australia.

[18] Department of Astronomy, University of Maryland, College Park, MD 20742, USA.

[19] Division of Physics, Mathematics and Astronomy, Caltech, Pasadena, CA 91125, USA.

[20] NASA Ames Research Center, Asteroid Threat Assessment Project, Mail Stop 239-1, Moffett Field, CA 94035, USA.

[21] Centre for Planetary Science and Exploration, Western University, London, Ontario, N6A 5B7, Canada.

[22] Astronomical Society of Southern Africa, Suite 617, Private Bag X043, Benoni 1500, South Africa.

[23] Royal Dutch Meteorological Institute, R&D Seismology and Acoustics, P. O. Box 201, NL-3730 AE De Bilt, The Netherlands.

[24] Delft University of Technology, Department of Geoscience and Engineering, P. O. Box 5048, NL-2600 GA Delft, the Netherlands.

[25] Gran Sasso National Laboratory, National Institute for Nuclear Physics, Via G. Acitelli 22, I-67100 Assergi, Italy.

[26] University of Helsinki, Department of Physics, P. O. Box 64, FI-00014 Helsinki, Finland.

[27] NASA Johnson Space Flight Center, ARES Division, Houston, TX 77058, USA.

[28] Brown University, Reflectance Experiment Laboratory, Department of Earth, Environmental and Planetary Science, Providence, RI 02912, USA.

[29] NASA Ames Research Center / Bay Area Environmental Research Institute, Mail Stop 245-3, Moffett Field, CA 94035, USA.

[30] Institute for Planetary Research, German Aerospace Center DLR, Rutherfordstrasse 2, D-12489 Berlin-Adlershof, Germany.

[31] University of New Mexico, Institute of Meteoritics, 221 Yale Blvd NE, 331 Northrop Hall, Albuquerque, NM 87131, USA.

[32] Institute of Geochemistry and Petrology, ETH Zürich, Clausiusstrasse 25, CH-8092 Zürich, Switzerland.

[33] Naturmuseum St. Gallen, Rorschacher Strasse 263, CH-9016 St. Gallen, Switzerland.

[34] University of California Berkeley, Space Science Laboratory, Berkeley, CA 94720, USA.

[35] Purdue University, Dept. Physics and Astronomy, 525 Northwestern Avenue, West Lafayette, IN 47907, USA.

[36] National Astronomical Observatories, Beijing, Chinese Academy of Sciences, Beijing 100012, China.

[37] State Key Laboratory of Lithospheric Evolution, Institute of Geology and Geophysics, Chinese Academy of Sciences, Beijing 100029, China.

[38] Catholic University of America, Department of Chemistry, 620 Michigan Ave, N.E., Washington, DC 20064, USA.

[39] NASA Goddard Space Flight Center, 8800 Greenbelt Rd., Greenbelt, MD 20771, USA.

[40] Helmholtz Zentrum München, Research Unit Analytical BioGeoChemistry, Ingolstädter Landstr. 1, D-85764 Neuherberg, Germany.

[41] Technische Universität München, Analytical Food Chemistry, D-85354 Freising-Weihenstephan, Germany.

[42] IRAP, Université de Toulouse, CNRS, CNES, Université de Toulouse (UPS), F-31028 Toulouse Cedex 4, France.

[43] Botswana National Museum, 161 Queens Rd., Gaborone, Botswana.

[44] Asteroid Engineering Laboratory, Onboard Space Systems, Lulea University of Technology, Box 848, S-981 28 Kiruna, Sweden.

*Correspondence to: Petrus.M.Jenniskens@nasa.gov.

**Abstract** – The June 2, 2018, impact of asteroid 2018 LA over Botswana is only the second asteroid detected in space prior to impacting over land. Here, we report on the successful recovery of meteorites. Additional astrometric data refine the approach orbit and define the spin period and shape of the asteroid. Video observations of the fireball constrain the asteroid's position in its orbit and were used to triangulate the location of the fireball's main flare over the Central Kalahari Game Reserve. 23 meteorites were recovered. A consortium study of eight of these classifies Motopi Pan as a HED polymict breccia derived from howardite, cumulate and basaltic eucrite, and diogenite lithologies. Before impact, 2018 LA was a solid rock of ~156 cm diameter with high bulk density ~2.85 g/cm$^3$, a relatively low albedo $p_V$ ~ 0.25, no significant opposition effect on the asteroid brightness, and an impact kinetic energy of ~0.2 kt. The orbit of 2018 LA is consistent with an origin at Vesta (or its Vestoids) and delivery into an Earth-impacting orbit via the $\nu_6$ resonance. The impact that ejected 2018 LA in an orbit towards Earth occurred 22.8 ± 3.8 Ma ago. Zircons record a concordant U-Pb age of 4563 ± 11 Ma and a consistent $^{207}$Pb/$^{206}$Pb age of 4563 ± 6 Ma. A much younger Pb-Pb phosphate resetting age of 4234 ± 41 Ma was found. From this impact chronology, we discuss what is the possible source crater of Motopi Pan and the age of Vesta's Veneneia impact basin.

## INTRODUCTION

Howardite-Eucrite-Diogenite (HED) meteorites are basaltic achondrites with characteristic prominent absorption bands in near-IR reflectance spectra that suggest they originate from asteroid 4 Vesta (McCord *et al.* 1970; Hiroi *et al.* 1994; DeSanctis *et al.* 2012). NASA's Dawn mission found from reflectance that much of the surface of Vesta is covered in an eucrite-rich howardite surface (e.g., Buratti *et al.* (2013), with the K/Th ratio and bulk chemical data such as Fe/Si versus Fe/O of Vesta resembling that of howardite meteorites (Prettyman *et al.* 2012, 2015). Other observations linking HEDs to Vesta are summarized by McSween *et al.* (2013).

The Rheasilvia impact basin on the southern hemisphere of Vesta, and the smaller underlying Veneneia basin, are the origin craters of a family of 0.8–8-km sized V-class asteroids called the Vestoids (Marchi *et al.* 2012; Ivanov & Melosh 2013). The location of Vestoids astride a prominent resonance escape hatch (the 3:1 resonance) argues for Vestoid sources of the larger V-class Near-Earth Objects (Binzel & Xu 1993).

There are also rare HED meteorites that have anomalous isotopic signatures (Bland *et al.* 2009), which may originate from asteroids on orbits outside the Vesta family (e.g., Licandro *et al.* 2017).

The idea that a crater on Vesta itself is the direct source of most HEDs, rather than Vestoids, is not the consensus position of the community currently. However, one-third of all HED meteorites originated from one or more significant collisions that occurred ~22 Ma ago, according to their cosmic ray exposure (CRE) age (Eugster & Michel 1995; Eugster *et al.* 2006). In an earlier paper, Unsalan *et al.* (2019) argued that few Vestoids are big enough (≥ 0.3 km) to generate the amount of 10 cm to 1 meter sized material needed to explain the HED 22-Ma meteoroid impact flux at Earth and those that do, have a combined cross section for collisions that is five times smaller than that of Vesta.

The only other photographed HED fall prior to Motopi Pan, the howardite Sariçiçek, belonged to this 22-Ma group and had an Earth-approaching orbit with the short semi-major axis expected for meteoroids ejected from Vesta, after enduring numerous encounters with Earth before impacting (Unsalan *et al.* 2019). Vesta crater Antonia was formed ~22 Ma ago in the lunar-based chronology scheme. The excavation of this 16.7-km diameter crater could have lifted sufficient meter-sized rocks from the gravity well of Vesta (Melosh 1989). Antonia is located in the Rheasilvia impact basin and Sariçiçek's Ar-Ar age matches that terrain's cratering age (Unsalan *et al.* 2019).

Less well observed HED falls linked to the 22-Ma group include the basaltic eucrite Puerto Lápice with an approximate radiant from visual observations and persistent train photographs, but no information on the entry speed (Trigo-Rodríguez *et al.* 2009). It has a CRE age of 19 ± 2 Ma and oxygen isotope compositions similar to Sariçiçek (Llorca *et al.* 2009).

In this paper, we report on the recovery of asteroid 2018 LA, which was detected by the Catalina Sky Survey on an impact trajectory with Earth on June 2, 2018, at 08:14 UTC (Fig. 1). This asteroid was only the second asteroid detected in space prior to impacting over land, following asteroid 2008 $TC_3$ ten years earlier (Jenniskens *et al.* 2009). The event offered a rare opportunity to recover samples from an asteroid observed in space and successful efforts were undertaken to recover the meteorites. When those proved to be HEDs, the question arose whether Sariçiçek and Motopi Pan shared the same source crater. New astrometric data to refine the trajectory of asteroid 2018 LA, a description of its impact and recovery, a detailed analysis of the recovered meteorites, their petrography and mineralogy, their cosmochemistry, and their physical properties and organic content are presented. From this data, we derive the initial size of the meteoroid and its collision history and discuss what might be the source crater of this meteorite on Vesta.

## SAMPLES AND METHODS OF METEORITE ANALYSIS

Twenty-one days after the fall, foot searches in the Central Kalahari Game Reserve (CKGR), Botswana, initially recovered a single 17.92-g meteorite, named "Motopi Pan" MP-01 after a nearby watering hole, during a search conducted on June 18–23, 2018, by members of the Botswana International University of Science and Technology (BIUST), the Botswana Geoscience Institute (BGI), and the University of Botswana's Okavango Research Institute (ORI) assisted by the SETI Institute (Fig. 2a).

MP-01 was measured by gamma-ray spectrometry in the STELLA low background laboratory of the Gran Sasso National Laboratory 80 days after the fall from August 21 to 31, 2018, using methods described in Unsalan *et al.* (2019). X-ray micro-CT was performed at the University of Helsinki. The meteorite was scanned twice at 14 and 7µm/voxel using custom-built Nanotom 180 NF tomography equipment (Phoenix X-ray Systems and Services, part of GE Measurement Systems and Solutions, Germany). In addition, measurements of magnetic susceptibility, bulk- and grain density, and X-ray fluorescence were performed. Magnetic remanence was measured using a 2G Enterprises Model 755 DC SQUID SRM, a superconducting rock magnetometer. Bulk volume was measured using a NextEngine 3D Scanner Ultra HD model 2020i laser scanner. Grain volume was measured using a Quantachrome Ultrapyc 1200e pycnometer with $N_2$ gas. X-ray fluorescence (XRF) was measured on a bulk sample using Thermo Scientific Niton XL3t hand held XRF and measuring table. Preliminary reflectance spectra measurements were done using an OL 750 automated spectroradiometric measurement system by Gooch & Housego equipped with a polytetrafluoroethylene (PTFE) and a gold integrating sphere, and with a specular reflection trap under atmospheric conditions. The illumination was provided using collimated deuterium (UV) or tungsten (VIS-NIR) lamps. Sample spectra were measured relative to the PTFE (UV-VIS) or the gold (NIR) standards.

22 additional meteorites (Fig. 3) were recovered between 9$^{th}$ and 12$^{th}$ October 2018, by a team comprised of the BGI, ORI, the Department of National Museum and Monuments (DNMM), the Department of Wildlife & National Parks (DWNP), and the Astronomical Society of Southern Africa (ASSA), now again assisted by the SETI Institute (Fig. 2b). The meteorites were curated at BGI (including weighing, documentation and storage) and photographed at DNMM.

Portions of eight Motopi Pan meteorites MP-04, -06, -09, -12, -13, -17, -18, and -19 (Fig. 3) were selected for petrographic and mineral chemical analysis. Based on their outward appearance, they seemed to represent most sample diversity in the collection. The eight stones were taken to NASA Johnson Space Flight Center, where ~1g fragments were cut using a new slowly rotating diamond blade and no lubricating coolant. Slices, fragments and powder were distributed among the consortium members.

For petrographic and mineralogic analysis, small fragments of MP-06, -09, -12, -18 and -19 were mounted in resin blocks. The polished mounts were studied at the University of the Witwatersrand by Backscattered Scanning Electron Microscopy (BSEM). 492 wavelength-dispersive mineral compositional analyses were collected using the *Cameca SX-Five* field emission Electron Micro Probe Analyzer (EMPA) in the Microscopy and Microanalysis Unit (Operator: A. Ziegler). Semi-quantitative mineral phase analysis was performed using a *Tescan Integrated Mineral Analyzer-X*, courtesy of Wirsam Scientific Solutions, Johannesburg (Operator: C. Stewart).

Reflection spectroscopy was performed at Brown University on both fragments (MP-04 and -06) and powders of MP-04, -06, -09, -12, -13 and -18, using methods described in Unsalan *et al.*

(2019). Thermal emissivity was determined on powdered samples MP-06 (diogenite), MP-09 (eucrite), and MP-18 (howardite) from Kirchhoff's law as $1 - R$, where $R$ is the hemispherical reflectance measured at room temperature at the Planetary Spectroscopy Laboratory of the German Aerospace Center (DLR) in Berlin, Germany using a *Bruker Vertex80V* Fourier-transform infrared instrument in hemispherical geometry (Maturilli *et al.* 2006). The illumination was provided using a collimated silicon carbide rod *Globar* lamp. Sample spectra were measured relative to a gold standard.

Thermoluminescence was performed at NASA Ames Research Center on part of a slice removed from MP-09. A chip was taken about 6 mm from visible fusion crust, and broken in two. These fragments were gently crushed, the magnetic fraction removed, and then gently crushed again to produce ~200 μm grains. The natural Thermoluminescence (TL) and induced TL were measured by the usual methods (Batchelor & Sears 1991).

The bulk elemental composition was measured at UC Davis in three sub-aliquots of Motopi Pan: MP-06 (diogenite lithology, 18.47 mg useful powder after crushing and homogenizing), MP-09 (eucrite, 20.29 mg), and MP-17 (howardite, 21.04 mg). The sample sizes are atypically smaller than what should have been used for the whole rock bulk compositional analyses (Mittlefehldt 2015), therefore a sampling bias could have been introduced. Each of the powder aliquots were placed into PTFE Parr capsules with a 3:1 mixture of concentrated $HF/HNO_3$. The PTFE capsules were sealed in stainless steel jackets and placed in a 190°C oven for 96 hours. The digested samples were then dried and re-dissolved in 1.1 mL of 6M HCl. 10% of this was used for elemental analysis. A further sub-sample was utilized for bulk isotopic measurements. The bulk samples of MP-06, -09, and -17 were diluted further by 2% nitric acid to dilution factors of 30249, 15262, and 4106 respectively for minor and trace elements and 60108, 59779, and 58114 respectively for major elements. The analytical procedures for bulk elemental composition measurements are described in Popova *et al.* (2013) and Unsalan *et al.* (2019).

Oxygen isotopes of all eight sampled Motopi Pan meteorites were measured at the University of New Mexico using methods described previously in Unsalan *et al.* (2019). At UC Davis, the Cr isotopic compositions were determined for aliquots of MP-06, -09, and -17 (90% of the sample dissolved). The samples were processed through a 3-stage column chromatography procedure to separate Cr from the sample matrix, as detailed in Yamakawa *et al.* (2009). The analytical procedure for measuring the Cr isotopic composition is detailed in Popova *et al.* (2013), Yamakawa *et al.* (2009), and Wiechert *et al.* (2004).

All isotopes of the noble gases He–Xe were analyzed in aliquots of MP-06, -09, -12 and -18 at ETH Zurich according to standard procedures (Riebe *et al.* 2017). Howardites often contain abundant solar wind ("sw") (Cartwright *et al.* 2013). To protect the system from gas overload in case howardite MP-18 contained SW, the sample was first measured as a small test sample. Gases were extracted at around 1700 °C in one step by fusion. Blanks for the main isotopes and typical corrections are given in the footnotes presented below the tables. Uncertainties of the concentrations include those of counting statistics, sample masses, blanks and detector sensitivity. Uncertainties of isotopic ratios include those of counting statistics, blank corrections, and instrumental mass discrimination. Those of cosmogenic and trapped concentrations include the uncertainties of the deconvolution, i.e., the range of chosen endmember components and all experimental uncertainties.

At UC Berkeley, 28–75 mg of each sample from MP-06, -09, -12, and -18 was dissolved in concentrated $HF/HNO_3$, along with Be and Cl carrier. Small aliquots of the dissolved samples that were analyzed for cosmogenic radionuclides were also analyzed by ICP-OES to obtain compositions of the main elements. Be, Al, and Cl were separated using methods given in Unsalan *et al.* (2019) and Welten *et al.* (2012). The $^{10}$Be/Be, $^{26}$Al/Al and $^{36}$Cl/Cl ratios of the samples were measured by Accelerator Mass Spectrometry at Purdue University's PRIME Lab (Sharma *et al.* 1990). The ratios were corrected for blanks (<1% of measured values) and normalized to $^{10}$Be, $^{26}$Al and $^{36}$Cl AMS standards (Sharma *et al.* 1990; Nishiizumi 2004; Nishiizumi *et al.* 2007). The $^{10}$Be, $^{26}$Al and $^{36}$Cl concentrations in the samples (in atoms/g) are converted to activities (in disintegrations per minute, dpm, per kg).

In-situ U-Pb analysis of MP-06 (diogenite), -09 (eucrite) and -17 (howardite) followed methods described in Popova *et al.* (2013), Unsalan *et al.* (2019), Liu *et al.* (2011), and Zhou *et al.* (2013). Backscattered electron (BSE) images of polished and carbon coated petrographic thick sections were obtained by the field emission scanning electron microscope (FESEM) of Carl Zeiss SUPRA-55 at the National Astronomical Observatories, Chinese Academy of Sciences in Beijing. U-bearing minerals grains, including zircon, merrillite, and apatite, were identified and located with an energy dispersive spectrometer (EDS). In-situ isotopic analysis of U-Pb was performed on a large-geometry, double-focusing secondary ion mass spectrometer, CAMECA IMS-1280HR ion microprobe at the Institute of Geology and Geophysics, Chinese Academy of Sciences. Here, U-Pb dating for zircon was conducted with a small primary beam of $O_2^-$ with a diameter both of ~5 µm and ~2 µm under mono-collector mode (Liu *et al.* 2011; Zhou *et al.* 2013). The primary ion beam of $O_2^-$ was accelerated at -13 kV potential, with an intensity of ~0.7 nA for ~5 µm diameter and ~0.1 nA for ~2 µm diameter, respectively. Before analysis, each spot was pre-sputtered using a ~3 nA primary beam on a square area of 25 × 25 µm for 120 s to remove the surface contamination and to enhance the secondary ions yield. $^{180}Hf^{16}O^+$ peak was used as reference for peak centering. $^{94}Zr_2{}^{16}O^+$, $^{204}Pb^+$, $^{206}Pb^+$, $^{207}Pb^+$, $^{208}Pb^+$, $^{238}U^+$ and $^{238}U^{16}O_2{}^+$ were measured on axial electron multiplier in peak jumping mode. Mass resolving power was set at 7,000 (50% peak height definition). For the ~5 µm spot size, each measurement consists of 12 cycles (taking nearly 18 minutes) and Pb/U fractionation was calibrated with the empirically established power law relationship between $^{206}Pb/^{238}U$ and $^{238}U^{16}O_2/^{238}U$ against standard M257 zircon with U ~840 ppm (Nasdala *et al.* 2008). For the spot size of ~2 µm, each measurement consists of 18 cycles and Pb/U fractionation was calibrated against standard Plesovice zircon (Sláma *et al.* 2008).

U-Pb dating for phosphate in MP-17 was performed with a 20 × 30 µm beam spot size. The $O_2^-$ primary ion beam was used with an intensity of ~7 nA. Positive secondary ions were extracted with a 10 kV potential. A mono-collector electron multiplier was used as the detection device to measure secondary ion beam intensities of $^{204}Pb^+$, $^{206}Pb^+$, $^{207}Pb^+$, $^{208}Pb^+$, $^{232}Th^+$, $^{238}U^+$, $^{232}Th^{16}O^+$, $^{238}U^{16}O^+$, $^{238}U^{16}O_2{}^+$ and a matrix reference peak of $^{40}Ca_2{}^{31}P^{16}O_3{}^+$ at a mass resolution of ~9000 (defined at 50% height). The $^{40}Ca_2{}^{31}P^{16}O_3{}^+$ signal was used as reference peak for tuning the secondary ions, energy, and mass adjustments. Pb/U ratios were calibrated with a power law relationship between $^{206}Pb^{*+}/^{238}U^+$ and $^{238}U^{16}O_2{}^+/^{238}U^+$ relative to an apatite standard of NW-1 (1160 Ma) that comes from the same complex of Prairie Lake as that of the Sano *et al.* (1999) apatite standard (PRAP). U concentration is calibrated relative to the Durango apatite which has U ~9 ppm (Trotter & Eggins 2006). The $^{206}Pb/^{238}U$ standard deviation measured in the standard was propagated to the unknowns. Each measurement for apatite U-Pb dating consists of 10 cycles, taking nearly 22 minutes.

Amino acids were analyzed in hot-water extracts of the meteorites MP-04, -06, -12, -17, and -19 and associated recovery site sands by liquid chromatography mass spectrometry at the NASA Goddard Space Flight Center using the same methods described in Unsalan *et al.* (2019). The total mass extracted of each sample is shown. Uncertainties in the abundances are based on the standard deviation of the average value of three separate measurements.

At the Helmholtz Zentrum Muenchen, fragments MP-04, -06 and -18 were extracted with LC/MS-purity grade methanol and the soluble organic fraction analyzed with direct infusion electrospray ionization Fourier Transformed Ion Cyclotron Mass Spectrometry (FTICR-MS) with ppb-level mass accuracy (with the precision of the mass of an electron), enabling the conversion of thousands of exact masses into their corresponding elementary compositions (Unsalan *et al.* 2019).

Finally, an analysis of Poly-Aromatic Hydrocarbons of MP-04, -06, -12, -13, -17 and -18 was conducted at the University of Toulouse. Grains were attached to sticky tape and analyzed with AROMA – the Astrochemistry Research of Organics with a Molecular Analyzer, following methods described in Sabbah *et al.* (2017).

## RESULTS: ASTEROID IN SPACE, IMPACT, AND METEORITE RECOVERY

### Asteroid photometry, shape, and spin period

Initial astrometry of asteroid 2018 LA was reported in the Minor Planet Electronic Circular 281-L04 (anonymous 2018a). Twelve asteroid positions were measured against the star background from Catalina Sky Survey observations, while the ATLAS survey added two more positions, extending the observation arc to 3.78 h. We searched archival data and found four more positions in SkyMapper Southern Survey *uvgriz* photometric bands (Wolf *et al.* 2018), which further extends the observing arc to 5.54 h (Fig. 1, Table 1). The time and location of the impact also refines the asteroid's orbit, as discussed below.

The SkyMapper observations provide insight into the shape and spin of the asteroid by showing a strong brightness variation in a single 20 s exposure and over the full 110 s interval. After the narrow-band SkyMapper data photometry was corrected for wavelength dependent albedo assuming a V-type spectrum, all 22 brightness measurements (Table 1) can be fitted by a sinusoid fit with a period of 112 ± 20s, with relative brightness amplitude 0.317 ± 0.050 (Fig. 4). A spinning asteroid has two brightness maxima per rotation, so the spin period is 224 ± 40s.

A triaxial ellipsoid was fitted assuming a 224s period. The minimum least squares error versus spin axis and axis ratios is shown in Fig. 5 for 3 orientations of the spin axis. The presence of valleys of equally most-likely axis-ratios in this diagram means the shape and orientation cannot be completely determined. If rotation is about the long axis any orientation is equally likely with c/b = 0.58, but on that line both cigar-shaped (b/a, c/a ≪ 1) and almost spherical (b/a, c/a ≈ 1) are equally likely. If the spin is about the short axis then spin axis orientations closer to R.A.= 0° or Dec. = 90° are more likely. The minimum aspect ratio of 1/0.58 = 1.7 occurs when a = b and the spin axis is oriented to R.A. = 0° or Dec. = 90°.

**Video and visual observations of the meteor**

Based on the measured orbit, 2018 LA reached Earth's atmosphere at an altitude of 100 km with a speed of 16.999 ± 0.001 km/s. The resulting bolide disrupted in a brief red flare, with fragments continuing further down, according to eyewitnesses.

Video security cameras in distant Maun, Rakops, Ghanzi, and Gaborone in Botswana, and Ottosdal in South Africa detected the fireball (Fig. 6). Directions were calibrated against that of stars using methods described in Popova *et al.* (2013), by placing the camera on the shadow (or along a straight line in the opposite direction) and imaging the foreground obstructing object that created the shadow against a star background. Results are tabulated in Table 2.

Triangulation of these directions puts the flare at Lon. = 23.287 ± 0.057°E, Lat. = 21.251 ± 0.007°S, and altitude = 27.8 ± 0.9 km, above a location and with uncertainty intervals outlined by the red ellipse in Fig. 7. The synchronized time of the flare is 16:44:11.5 ± 3.0s UTC, based on 6 video security systems with low time drift.

The light curve of the meteor was calibrated against the brightness of foreground lamps in the Gaborone video, using reverse binocular and aperture photometry. Multiple exposures of the lamps were made at a later date when the Moon was also in the field of view, using a Nikon DSLR camera (Fig. 8), and then the Moon was used as a comparison to determine the lamp brightnesses in the fireball video since its magnitude at the time of exposure was known (Table 3). The fireball's flare peaked at -23 magnitude (Fig. 9).

Fig. 9 compares the resulting lightcurve to that of the Sariçiçek bolide. When 2018 LA disrupted, it generated more light than Sariçiçek. The disruption altitudes are the same: While Motopi Pan disrupted at 27.8 ± 0.9 km, Sariçiçek disrupted at 27.4 ± 1.4 km altitude (Unsalan *et al.* 2019).

U.S. Government sensors (anonymous 2018b) detected the bolide from space, peaking in brightness at 28.7 km altitude at 16:44:12 UTC, when its speed was 16.9 km/s (no uncertainty ranges provided). The calculated total radiated energy was $37.5 \times 10^{10}$ J, and given as equivalent to 0.98 kt kinetic energy (i.e., luminous efficiency 9.1%) (Tagliaferri *et al.* 1994). However, a lower 0.3–0.5 kt energy was derived from the oscillation frequency of the shock wave when it arrived at the infrasound station I47ZA in South Africa (see next paragraph), suggesting the luminous efficiency may have been higher in this case.

**Infrasound observations of the meteor**

Fig. 10 shows details of the infrasound signal. Examining all detections above background from the Comprehensive Nuclear-Test-Ban Treaty Organization infrasound station I47ZA on June 2, the northward directed signal from this event arriving at 17:30 UTC stands out from the background of infrasound coherent sources typically detected at I47ZA, based on back-azimuth directions alone. The signal shows high correlation and a signal over noise ratio close to 85. Intersection with the ground path of 2018 LA (given predicted impact times) yields a celerity of 0.29 km/s, consistent with a stratospherically ducted infrasound return.

Important features are the broadband nature of the signal and dominant periods between 3–4 seconds (Fig. 10), both of which indicate a 0.3–0.5 kt explosive atmospheric source at moderate (<1000 km) range, using the calibration in Ens *et al.* (2012).

The infrasound begins arriving at I47ZA from lower azimuths (around 352–353°, depending on the frequency band) and moving to progressively higher azimuths during the 5 minutes of the signal. This suggests the earliest arrivals are from locations along the fireball trajectory furthest west (possibly point of fragmentation) and then sweeping backward along the trajectory to higher elevations. The best estimate for the origin time (correlating with the 2018 LA trajectory) for this back azimuth is 16:44 UTC, in agreement with that of video observations of the flare.

**Asteroid trajectory and orbit**

To better define the asteroid position in its orbit, the timing from USG satellite data (assuming an uncertainty of ± 1s) was combined with the video-derived location of the disruption. From the astrometry of the asteroid (anonymous 2018a), using methods described in (Farnocchia *et al.* 2016), and combining with the position of the main disruption from video observations (at Lon. =

23.287 ± 0.057 °E, Lat. = 21.251 ± 0.007 °S, Alt = 27.8 ± 0.9 km, 2σ errors) and the time from USG satellite data (16:44:12.0 ± 1.0s UTC, assumed 1 σ) results in the atmospheric trajectory tabulated in Table 4, with residuals given in Table 5 and orbital elements listed in Table 6 (JPL Solution 8).

Based on this solution, on June 2 at 16:44:11.74 ± 0.98s UTC (Julian Date 2458272.1981589 TDB), the asteroid was at an altitude of 27.98 ± 0.87 km. For a fixed altitude of 27.98 km, the geodetic coordinates are Long. = 23.2804 ± 0.0200 °E, Lat. = 21.2457 ± 0.0056 °S, and the corresponding footprint has major / minor uncertainties of 2.14 / 0.34 km, respectively, and a major axis to azimuth of 104.2°. This is shown as a black ellipse in Fig. 7 above. At that altitude, 2018 LA moved with a speed of 17.040 ± 0.001 km/s towards Az. = 275.70 ± 0.01° and El. = 24.12 ± 0.01° (1σ errors).

The calculated orbit has a short semi-major axis a = 1.37640 ± 0.00011 au, similar to Sariçiçek's a = 1.454 ± 0.083 au (Unsalan *et al.* 2019), but was measured 750 times more precisely. The inclination i = 4.29741 ± 0.00043° is also much more precise than Sariçiçek's higher i = 22.6 ± 1.6°. As a result, Motopi Pan's position in the a-i diagram of Fig. 11 is well defined. Puerto Lápice arrived from a > 1 AU and i < 12° (Trigo-Rodríguez *et al.* 2009).

**Meteorite recovery and strewn field**

Falling from 27.8 km, surviving fragments drifted off course by upper atmosphere winds, strength and direction of which were taken from the European Centre for Medium-Range Weather Forecasts (ECMWF) model (Fig. 12). The wind direction on the ground in Maun at the time of the meteor was mild (2 m/s) and mainly from ENE (70° Az from North). For higher altitudes, the ECMWF model was used to calculate the along-track and cross-track wind speed and direction, as well as the air pressure, humidity and temperature. Strong along-track westerly winds of ~40 m/s were present between 5 and 18 km, which blew the meteorites back on-track.

From the derived disruption point at 27.8 km altitude, a dark flight simulation was performed for different assumed size fragments and initial speeds at disruption. A Finnish Meteor Network team calculated the strewn field location (Fig. 13) based on the asteroid astrometry and directions from the video at Ottosdal (Moilanen *et al.* 2021). Independently, the SETI Institute and ORI team determined the location of the strewn field (Fig. 14) by triangulating local video camera observations from Maun and Rakops, later adding calibrated observations from Ottosdal and Gaborone using methods in Popova *et al.* (2013).

The find location of recovered meteorites in the strewn field are shown in Fig. 14 and tabulated in Table 7. A brief description of each meteorite's appearance is given in Table 8. The fall area is located in the CKGR, and consists of sandy dunes with sporadic, sometimes dense, tall grass and shrubs, located 8–20 km from the nearest road, and is frequented by large game animals.

## METEORITE PETROGRAPHY AND MINERALOGY

### Non-destructive analysis

Two months after the first find of MP-01 (close to the location of the second camping site in Fig. 14), gamma-ray emissions of cosmogenic radionuclide $^{46}$Sc, with a half-life of 83.8 days, were detected at the Gran Sasso National Laboratory, confirming that Motopi Pan originated from 2018 LA (Table 9). Hence, the meteorite cannot be a relic of the Kalahari 008/009 lunar achondrites that were found 40 km further to the north-west in 1999 (Sokol *et al.* 2005).

Characteristic absorption bands of HEDs were measured in 400–2500 nm reflectance spectroscopy on the bulk meteorite at a spot where the interior was largely exposed, and handheld X-ray fluorescence showed element values of Ni versus Fe/Mn indicative of a HED-like composition. Other non-destructive studies of MP-01 at the University of Helsinki established a bulk density of 2.85 ± 0.01 g/cm$^3$ based on a laser-scan derived volume, compared to 2.86 g/cm$^3$ for eucrites (Britt & Consolmagno 2003). Combined with gas pycnometry, this provided a 3.26 g/cm$^3$ grain density and 13% porosity. The log of magnetic susceptibility (in $10^{-9}$ m$^3$/kg) log $\chi_m$ = 2.66 ± 0.05 is characteristic of low-Fe content eucrites (2.75 ± 0.38) (Rochette *et al.* 2008). The meteorite is weakly magnetized at 5.36 ± 0.27 (vol) mA/m, below values typically found in terrestrial basalts.

X-ray computed micro-tomography at 14 and 7 μm/voxel indicated a composite of fine-grained material and coarse-grained clasts of denser material (Fig. 15). The clasts are not distributed homogeneously within the sample, being more abundant at the upper part of the meteorite. The cross-sectional contours of 11 major coarse-grained clasts were manually defined by polygonal selections in every ~10th slice. The contours were then linearly interpolated for the rest of the slices to calculate the volumes of the clasts. The single-crystal clasts in the fine-grained matrix were not included in the total clast volume. The ratio of the clast to bulk volume is 15%. A conservative uncertainty is ±5%. The coarse-grained clasts are likely the diogenitic components (mainly pyroxene) in this HED. If the single crystal clasts are diogenite, then the 15% abundance identifies the meteorite as a howardite (defined as an HED with ≥ 10% diogenitic component).

**Destructive analysis**

The variety in surface color and textures of individual meteorites is large (Table 8, Fig. 3 above). Based on outward appearance alone, at least eight different lithologies appear to be present. One example of each was sampled. Five meteorites were analysed for petrography and mineralogy.

*Sample MP-06: Diogenite*

The BSE image of MP-06 is shown in Fig. 16. This is a relatively coherent fragment (7×4 mm, 0.04 g) with moderate-to-low subplanar to curved fractures. **Mineralogy:** Low-Ca $Pyx_{94}$-$Pl_{5}$-$chromite_{1}$, accessory olivine, troilite and kamacite. In this and other descriptions below, "Pyx" refers to Pyroxene, "Pl" to Plagioclase, "Ol" to Olivine. **Texture:** well-equilibrated, phaneritic Pyx (≤ 1 μm grain size) with absence of compositional zoning or exsolution; ≤ 0.5 μm interstitial, anhedral Pl showing strongly cuspate grain boundaries; anhedral Ol (< 0.1 μm) occurs associated with troilite (0.05–0.15 μm) and kamacite (≤ 0.1 μm) as inclusions in Pyx; subhedral, rounded-elongate chromite (0.1–0.2 μm) occurs interstitially with Pl and anhedral, irregular-elongate troilite (0.1-0.15 μm). **Mineral compositions:** Pyx, Pl and Ol are unzoned; Pyx ($Wo_{3.4}En_{70}$) is the most magnesian (Mg# = 72; Mg# = 100* molar Mg/[Mg+Fe]) and (marginally) least calcic low-Ca Pyx analyzed in the sample suite (Table 10); plagioclase is only very slightly less calcic ($An_{94.7}$) than in MP-09 and MP-19; chromite is the most magnesian (Mg# = 17) in the sample suite (Table 10); **Interpretation and name:** Based on the previous, this meteorite is an Orthopyroxenite (diogenite).

*Sample MP-09: Cumulate eucrite*

A BSE image of MP-09 is shown in Fig. 16. This is a highly friable, fractured, elongate 7×3 mm 0.10g fragment with dominant fracture set parallel to length (< 0.4 μm spacing); no fusion crust visible. **Mineralogy:** low-Ca $Pyx_{40}$-$Pl_{60}$, trace chromite and troilite. **Texture:** well-equilibrated, equigranular, phaneritic (~1 μm grain size) Pyx and Pl with no compositional zoning or exsolution; anhedral, fine-grained (< 0.05 μm) chromite occurs within and adjacent to Pyx; troilite (≤ 0.02 μm) occurs within Pyx in small clusters. **Mineral compositions:** Compositionally homogenous low-Ca Pyx ($En_{62}Wo_5$, Mg# = 66) and Pl ($An_{95.6}$); compositions similar to MP-19. **Interpretation and name:** Norite (cumulate eucrite).

*Sample MP-19: Cumulate eucrite*

A BSE panorama is shown in Fig. 17. This is a triangular 10×8 mm 0.52g fragment displays a vesiculated fusion crust usually < 0.05 μm thick but locally showing lobes up to 0.2 μm; sample is moderately to strongly fractured, with fractures locally exploiting mineral cleavages. **Mineralogy:** low-Ca $Pyx_{44}$–$Pl_{55}$-$chromite_1$. **Texture:** well-equilibrated, equigranular, phaneritic (~1 μm grain size). Pyx shows no exsolution. Chromite occurs as subrounded-elongate subhedral interstitial grains of ≤ 0.6 μm length, usually in contact with Pl. A ≤ 1 μm wide, fracture-controlled melt vein displays a heterogeneous composition in BSE, and rounded to elongate flow-aligned Pyx±Pl microclasts. **Mineral compositions:** low-Ca Pyx ($Wo_4En_{61}$, Mg# = 64) and Pl ($An_{95.4}$) are homogenous (Table 10); textures and compositions resemble those in MP-09. **Interpretation and name:** Norite (cumulate eucrite) with shock/friction melt vein.

*Sample MP-12: Eucrite breccia*

A BSE panorama is shown in Fig. 17. This is an elongate-rhomboid fragment (8×5 mm, 0.22g) with long edges parallel to spaced branching-planar fracture set; fractures are most visible in large clasts. Vesicular fusion crust (≤ 0.2 μm) extends along ~50% of the fragment edge. **Mineralogy:** low-Ca $Pyx_{45}$-high-Ca $Pyx_9$–$Pl_{42}$–silica $phase_3$–$Ilm_{0.5}$–$troilite_{0.5}$ + accessory chromite, phosphate. Low-Ca pyroxene dominates; high-Ca pyroxene mainly occurs in fine exsolution lamellae and in poikilitic Pl-Pyx-silica phase-troilite-Ilm clasts. **Texture:** breccia comprising mostly angular to subrounded mineral clasts (low-Ca Pyx, Pl mostly > 0.2 μm; Ilm, chromite < 0.2 μm) and several polymineralic (lithic) clasts (up to 1–3 μm in size) displaying cumulate Pyx-Pl or elongate lath textures. Texturally-complex fine-grained Pl-Pyx-silica phase clasts with fine-grained euhedral Pl and high-Ca Pyx laths (≤ 0.1 μm) contain disseminated very fine-grained troilite + Ilm (0.02–0.1 μm). The largest (3 μm) lithic clast is cut by a fracture-hosted melt vein with a compositionally heterogeneous matrix that is truncated by the clast edge. A subangular clast (0.5 μm) consists of low-Ca Pyx-Pl microbreccia. The matrix of the overall breccia is fragmental and comprises irregular, angular to conchoidally-fractured and locally elongate mineral fragments < 0.05 μm in size, with local *in situ* disaggregation of larger mineral fragments. Plagioclase dominates over pyroxene in the finest fraction, with oxides and sulphides being rare. **Mineral compositions:** The compositions of both Pyx and Pl are highly variable (Table 10), consistent with the textural evidence of derivation from both plutonic/cumulate and hypabyssal/basaltic sources; however, some low-Ca Pyx clasts additionally show internal zoning in which Mg# decreases, and Wo content increases marginally, towards both grain edges and intragranular fractures. Cores are characterised by Mg# > 71 and $Wo_{<5}$ whereas rims display Mg# < 47 and $Wo_{2–8}$. Unzoned, coarse low-Ca Pyx grains with Mg# ~ 57 and $Wo_{<6}$ typically display fine (< 10 μm) high-Ca Pyx exsolution lamellae (Mg# ~ 58, $Wo_{\leq 38}$). Pl compositions vary in the range $An_{70-95}$, with lowest values in the clasts with finer-grained lath-shaped grains. Matrix Pl and Pyx compositions are consistent with derivation from similar sources to those of the clasts. Fine-grained plagioclase laths in the Pl-Pyx-silica phase-troilite-Ilm clast lie within the same range but have Ca contents towards the higher end of this range ($An_{86–91}$). Chromite clasts have an average Mg# ~ 4.7, consistent with a less magnesian source than the cumulate eucrite and diogenite fragments. **Interpretation and name:** Eucrite breccia derived from lithology/lithologies that are less magnesian and slightly less calcic than MP-06, -09 and -19. Evidence of shock/friction melt in the largest fragment and diffusional re-equilibration in low-Ca Pyx indicate post-crystallization modification prior to breccia formation.

*Sample MP-18: Eucrite breccia*

A BSE panorama is shown in Fig. 17. This is a slightly elongate angular fragment of 0.03 g with highly vesicular fusion crust up to 0.5 mm thick; triangular internal fracture defines a large breccia clast that displays a smoother overall polish than surrounding material. **Mineralogy:** low-Ca $Pyx_{53}$–high-Ca $Pyx_{3}$ –$Pl_{40}$-silica $phase_{3}$-$Ilm_{0.5}$-$troilite_{0.5}$, accessory chromite**. Texture:** Breccia dominated by < 0.5 μm mineral clasts containing a 4×3 μm triangular, older, breccia clast (Breccia A) within a slightly finer-grained breccia (Breccia B; Table 11).

Breccia A matrix is more strongly sintered/recrystallized than Breccia B. Some Pl clasts in Breccia A display < 0.01 μm aligned Pyx inclusions. Pyx clasts display more abundant and coarser (≤ 30 μm wide) exsolution lamellae than in MP-12; this Pyx texture is present in both breccias. Ilmenite (≤ 0.6 μm) is the primary opaque mineral clast phase. **Mineral compositions:** Both Pl and Pyx clasts and matrix show similar compositional ranges to those found in MP-12. For Pl, the coarser clasts in both Breccia A and Breccia B and the poikilitic Pl-Pyx-Ilm-troilite clasts lie in the range $An_{81-90}$, with finer-grained laths in lithic clasts being slightly less calcic ($An_{76-82}$). Matrix Pl ranges from $An_{78-92}$. Compared with MP-12, a similar range of Pyx textural types has been identified. Zoned Pyx cores range from $En_{70}Wo_{2.5}$; Mg# > 70 to rims of $En_{54}Wo_{3}$; Mg# ~ 50. Some low-Ca Pyx compositions are similar to those noted in MP-06, MP-09 and MP-19. As in MP-12, the Pyx displaying exsolution lamellae is less magnesian and more calcic than the zoned Pyx. High-Ca Pyx lamellae display Wo values as high as $Wo_{40}$, however, the generally fine nature of lamellae commonly leads to mixed EPMA results; Mg# in both lamellae and the low-Ca Pyx host typically lie in the range 40-50. Low-Ca Pyx in the poikilitic Pyx-Pl-silica phase-Ilm-troilite texture displays $Wo_{>40}$ and Mg# ~ 60. **Interpretation and name:** Eucrite breccia comprising components exclusively / predominantly derived from cumulate eucrite, but via at least two brecciation events.

*Overall assessment of meteorite type*

The eucritic breccias MP-12 and MP-18 were examined more closely to determine whether they are howardites, wich are defined by containing more than 10% diogenetic clast material (Delaney *et al.* 1983). Fig. 18 shows Mg-K X-ray maps of these fragments. Both contain low-Ca Pyx clasts with Mg-zoning whose cores display Mg abundance values consistent with the diogenite sample MP-06. The lack of distinctive differences in Pl composition between diogenites and eucrites preclude its use as an indicator. However, if the pyroxene population alone is considered as a proxy for the sampled lithologies, MP-18 has > 10% diogenite component, classifying it as a howardite. In a similar way, if the very small high-Mg cores in the largest lithic clast that constitutes ~20 vol% of sample MP-12 are considered relics of almost completely re-equilibrated pyroxenes that once had a uniform high-Mg composition, MP-12 would also be classified as a howardite.

Based on visual inspection of the other Motopi Pan meteorite samples collected, compared with those studied petrographically, they are classified as per Table 12. Factoring in their relative sizes, and assuming the other diogenite and eucrite samples are part of a larger breccia, we conclude that the Motopi Pan meteoroid constituted a howardite.

## METEORITE PHYSICAL PROPERTIES

### Visible and near-infrared reflection spectroscopy

The visible appearance of the fragments and powders studied for visible and near-infrared spectroscopy is shown in Fig. 19. Fragments of sample MP-06 (diogenite) are slightly green in color. Note the strong albedo differences between eucrite sample MP-09 and diogenite samples MP-06 and MP-13.

Fig. 20 shows the reflectance spectra over the 0.3–9 μm wavelength range of fragments MP-04 and MP-06 (top), and in powdered form with sizes <45 μm (middle). The visible to near-infrared spectra of most samples show the characteristic mineral absorption bands of HED meteorites in shape and wavelength position, which depend on the pyroxene composition and crystal structure (e.g., Bancroft & Burns 1967; Moriarty & Pieters 2016). The spectra of fine powders show higher reflectances and redder slope than their original fragments, which is a typical trend for rocks made of relatively transparent minerals. The Christiansen Feature (CF) wavelengths of MP-04 (eucrite) at 8.1 μm and MP-06 (diogenite) at 8.5 μm exhibit a clear distinction due to their difference in mineral composition, with eucrite having abundant plagioclase whereas diogenite is dominated by orthopyroxene. The bottom panel of Fig. 20 shows the reflectance spectra of unsorted fine powders from the cutting of MP-09, -12, -13, and -18. Howardites MP-12 and MP-18 are spectrally very similar to each other over the entire wavelength range, except that MP-18 shows a closer CF wavelength to diogenite MP-13, suggesting a higher diogenite mineral fraction.

Fig. 21 shows the corresponding mid-infrared reflectance spectra. Freshly-ground powder samples of eucrite MP-04 and diogenite MP-06 show very distinct spectra (left panel) over this entire wavelength range. On the right panel, eucrite powder MP-09 shows a similar spectrum to the other eucrite MP-04. Howardites powders MP-12 and MP-18 show very similar spectra to each other in the same manner as in the shorter wavelength range in Fig. 20. However, the unsorted cutting dust of diogenite MP-13 shows a strange spectrum that is similar to diogenite powder MP-06 over the wavelength range of 16–22 μm but is more similar to eucrite powders M-04 and MP-09 over the wavelength range of 9–15 μm. The cause for this is not clear. The MP-13 spectrum shows low reflectances and very weak 1, 2, and 3 μm absorption bands. The MP-13 sample may have some contaminations or its extremely small particle size may affect the spectrum.

The 3 μm band due to adsorbed or absorbed water is very strong in MP-09 (Fig. 20), while the 3.4–3.5 μm bands due to organics are very strong in MP-12 and MP-18 spectra. Both bands are most likely due to terrestrial contamination. The distorted 2 μm band of MP-09 also indicates that the narrow 1.9 μm $OH/H_2O$ is overlapped with the broad absorption band due to $Fe^{2+}$ in the M2 site of pyroxene crystal structure (*e.g.*, Bancroft & Burns 1967).

Fig. 22 compares the reflectance spectra of the Motopi Pan samples (black solid and dashed lines) to that of other HED meteorites measured with the same facilities and archived in the Reflectance Experiment Laboratory (RELAB) spectral database. Meteorites classified as diogenites are shown with a thin red line. Those classified as basaltic eucrites are shown in blue, and tend to have that band shifted to longer wavelength (~2.05 μm, Fig. 22). Howardites are in green and have a range of band positions in between those two extremes, as have cumulate eucrites. Diogenite MP-06 has a typical diogenite band center wavelength (1.89 μm). MP-13 has a band center at 1.93 μm. Howardites MP-18 (1.97 μm) and MP-12 (1.99 μm) resemble other howardites. Cumulate eucrite MP-09 resembles that of cumulate eucrites (1.95 μm). MP-04 (solid line), based on mid-IR features and magnetic susceptibility (see below), is classified as an eucrite (Table 12), but resembles diogenites instead in its 2 μm band position (1.92 μm). In good agreement, spectra of Vesta and its Vestoids span the range 1.90–2.02 μm, with most sites in the howardite range 1.93–1.99 μm (De Sanctis *et al.* 2013). Some V-class asteroids that do not belong to the Vesta family have lower or higher band positions (Migliorini *et al.* 2017).

**Thermal emissivity spectroscopy**

The three different HED types may have differently ablated during entry. Results from thermal emissivity measurements in Fig. 23 show that diogenites have a lower thermal emissivity. Less material of this type may have survived to the ground. If so, this result only strengthens the classification of howardite.

Note that in the emissivity spectra of all three samples, the Christiansen frequency and transparency features can be identified around 8–9 μm and 12–18 μm, respectively (Fig. 23). The transparency feature is a minimum in emissivity that occurs in fine powders between the silicate stretching and bending modes. The Christiansen frequency features occur where the refractive index of the mineral samples undergoes rapid change and approaches the refractive index of air, resulting in minimal scattering. They occur in silicates at wavelengths just short of the Si-O band.

The Si-O Reststrahlen bands are partially visible in the diogenite MP-06. Reststrahlen is a reflectance phenomenon in which the light cannot propagate a medium due to a change of refractive index concurrent with a specific absorbance band (here: Si-O). Their partial appearance in the MP-06 sample may be explained by its relatively coarser diogenite crystalline structure compared to eucrites and howardites.

**Thermoluminescence**

The natural TL glow curves of MP-09 has two or three visible peaks and thus the LT/HT ratio can be determined (Table 13). The Natural TL value (NTL) is exp[20303 × (log(LT/HT) + 0.884)/0.775] = 20 ± 2 krad. Anomalous fading causes values to be low by about 15% (Sears *et al.* 1991), which means NTL = 24 krad for MP-09. This value of natural TL is consistent with the meteoroid not having approached the sun much below 0.6 AU (Benoit & Sears 1997).

The induced Thermoluminescence Sensitivity is also listed in Table 13. When normalized to Dhajala equals 1000, the resulting TL Sensitivity is 6,050 ± 968. Fig. 24 compares that result to other HED meteorites (Sears *et al.* 2013), including Sariçiçek (Unsalan *et al.* 2019). MP-09 has the highest TL Sensitivity measured to date for a HED meteorite, significantly different from Sariçiçek.

**Magnetic susceptibility**

A portable SM30 magnetic susceptibility meter of ZH Instruments was used to measure the magnetic susceptibility of the meteorites. Repeated measurements of bulk meteorites were made using the method in Gattacceca & Rochette (2004). Results are tabulated in Table 14. Diogenites have the higher magnetic susceptibility, eucrites the lowest. The measured values correlate well with petrographically determined meteorite types.

## METEORITE COSMOCHEMISTRY

**Elemental composition**

The elemental composition results are listed in Table 15. Compared to data reported in Mittlefehldt (2015), MP-06 and MP-09 have abundances similar to polymict breccias rather than their diogenite and cumulate eucrite end members, respectively. The abundances of howardite MP-17 are generally consistent to those of Sariçiçek SC14, while those of eucrite MP-09 are close to SC12 (Unsalan *et al.* 2019). Differences could point to different amounts of exogenous material in Motopi Pan than Sariçiçek, to terrestrial contamination of the sample in the field or during handling (Na, K), or due to lack of phosphate in this small sample, an observation supported by our extensive search of phosphate minerals for U-Pb dating (see below), but in vain in both MP-06 and MP-09.

### Oxygen isotopes

Oxygen isotopes were measured for all eight meteorites sampled (Table 16). Figure 25 compares the $^{17}O/^{16}O$ and $^{18}O/^{16}O$ oxygen isotope ratios for Motopi Pan to those of Sariçiçek. The normalized values $\delta^{17}O'$ and $\delta^{18}O'$ scatter along a line parallel to the Terrestrial Fractionation Line (TFL). Colored symbols mark the different meteorite types (Table 12). The howardites in Motopi Pan span the full range of values of their end members. On average, the Motopi Pan samples plot to higher values than Sariçiçek (Unsalan *et al.* 2019).

### Chromium isotopes

Chromium isotopes for Motopi Pan fractions MP-06, -09, and -17 were respectively $\varepsilon^{54}Cr$ (± 2SE) = -0.60 ± 0.13, -0.65 ± 0.08, and -0.40 ±0.10. Fig. 26 shows the chromium-oxygen isotope diagram. Mixing lines are shown indicating the mixing trajectory from average eucrite composition to CR, CM, and CV carbonaceous chondrite end-members. Each tick mark on the mixing line represents a 2% incremental increase in the amount of carbonaceous chondrite component added to the average eucrite composition. Unlike Sariçiçek, howardite MP-17 results suggest an about 7% admixture of CR-type material.

### Noble gasses and Cosmic Ray Exposure Age

Helium in all samples is entirely cosmogenic (“cos”, all $^{3}He$, some $^{4}He$) and radiogenic (“rad”, $^{4}He$) (Table 17). There is no evidence for a trapped (“tr”) component. None of the samples contained solar wind. This is perhaps somewhat surprising, as the polymict breccia nature of Motopi Pan may suggest an origin from the surface of its parent body. Cartwright *et al.* (2013) determined that about a third of all howardites contain a solar wind component and, hence, suggested a regolith origin for them.

Neon isotopes (Table 17) are shown in Fig. 27. While the three howardite samples MP-18S, -18L and MP-12 show identical, purely cosmogenic Ne isotopic compositions, consistent with similar chemistry and shielding of these samples, the diogenite MP-06 shows a higher $^{21}$Ne/$^{22}$Ne, in agreement with the much higher Mg content of diogenites compared to eucrites and howardites (Table 22). The eucrite MP-09 is the only fragment with a very small $Ne_{tr}$, possibly an atmospheric component (see below).

Combining the samples' chemistry with the shielding parameter ($^{22}$Ne/$^{21}$Ne)$_{cos}$, as measured in each sample, constrains the shielding of each sample and the meteoroid's pre-atmospheric radius. We used the physical model by Leya & Masarik (2009) and the bulk chemistry for the four fragments given in Table 22 below (here assumed to be more representative for the major target element chemistry than the data in Table 15). We used the Na concentrations from Table 15, as these are not given in Table 22. Minimum radii based on ($^{22}$Ne/$^{21}$Ne)$_{cos}$ are 37 cm for the diogenite and 25 cm for the other samples, close to the minimum radius of 40 cm determined by radionuclides (see below). Cosmogenic Ne alone would allow much larger radii than the radionuclides, as the ratio ($^{22}$Ne/$^{21}$Ne)$_{cos}$ is not very sensitive to radii >50 cm. If restricting ourselves to radii of 40–80 cm based on the radionuclides (next section), the samples would originate from 23–49 cm (diogenite), 17–66 cm (eucrite) and 16–80 cm (howardites) depth.

Argon, Krypton, and Xenon isotopes are given in Tables 18–20. Only the eucrite sample contains significant trapped Ne-Xe (Tables 17,20). Trapped $^{36}$Ar/$^{132}$Xe, $^{84}$Kr/$^{132}$Xe, $^{40}$Ar/$^{36}$Ar and $^{129}$Xe/$^{132}$Xe ratios suggest that this trapped gas is air, implying that this eucrite fraction of Motopi Pan was more affected by weathering. Cosmogenic Kr and Xe is hence not detectable in the eucrite sample. This is true also for the diogenite, whereas the three howardites contain detectable cosmogenic Kr and Xe. This is consistent with the larger concentrations of the target elements Rb, Sr, Y, Zr, Ba and REEs in howardites compared to in diogenites (Mittlefehldt 2015). Cosmogenic neutron-induced excesses in $^{80,82}$Kr and $^{128}$Xe are not discernible, in line with the suggested low depth of the samples in the meteoroid and low abundances of the halogens Br and I in HED. Likewise, excesses in short-lived $^{129}$I-derived radiogenic $^{129}$Xe are not detected. Fission Kr and Xe are detectable in all howardite samples, consistent with high U concentrations in howardites compared to diogenites. Eucrites and howardites have similar Rb, Sr, Y, Zr, Ba, REEs and U concentrations and we expect the eucrite to have concentrations of cosmogenic and fission Kr and Xe similar to those of the howardites. However, these Kr and Xe signatures are overprinted by the large amounts of air, Kr, and Xe contamination.

The resulting production rate and cosmic ray exposure (CRE) age ranges are given in Table 21. Apart from the CRE ages for $^{3}$He, and that for $^{21}$Ne in the eucrite sample, which appear to have been lowered by diffusive loss, the CRE ages in the scheme of Leya & Masarik (2009) average to 19.2 ± 2.4 Ma. For comparison, we also give between brackets in Table 21 the CRE ages based on Eugster & Michel (1995). These are about 23% higher and range from 16 to 26 Ma with a mean of 22.8 ± 3.8 Ma.

In comparison, the 4π age for Sariçiçek was ~22 Ma (Unsalan *et al.* 2019). The latter was determined with the production rate systematics given by Eugster & Michel (1995) for howardites after removing a 2π exposure component as measured by its high solar wind content. The uncorrected values were in the range 27–31 Ma. The lack of solar wind in Motopi Pan suggests no such 2π-exposure correction is warranted.

## Cosmogenic nuclides and preatmospheric size

### *Chemical Composition*

The mixing ratio of eucrite to diogenite in HED breccias can be expressed as the mass Percentage Of Eucrite Material (POEM), which is based on the measured Al and Ca concentrations of the samples (Fig. 28), using the average of the two calculations (Mittlefehldt *et al.* 2013). The POEM values range from 12 wt% for the diogenite-rich sample (MP-06), consistent with the classification as a diogenite rich clast, to 97 wt% for the eucrite-rich sample (MP-09), Table 22. For the latter only the Ca-based POEM was used, as the unusually high Al concentration would result in an unrealistic value of 136 wt%.

### *Cosmogenic Radionuclides*

The concentrations of $^{10}$Be, $^{26}$Al and $^{36}$Cl are near expected saturation values, consistent with a noble gas derived CRE age >10 Ma as discussed in the previous section. Since the radionuclide concentrations represent saturation values, they can directly be compared to calculated production rates from the model of Leya & Masarik (2009) to estimate the pre-atmospheric size and depth of the Motopi Pan samples, and thus the size of asteroid 2018 LA before impact. We used the elemental production rates of Leya & Masarik (2009) in ordinary chondrites as a function of size and depth and adjusted the radius and depth (in cm) for the 20% lower density of the MP samples relative to ordinary chondrites (2.85 vs. 3.55 g/cm$^3$).

The $^{36}$Cl concentration variations are mainly due to variations in chemical composition of the samples, since cosmogenic $^{36}$Cl in HED meteorites is mainly produced from Fe and Ca, with minor contributions from K, Ti, Cr and Mn. Fig. 29a shows the measured $^{36}$Cl concentrations as a function of the chemical composition of each sample, which is expressed as Fe + 10 Ca + 50 K [in wt%]. This normalization is based on the concentrations of major target elements (K, Ca, Fe) for the production of $^{36}$Cl and the assumption that in meter-sized objects the $^{36}$Cl production rates from Ca and K are approximately 10× and 50× higher than from Fe (Leya & Masarik 2009). Contributions from Ti, Cr and Mn probably comprise another 1–3% of the total $^{36}$Cl production, but are insignificant relative to the uncertainty in the relative contribution from Ca (which is shielding dependent).

The dashed line in Fig. 29a represents a linear fit through the data, corresponding to an average $^{36}$Cl concentration of 24.5 dpm/kg[Fe+10Ca+50K]. The $^{36}$Cl concentrations in MP are up to ~5% higher than the maximum calculated $^{36}$Cl production rates of 23–24 dpm/kg[Fe+10Ca+50K], which occur in the center of HED-like objects with R = 38–62 cm (Leya & Masarik 2009). The uncertainty in the model production rates is ~10–15%. Fig. 29b shows the model depth profiles of Leya & Masarik (2009) if we increase the $^{36}$Cl production rates by 10% as we did previously for Sariçiçek (cf., Unsalan *et al.* 2019). $^{36}$Cl concentrations are consistent with sample depths of 10–25 cm in an object with a pre-atmospheric radius of 40–60 cm, or with depths of 15 cm to the center of an object with a radius up to ~80 cm.

Cosmogenic $^{10}$Be in stone meteorites is mainly produced from O, Mg, Al and Si, with minor contributions from heavier elements (Ca through Fe). Since reaction on O dominate the $^{10}$Be production and the O content is relatively constant at 43–45 wt% in HED meteorites, the $^{10}$Be concentrations are relatively independent of the chemical composition of the sample within the same meteorite; any variations in $^{10}$Be mostly reflect different shielding depth. Comparison of the measured values with calculated $^{10}$Be production rates from the model of Leya & Masarik (2009) are shown in Fig. 30. Ca concentrations are relatively high in the Motopi Pan howardites (6–7 wt%), but production rates of $^{10}$Be from Ca are not included in the model. We estimated elemental production rates from Ca by averaging those from Si and Ti. This increases the total $^{10}$Be production rates in howardites by 1–3% depending on shielding conditions and <1% for the diogenite composition. The $^{10}Be$ production rate varies by <2% for the four compositions listed in Table 23, so we compare the measured $^{10}Be$ concentrations with the average production rate of the four compositions. The measured $^{10}Be$ concentrations in Motopi Pan either indicate irradiation depths of <15 cm in an object with a radius of 38–62 cm or depths up to ~40 cm in an object with a radius of ~80 cm.

The cosmogenic radionuclide data do not provide an absolute constraint of the radius, but rather on the product of radius and density. For example, a radius of 80 cm and density of 2.85 g/cm$^3$ yield the same production rates as a radius of 120 cm and 1.90 g/cm$^3$. The former object has a mass of ~6,000 kg, while the latter has a mass of 14,000 kg. Given that the surviving meteorites have a density of 2.85 g/cm$^3$, this larger object with density of 1.9 g/cm$^3$ would have a macroporosity of ~33%.

To get to a better CRE age, noble gas production rates in HED can be refined from this data, since many cosmogenic nuclides show similar depth and size dependencies. The best correlation is probably between $^{38}Ar$ and $^{36}Cl$, since both of these nuclides are mainly produced from Ca and Fe. The model calculation of Leya & Masarik (2009) shows that for howardites with radii up to ~50 cm, the $^{38}Ar$ and $^{36}Cl$ production rates are linearly correlated (Fig. 31). This implies that the $^{38}Ar$ production rate can simply be derived from the measured $^{36}Cl$ concentration if we assume that all of the $^{38}Ar$ was produced under the same shielding conditions as the $^{36}Cl$ that was produced in the last ~1 Ma, meaning that the meteorite had a simple exposure history. Since the combined radionuclides and noble gas data show no evidence of a complex exposure history, this assumption seems justified.

Table 24 shows the $^{38}Ar$ production rates of the four MP samples based on the $^{36}Cl$ concentration. For objects with a radius <50 cm (blue dots), the $^{38}Ar$ and $^{36}Cl$ production rates are linearly correlated (dashed line), while objects with radii of 60–100 cm show only small deviations from this simple correlation. Relationship between $^{38}Ar$ and $^{36}Cl$ production rates in MP-09 and MP-18 are very similar to MP-12, while the production rates in MP-06 are a factor of ~3 lower, but with a similar slope.

Based on the cosmogenic $^{38}Ar$ concentration in Table 24, we calculated CRE ages of 15.7–20.2 My, similar to the ages given in Table 21. If we exclude the low value for MP-09, which may have lost some of its cosmogenic noble gas inventory, we find an average age of 19 ± 2 Ma for Motopi Pan. This is slightly lower than the CRE age of ~22 Ma for Sariçiçek (Unsalan *et al.* 2019).

## Gamma-ray spectroscopy

Positively identified are short- and medium-lived cosmogenic radionuclides $^{7}Be$, $^{46}Sc$, $^{51}Cr$, $^{54}Mn$, $^{22}Na$, $^{60}Co$ and $^{26}Al$. Only upper detection limits are reported for $^{56}Co$, $^{57}Co$, $^{58}Co$ and $^{44}Ti$ (Table 9, above). Activities of the short-lived $^{52}Mn$ (half-life = 6 days) and $^{48}V$ (half-life = 16 days) were below the detection limit. The activity of naturally occurring radionuclides was similar in howardite Sariçiçek (Unsalan *et al.* 2019) and eucrite Bunburra Rockhole (Welten *et al.* 2012).

The activities of the short-lived radioisotopes, with half-lives less than the orbital period, represent the production integrated over the last segment of the orbit. The fall of the Motopi Pan howardite occurred during the Solar Cycle 24 minimum (Modzelewska *et al.* 2019). The cosmic ray flux was high in the year before the fall. So, the activities for the short-lived radionuclides are expected to be high. In fact, compared to Sariçiçek the values are all about 1.5 times higher.

## U-Pb chronology

No suitable minerals for U-Pb dating were found in the diogenite section MP-06. The eucrite section MP-09 contained some phosphate minerals along cracks, but they were all small. In howardite MP-17, there were single grains of ~10 µm zircon and phosphates (Fig. 32). Also found was a baddeleyite surrounded by zircon (Fig. 32). Zircons in MP-17 (Table 25) record a concordant U-Pb age of 4563 ± 11 Ma and a $^{207}Pb/^{206}Pb$ age of 4563 ± 6 Ma (Fig. 26a, above). In comparison, zircons in Sariçiçek recorded similar ages of 4550.4 ± 2.5 Ma and 4553.5 ± 8.8 Ma, respectively (Unsalan *et al.* 2019).

The howardites also contained numerous phosphate grains (Table 26). Phosphates give a much younger and multiple resetting age spectrum with an average concordia age of 4234 ± 41 Ma (2σ, 10 grains) (Fig. 26b). This compares to 4525 ± 17 Ma for Sariçiçek (Unsalan *et al.* 2019).

## Amino acids

All four meteorite samples analyzed in this study have an amino acid distribution that is similar to the recovery site sands and thus can be explained by terrestrial contamination after their fall to Earth (Table 27a,b). The higher total abundance of amino acids compared to free is consistent with the hydrolysis of terrestrial proteins. α-Aminoisobutyric acid (AIB), a non-protein amino acid detected at trace levels in both the meteorites and sands, is also found in some terrestrial fungal peptides (Brückner *et al.* 2009; Elsila *et al.* 2011).

The presence of elevated abundances of bound ε-amino-*n*-caproic acid in the meteorites but not in the sand, indicates that this amino acid is not a likely contaminant from the fall site. However, ε-amino-*n*-caproic acid is the hydrolysis product of nylon-6 and its presence in predominately bound form in carbonaceous meteorites has previously been attributed to nylon-6 contamination during or after collection (Glavin *et al.* 2006). Since the recovery, handling and shipping of the meteorites and sand were different, it is reasonable that the meteorites were exposed to additional nylon contamination that the sands did not witness.

**Methanol soluble compounds**

Van Krevelen diagrams of CHNOSMg-space show an abundance profile typical of thermostable meteoritic soluble organic matter with a high degree of saturation (Fig. 33). Sample MP-04 shows a higher chemical diversity than MP-06 and MP-18, as rich as the howardite Sariçiçek (fall in 2015) (Unsalan *et al.* 2019) and the eucrite Tirhert (fall in 2014).

In the cluster analysis of Fig. 34, samples MP-06 and MP-18 show a lower abundance of signals and many signals with negative mass defects typical of oxygen and sulfur rich molecules. The electrospray ionization approach complements the Poly-Aromatic Hydrocarbon (PAH) analysis of Fig. 35 with profiles of the corresponding oxygenated hydrocarbons in the higher molecular weight range.

**Insoluble Poly-Aromatic Hydrocarbons**

Aromatics present in the samples are mostly PAHs in a simple distribution such as peaks at m/z = 128 ($C_{10}H_8$), 178 ($C_{14}H_{10}$), and 202 ($C_{16}H_{10}$) (Fig. 35). They represent the majority of detected carbonaceous molecules in all samples, except for the MP-04 (eucrite-looking) which contains more intense peaks corresponding to carbon clusters ($C_9$, $C_{10}$, $C_{11}$,$C_{12}$, $C_{14}$ and $C_{15}$). In addition of the mass spectra, double-bond equivalent (DBE) values as a function of carbon number are provided (Fig. 36).

# DISCUSSION

## Meteoroid size

In transit to Earth, 2018 LA was exposed to cosmic rays, with cosmogenic radionuclides $^{10}Be$, $^{26}Al$, and $^{36}Cl$ constraining the product of radius (r) and density (ρ). Results suggest that all Motopi Pan samples originated from near the surface (<20 cm deep) of a meteoroid with diameter 80–120 cm for a density 2.85 g/cm$^3$ (product rρ = 143 ± 28 g/cm$^2$), or alternatively from a depth of 15–40 cm in a larger ~160 cm diameter meteoroid (rρ = ~228 g/cm$^2$).

The asteroid's spectral type and impact kinetic energy provide further constraints. V-class asteroids have a high visual geometric albedo $p_V$ = 0.37 ± 0.12 (Licandro *et al.* 2017). A homogeneous sphere of this albedo and absolute magnitude of either +31.08 (G = 0.0) or +31.78 (G = 0.15) corresponds to an asteroid diameter of 133 ± 23 cm or 96 ± 17 cm, respectively. The U.S. Government satellite derived kinetic energy of 0.98 kt is likely overestimated, given the unusually bright flare in Motopi Pan's lightcurve and the much lower infrasound-derived kinetic energy of about 0.3–0.5 kt. Given the precisely known entry speed, the latter would correspond to masses in the range of about 8,700–14,500 kg.

All these constraints together are best approached if 2018 LA was a solid rock with high bulk density ~2.85 g/cm$^3$, a relatively low albedo $p_V$ ~ 0.25, no significant opposition effect of scattered sunlight adding to its brightness (G ~ 0.0), and a kinetic energy of ~0.2 kt. In that case, the mass was about 5,700 kg and the diameter about 156 cm. For comparison, Sariçiçek was 100 ± 20 cm in diameter (Unsalan *et al.* 2019).

## Possible source crater

The impact that ejected 2018 LA in an orbit towards Earth occurred 22.8 ± 3.8 Ma ago. Fig. 37 shows the CRE ages derived from production rates of the Eugster & Michel (1995) exposure model, which are slightly higher than those from Leya & Masarik (2009). The ages derived from helium are slightly lower due to diffusive losses (Fig. 37b). In that same scheme, Sariçiçek was ejected 21.7 ± 1.5 Ma ago (Unsalan *et al.* 2019). Unlike Sariçiçek, none of the MP samples contain solar wind noble gases, which implies that Motopi Pan was buried deeper prior to this impact and the MP ages do not need to be corrected for 2π exposure while still a surface regolith.

The semi-major axis and inclination of the approach orbit prior to impact (Table 6, Fig. 11) still preserve some information about the the delivery resonance and the inclination of the source region, respectively, which can be used to identify the source region (e.g., Jenniskens et al. 2012, 2014; Jenniskens 2020). Like Sariçiçek, the approach orbit of 2018 LA is consistent with an origin at Vesta or one of its Vestoids. Based on its size and impact orbit, we estimate a 73 ± 8% probability that 2018 LA was delivered by the $\nu_6$ resonance on the inside of the inner main belt, where Vesta is located, based on the model by Granvik *et al.* (2018).

Dynamical modelling shows that the evolution from Vesta to the $\nu_6$ resonance and evolution to a $q < 1.3$ au orbit was fast, while the CRE age is mostly on account of interactions with the terrestrial planets until impact with Earth, and consistent with Earth experiencing now the peak of the influx pulse (Unsalan *et al.* 2019). 2018 LA's interactions with the terrestrial planets lowered the semi-major axis as much as that of Sariçiçek (Fig. 11). The inclination of the orbit ended up lower than that of Sariçiçek. The natural thermo-chemiluminescence of 24 krad for MP-09 implies that 2018 LA approached the Sun in the past to just below ~0.6 au (Benoit *et al.* 1991; Sears *et al.* 2013), only slightly less than the perihelion distance of 0.78 au of the impact orbit.

If, indeed, the surface area of Vesta dominates that of its ≥ 0.3 km Vestoids by a factor of five (Unsalan et al. 2019), then there is a reasonable probability that Motopi Pan originated from one of the impact craters on Vesta mapped by the Dawn mission. This calculation assumed a collisional equilibrium size frequency distribution from the observed >1–2 km frequency of Vestoids to the smaller ~0.3 km regime.

On Vesta, there are only a few candidate source craters. Fig. 37a plots the crater diameter and age of all Vesta craters for which such information is available in both the lunar-based chronology scheme, derived from crater counts on lunar terrains of known age (Schmedemann *et al.* 2014), and in the asteroid-based chronology scheme that is based on a modelled population of small asteroid impactors (Marchi *et al.* 2012). A recent evaluation of the lunar- vs. asteroid-based chronologies for Vesta and Ceres, in favor of the latter, can be found in Roig & Nesvorny (2020).

The largest possible source craters in the lunar-based scheme are the 16.7-km diameter Antonia (21.1 ± 3.7 Ma – Unsalan *et al.* 2019) on a slope of the Rheasilvia basin near its floor and the smaller 10.3-km sized Rubria (18.8 ± 3.2 Ma – Krohn *et al.* 2014) on Rheasilvia ejecta superimposed on the equatorial troughs of the Divalia Fossae Formation (Fig. 37c). Rubria excavated only 38% as much material as Antonia. There are no other rayed craters on Vesta of this 10–20 km size range that have not been dated. Some of the smaller craters remain undated, but also produce much less debris.

In the alternative asteroid-based chronology scheme, no rayed crater is a good candidate (Fig. 37a), with 10.5-km Arruntia being excluded because its ejecta has olivine-like reflection spectra, unlike Motopi Pan (Cheek & Sunshine 2020).

There are some indications that Motopi Pan may have originated from Rubria. The different phosphate resetting ages point to Sariçiçek and Motopi Pan having originated from pre-impact sites that experienced a different collision history. Motopi Pan was more subjected to larger impacts than Sariçiçek. While both Sariçiçek and Motopi Pan sample Rheasilvia ejecta, Rubria is outside the basin (Fig. 37c), requiring a higher ejection speed from Rheasilvia to bring ejecta to this site. The lack of solar wind noble gasses implies that Motopi Pan was buried deeper at the time of ejection than Sariçiçek and was not exposed to solar wind in recent times. Hence, there was not much regolith gardening and landslides were not significant, unlike where Antonia is located. Rubria is on a topographic high.

Surviving zircon grains in MP-17 probe the oldest primordial crust on Vesta, having crystallized only 5.5 ± 6.4 Ma after Calcium Aluminum Inclusion (CAI) formation. The Veneneia impact reset the phosphate Pb-Pb age of the basement rock, but not the zircons. The later overlapping Rheasilvia impact created the Divalia Fossae Formation and spread this material over the troughs. If Motopi Pan samples Rheasilvia ejecta some distance from the crater center, then the low Pb-Pb resetting age of 4234 ± 41 Ma may well measure the age of the Veneneia impact basin (Fig. 37a). This age marks the beginning of the Late Heavy Bombardment period on Vesta (4.2–3.4 Ga) seen in Ar-Ar ages. The old terrain cratering age of 3.43–3.78 Ga has been interpreted as ~3.5 Ga being the formation age of the younger Rheasilvia impact basin (Kennedy *et al.* 2019). These ages are significantly older than the ~1.0 ± 0.2 Ga and 2.1 ± 0.2 Ga ages derived from crater counts on the floor of the Rheasilvia and Veneneia basins based on the asteroid-flux based chronology scheme (Schenk *et al.* 2012), but are consistent with gravity-driven mass wasting resurfacing the basin floors (Kneissl *et al.* 2014).

## CONCLUSIONS

Combining the asteroid's astrometric positions over time with the time and impact location derived from ground-based and space-based observations of the impact has resulted in an orbit of 2018 LA that was measured 750 times more precisely than that of the meteoroid Sariçiçek, which was tracked only during passage in the Earth's atmosphere. The orbital elements are at a location in the semi-major axis vs. inclination diagram consistent with an origin from the inner main asteroid belt, where Vesta and its Vestoids are found, and delivery via the $\nu_6$ resonance.

2018 LA had an absolute magnitude of +31.08 ± 0.10 if lacking opposition backscattering (G = 0), or +31.78 for a standard G = 0.15, and oscillated in brightness with an amplitude of 0.317 ± 0.005 magnitude, suggesting principal rotation with a spin period of 224 ± 40 s and a minimum biaxial ellipsoid axis ratio of 1.93 ± 0.21. The constraints set by the absolute magnitude, the asteroid's impact kinetic energy, and the noble gas and cosmogenic nuclide data imply that the asteroid was a solid object of albedo ~0.25 and measured about 156 cm in diameter, with a mass of about 5,700 kg.

The petrography and mineralogy of five meteorites show them to be pieces of a HED polymict breccia derived from howardite, cumulate and basaltic eucrite, and diogenite lithologies. Three meteorites are single plutonic lithologies: the green MP-06 is dominated by low-Ca pyroxene, with minor plagioclase, and is interpreted as a diogenite, whereas the light-colored MP-09 and MP-19 comprise roughly equal portions of low-Ca pyroxene and plagioclase in noritic lithologies, interpreted as cumulate eucrites. Two other samples, the gray MP-12 and -18, are breccias typical of howardites, comprising predominantly sub-mm angular to subrounded pyroxene and plagioclase mineral clasts with subsidiary noritic to basaltic lithic clasts up to a few mm in size, set in a fused fragmental matrix. Mineral compositions vary, suggesting clasts are from multiple source rocks. MP-09 has the highest thermoluminescence of any HED measured to date, which suggests that this rock had suffered a high degree of parent body metamorphism.

Reflectance spectra are similar to those of other HED. Like Sariçiçek, Motopi Pan belongs to the isotopically normal group of HED meteorites, but the oxygen isotopes plot along a slightly different range of $\delta^{18}O'$, and there are small differences in the bulk geochemical composition. Oxygen and chromium isotopic compositions record a range of admixing with exogenous carbonaceous chondrite reservoirs, but less so than Sariçiçek. Despite the lack of solar wind gases, the iridium content, from metal grains in carbonaceous and ordinary chondrite impactors, is also relatively high.

The three HED subgroups had different thermal histories before being assembled in the current breccia. MP-09 has the highest thermoluminescence sensitivity measured for an HED to date, suggesting that it is highly metamorphosed. Methanol-soluble matter in MP-04 (eucrite) showed an organo-magnesium compound profile with almost 1000 different CHOMg formulae similar to that of Sariçiçek, but MP-06 (diogenite) and MP-18 (howardite) had profiles with interrupted chemical homologous series, suggesting higher temperature and pressure stresses. The insoluble poly-aromatic hydrocarbons (PAHs) also showed distinct patterns among the HED subgroups in the double-bond equivalent value (DBE) versus carbon atom number in the molecule. Eucrite MP-04 contained more high carbon number and high DBE carbon clusters, while howardites and diogenites contained a wider variety of PAHs. Most water-soluble compounds, like the amino acids, were found to be terrestrial contamination also found in recovery-site sands.

Even though their cosmic ray exposure ages are similar and in the 20–25 Ma range of one third of all HED falls on Earth, Sariçiçek and Motopi Pan did not originate from the same location on Vesta or one of its Vestoids. Motopi Pan experienced a U-Pb age resetting event in phosphates not seen in Sariçiçek. Arguments are presented that the impact crater Rubria on Vesta is a likely source crater of Motopi Pan. If so, the low concordant U-Pb resetting age of 4234 ± 41 Ma may well measure the age of the Veneneia impact basin

*Note added in proof* – An additional 92-g Motopi Pan meteorite was found by Thato Makgane at coordinates 21.22737S, 23.15885E during an expedition led by Fulvio Franchi from BIUST on 7 November 2020.

*Acknowledgments* – We thank the Department of Wildlife and National Parks in Botswana for park access and for Central Kalahari Game Reserve staff assistance during fieldwork. PJ and OM thank the Maun Lodge, Maun, Botswana, for assistance with measuring the meteor shadows. Robert D. Matson assisted EL. We thank Caiphas Majola for preparing the fragile samples in polished resin blocks, Alexander Ziegler for the EMP analyses, and Caitlin Stewart for the TIMA sample processing. – *Funding:* This work was made possible by funding from the Botswana Geoscience Institute, and the Department of National Museum & Monuments in Botswana.

SkyMapper was funded through ARC LIEF grant LE130100104 from the Australian Research Council. TK acknowledges support by the Academy of Finland (project no. 293975). LE is supported by the Dutch Research Council (NWO VIDI project 864.14.005). QZ acknowledges support from the National Natural Science Foundation of China (project 41403055). MEIR and HB thank the Swiss National Science Foundation for support through the framework of the NCCR “PlanetS”. HS and JB acknowledge support from the European Research Council (FP7) ERC-2013-SyG, G.A. Nº610256 NANOCOSMOS. Thermal emissivity measurements have been performed by AM, FF and TK with the support of Europlanet. Europlanet 2020 RI has received funding from the European Union's Horizon 2020 research and innovation programme (No 654208). TK acknowledges spuuport by the Academy of FInland (project nos. 293975 and 335595) and institutional support RVO 67985831 of the Institute of Geology of the Czech Academy of Sciences. Part of this research was carried out at the Jet Propulsion Laboratory, California Institute of Technology, under a contract with the National Aeronautics and Space Administration (80NM0018D0004). Several authors acknowledge support from NASA, NASA’s Planetary Defense Coordination Office, and the NASA SSERVI Center for Asteroid and Lunar Surface Science (CLASS). JPD, DPG, and HLM acknowledge a grant from the Simons Foundation (SCOL award 302497 to JPD) and support from the NASA Astrobiology Institute and the Goddard Center for Astrobiology. MEZ and QZY acknowledge support from the NASA Emerging Worlds Program (NNX16AD34G). MEZ also acknowledges support from the Hayabusa2 Program. PJ acknowledges support from the NASA NEOO program (NNX14-AR92G) and the SSO program (80NSSC18K0854).

*Editorial Handling – Josep Trigo-Rodríguez*

**Table 1.** Asteroid 2018 LA apparent brightness in monochromatic AB magnitudes (Oke & Gunn 1983), based on flux density and a zero point of 3631 Jansky at any frequency (or $3.62\times10^{-11}$ W/m$^2$/nm at 548.3 nm in the V band).

| Station | Day June '18 | Time (s) | Band (nm)§ | Photom. System | Brightness (magnitude) | Distance (au) | Phase Angle (°) | Reflectance if V-class† |
|---|---|---|---|---|---|---|---|---|
| I41 | 1.33652 | -86986 | r | V | >19.7 (5s) | 0.0098000 | 168.0 | 1.20 |
| G96 | 2.34330 | 0.0 | o§ | Gaia G | (18.76)* | 0.0025760 | 12.1 | 1.00 |
| G96 | 2.34851 | 450.1 | o | Gaia G | 18.35 ± 0.10* | 0.0025390 | 12.2 | 1.00 |
| G96 | 2.35373 | 901.1 | o | Gaia G | 18.53 ± 0.09* | 0.0025020 | 12.3 | 1.00 |
| G96 | 2.35895 | 1352.1 | o | Gaia G | 18.27 ± 0.09* | 0.0024651 | 12.4 | 1.00 |
| I52 | 2.39785 | 4713.1 | o | Gaia G | 18.20 ± 0.34* | 0.0021904 | 12.9 | 1.00 |
| I52 | 2.39804 | 4729.5 | o | Gaia G | 18.03 ± 0.26* | 0.0021891 | 12.9 | 1.00 |
| I52 | 2.39822 | 4745.1 | o | Gaia G | 18.25 ± 0.27* | 0.0021879 | 12.9 | 1.00 |
| I52 | 2.39840 | 4760.6 | o | Gaia G | 18.24 ± 0.25* | 0.0021866 | 12.9 | 1.00 |
| G96 | 2.40223 | 5091.5 | o | Gaia G | 18.28 ± 0.30* | 0.0021596 | 13.0 | 1.00 |
| G96 | 2.40233 | 5100.2 | o | Gaia G | 18.24 ± 0.33* | 0.0021590 | 13.0 | 1.00 |
| G96 | 2.40242 | 5108.0 | o | Gaia G | 18.20 ± 0.35* | 0.0021583 | 13.0 | 1.00 |
| G96 | 2.40251 | 5115.7 | o | Gaia G | 18.33 ± 0.34* | 0.0021577 | 13.0 | 1.00 |
| T08 | 2.49263 | 12902.1 | or | V | 17.50 ± 0.30 | 0.0014997 | 14.1 | 1.00 |
| T08 | 2.50096 | 13621.8 | or | V | 17.28 ± 0.30 | 0.0014408 | 14.3 | 1.00 |
| SM | 2.57265 | 19815.8 | g (510) | AB | 16.82 ± 0.10 | 0.0008990 | 15.1 | 1.00 |
| SM | 2.57271 | 19820.9 | g (510) | AB | 16.78 ± 0.10 | 0.0008986 | 15.1 | 1.00 |
| SM | 2.57299 | 19844.9 | r (617) | AB | 16.42 ± 0.07 | 0.0008965 | 15.1 | 1.20 |
| SM | 2.57304 | 19849.9 | r (617) | AB | 16.44 ± 0.07 | 0.0008960 | 15.1 | 1.20 |
| SM | 2.57331 | 19872.9 | i (779) | AB | 15.78 ± 0.11 | 0.0008940 | 15.1 | 1.26 |
| SM | 2.57343 | 19882.9 | i (779) | AB | 15.88 ± 0.11 | 0.0008931 | 15.1 | 1.26 |
| SM | 2.57369 | 19905.9 | z (916) | AB | 16.70 ± 0.19 | 0.0008911 | 15.1 | 0.87 |
| SM | 2.57392 | 19925.9 | z (916) | AB | 16.77 ± 0.19 | 0.0008893 | 15.1 | 0.87 |

Notes: G96 = Catalina Sky Survey; I52 = Catalina Sky Survey; T08 = ATLAS; SM = SkyMapper Survey. §) o = open (380-1000 nm); r = red; or = orange; g = green; i = near-infrared; z = near-infrared "z" band. †) Relative reflectivity compared to open pass band. *) To get $G_{AB}$ from Gaia G-band magnitudes, use $G_{AB} = G + 0.17 \pm 0.05$ magnitude (*anonymous 2017*).

**Table 2.** Video observations of the meteor and shadows.

| Station | Latitude (°S) | Longitude (°E) | Alt. (m) | Range (km) | Time† (UTC) | Azimuth (° from N) | Elevation (°) | Notes |
|---|---|---|---|---|---|---|---|---|
| Maun | 20.005011 | 23.42756 | 945 | 139 | 16:44:11.109 | 183.48±1.7 | 11.79±0.80 | sh., pre-flare |
| Rakops | 21.030190 | 24.402380 | 926 | 120 | 16:44:11.500 | 257.80±0.20 | (12.28) | flare, sh. |
| Ghanzi | 21.69500 | 21.653473 | 1143 | 175 | 16:44:11.500 | 73.13±1.10 | 7.80±0.20 | flare, sh. |
| Gaborone | 24.64617 | 25.815750 | 1063 | 457 | 16:44:05.800 | 335.08±0.10 | 7.54±0.10 | Meteor |
| Gaborone | 24.64617 | 25.815750 | 1063 | 457 | 16:44:06.800 | 333.32±0.10 | 6.34±0.10 | Meteor |
| Gaborone | 24.64617 | 25.815750 | 1063 | 457 | 16:44:07.800 | 331.67±0.10 | 5.21±0.10 | Meteor |
| Gaborone | 24.64617 | 25.815750 | 1063 | 457 | 16:44:08.800 | 329.88±0.10 | 4.15±0.10 | Meteor |
| Gaborone | 24.64617 | 25.815750 | 1063 | 457 | 16:44:09.800 | 328.25±0.10 | 3.05±0.10 | Meteor |
| Gaborone | 24.64617 | 25.815750 | 1063 | 457 | 16:44:11.500 | 325.63±0.20 | 1.27±0.20 | flare § |
| Ottosdal | 26.752972 | 26.18866 | 1575 | 680 | 16:44:06.650 | 339.34±0.10 | 2.76±0.10 | meteor |
| Ottosdal | 26.752972 | 26.18866 | 1575 | 680 | 16:44:07.650 | 338.12±0.10 | 2.12±0.10 | meteor |
| Ottosdal | 26.752972 | 26.18866 | 1575 | 680 | 16:44:08.650 | 336.93±0.10 | 1.55±0.10 | meteor |
| Ottosdal | 26.752972 | 26.18866 | 1575 | 680 | 16:44:09.650 | 335.77±0.10 | 1.00±0.10 | meteor |
| Ottosdal | 26.752972 | 26.18866 | 1575 | 680 | 16:44:10.650 | 334.72±0.10 | 0.48±0.10 | meteor |
| Ottosdal | 26.752972 | 26.18866 | 1575 | 680 | 16:44:11.500 | 333.79±0.20 | 0.12±0.20 | flare § |

Notes: †) Time from video synchronization 16:44:11.5 ± 3.0s UTC, from: 16:44:08.0 (Ottosdal), 11.5s (Maun, Caltex A3), 15.0s (Maun, Maun Lodge), 12.0s (Maun, Caltex Boseja), and 08.0s (Maun, Cresta Maun); §) Obstructed.

**Table 3.** Brightness of foreground lamps in Gaborone video (in magnitudes). Since the Moon was at an altitude 41.7°, we used an extinction-corrected visual magnitude of -10.8. Results are based on photometry of triplicate images taken on May 11, 2019.

| Lamp | A | B | C | D |
|---|---|---|---|---|
| Reverse binocular (±0.5) | -8.5 | -.- | -.- | -8.1 |
| Aperture photometry | -9.1 | -9.0 | -8.5 | -7.8 |

**Table 4.** Asteroid trajectory on 2 June 2018. HT = Height (km); Time is in Coordinated Universal Time; Latitude and Longitude are in the WGS84 reference frame; SEO = Sun-Earth-Object angle (°); LST = Local Solar Time angle (°); Vel. = speed relative to the atmosphere; Az = Radiant azimuth direction of motion relative to North; El. = Radiant elevation direction of motion relative to the ground. This is JPL solution 8.

| HT | Time (UTC) | Lat. (°N) | Lon. (°E) | SEO (°) | LST (°) | Vel. (km/s) | Az. (°N) | El. (°) |
|---|---|---|---|---|---|---|---|---|
| 100 | 16:44:01.59 | -21.378885 | 24.772478 | 103.380 | 83.729 | 16.999 | 95.19 | 25.14 |
| 90 | 16:44:02.98 | -21.361576 | 24.570251 | 103.201 | 83.925 | 17.005 | 95.26 | 25.00 |
| 80 | 16:44:04.37 | -21.343897 | 24.366475 | 103.021 | 84.123 | 17.011 | 95.33 | 24.86 |
| 70 | 16:44:05.78 | -21.325839 | 24.161123 | 102.839 | 84.323 | 17.016 | 95.40 | 24.72 |
| 60 | 16:44:07.18 | -21.307393 | 23.954159 | 102.656 | 84.524 | 17.022 | 95.47 | 24.58 |
| 50 | 16:44:08.60 | -21.288552 | 23.745556 | 102.471 | 84.727 | 17.028 | 95.54 | 24.44 |
| 40 | 16:44:10.02 | -21.269310 | 23.535323 | 102.285 | 84.931 | 17.033 | 95.61 | 24.30 |
| 30 | 16:44:11.45 | -21.249654 | 23.323378 | 102.097 | 85.137 | 17.039 | 95.69 | 24.15 |
| 20 | 16:44:12.89 | -21.229576 | 23.109705 | 101.908 | 85.345 | 17.045 | 95.76 | 24.01 |
| 10 | 16:44:14.34 | -21.209067 | 22.894282 | 101.717 | 85.554 | 17.050 | 95.83 | 23.86 |
| 0 | 16:44:15.79 | -21.188117 | 22.677066 | 101.524 | 85.765 | 17.056 | 95.91 | 23.71 |

**Table 5.** Residuals for asteroid trajectory and orbit. The table gives the astrometric assumed data uncertainties and residuals in Right Ascension (R.A.) and Declination (Dec.), and also in the Along-Track (AT, in both arcsec "AT" and s "$AT_{SEC}$") and Cross-Track (CT) components. The R.A. uncertainties and residuals include a cos(Dec.) factor. SkyMapper (observer code 247) is located at coordinates Long. = 149.06147°E, Lat. = 31.27222°S, Alt. = 1170 m.

| 2018 June (day, UTC) | R.A. hhmmss.sss | Dec. ddmmss.ss | Obs. Code | ΔR.A.cosδ (") | ΔDec. (") | σR.A. (") | σDec. (") | AT (") | CT (") | $AT_{SEC}$ (s) |
|---|---|---|---|---|---|---|---|---|---|---|
| 02.573368 | 160348.191 | -102534.11 | 247 | 1.42 | 0.09 | 1.8 | 0.50 | -1.42 | -0.07 | -0.87 |
| 02.573003 | 160351.440 | -102533.68 | 247 | -2.60 | -0.08 | 0.10 | 0.10 | 2.60 | 0.04 | 1.60 |
| 02.572668 | 160354.761 | -102532.97 | 247 | -1.03 | 0.08 | 0.10 | 0.10 | 1.03 | -0.09 | 0.64 |
| 02.500959 | 160519.961 | -114658.13 | T08 | 2.76 | 0.97 | 1.5 | 1.5 | -2.92 | -0.19 | -3.06 |
| 02.492633 | 160603.891 | -114401.46 | T08 | 1.22 | 0.28 | 1.0 | 1.0 | -1.25 | 0.06 | -1.40 |
| 02.402508 | 160850.275 | -112916.65 | G96 | -0.24 | 0.41 | 0.30 | 0.30 | 0.12 | -0.46 | 0.26 |
| 02.402417 | 160850.570 | -112916.11 | G96 | 0.53 | -0.06 | 0.30 | 0.30 | -0.49 | 0.20 | -1.06 |
| 02.402324 | 160850.778 | -112915.04 | G96 | -0.06 | -0.01 | 0.30 | 0.30 | 0.06 | -0.01 | 0.13 |
| 02.402231 | 160851.030 | -112914.14 | G96 | 0.00 | -0.14 | 0.30 | 0.30 | 0.04 | 0.13 | 0.08 |
| 02.398399 | 160901.154 | -112831.88 | I52 | -0.05 | 0.20 | 0.25 | 0.25 | -0.00 | -0.21 | -0.01 |
| 02.398218 | 160901.618 | -112830.25 | I52 | -0.21 | -0.13 | 0.25 | 0.25 | 0.24 | 0.07 | 0.52 |
| 02.398035 | 160902.122 | -112828.09 | I52 | 0.16 | 0.05 | 0.25 | 0.25 | -0.16 | -0.00 | -0.36 |
| 02.397853 | 160902.579 | -112826.22 | I52 | -0.13 | -0.05 | 0.25 | 0.25 | 0.14 | 0.02 | 0.30 |
| 02.358944 | 161036.604 | -112157.30 | G96 | -0.11 | -0.38 | 0.25 | 0.25 | 0.21 | 0.34 | 0.54 |
| 02.353723 | 161048.154 | -112108.95 | G96 | 0.53 | -0.16 | 0.25 | 0.25 | -0.47 | 0.31 | -1.23 |
| 02.348508 | 161059.352 | -112021.37 | G96 | -0.18 | 0.15 | 0.25 | 0.25 | 0.13 | -0.20 | 0.34 |
| 02.343295 | 161110.342 | -111934.92 | G96 | -0.31 | 0.13 | 0.25 | 0.25 | 0.26 | -0.21 | 0.71 |

**Table 6.** 2018 LA atmospheric trajectory and pre-impact orbit.

| Atmospheric trajectory: | 2018 LA | Orbit (Equinox J2000): | 2018 LA § |
|---|---|---|---|
| Date | 2018-06-02 | Epoch (date - TDB) | 2018-06-2.0 |
| Time (at 100 km altitude - UTC) | 16:44:01.59 ± 0.98 | Epoch (Julian date - TDB) | 2458271.5 |
| $V_g$ (geocentric speed - km/s) † | 12.3750 ± 0.0016 | Tp (perihelion time – TDB) | 2018-07-26.5090 ± 0.0026 |
| $RA_g$ (geocentric right ascension - º) † | 244.18619 ± 0.00023 | a (semi-major axis - au) | 1.37640 ± 0.00011 |
| $Dec_g$ (geocentric declination - º) † | -10.32063 ± 0.00019 | e (eccentricity) | 0.431861 ± 0.000061 |
| V (entry apparent speed - km/s)* | 16.999 ± 0.001 | q (perihelion distance - au) | 0.781986 ± 0.000020 |
| $a_z$ (entry azimuth angle from North - º)* | 95.19 ± 0.01 | ω (argument of perihelion - º) | 256.04869 ± 0.00055 |
| h (entry elevation angle - º)* | 25.14 ± 0.01 | Ω (longitude of ascending node - º) | 71.869605 ± 0.000012 |
| $H_b$ (beginning height - km) | >68.4 | i (inclination - º) | 4.29741 ± 0.00043 |
| $H_d$ (disruption height - km) | 27.8 ± 0.9 | M (mean anomaly - º) | 326.7298 ± 0.0056 |
| $V_d$ (disruption speed - km/s) | 17.040 ± 0.001¶ | n (mean motion - º/d) | 0.610362 ± 0.000074 |
| Time at disruption (UTC) | 16:44:11.77 ± 0.98 | Q (aphelion distance - au) | 1.970812 ± 0.00024 |
| $H_e$ (end height - km) | -.- | $T_J$ (Tisserand parameter w.r.t. Jupiter) | 4.70623 ± 0.00030 |

Notes: †) At one Hill sphere from Earth, Date = 2018-06-01, Time = 07:27 TDB; *) At 100 km altitude; ¶) Assuming no deceleration; §) Orbit solution JPL8 (heliocentric ecliptic J2000).

**Table 7.** Location of recovered meteorites.

| MP-# | Latitude (ºS) | Longitude (ºE) | Alt. (m) | Date (2018) | Time (UTC) | Finder | Affil. † | Mass (g) |
|---|---|---|---|---|---|---|---|---|
| 01* | 21.24848 | 23.23866 | 1002 | 6/23 | 07:00 | Lesedi Seitshiro | BIUST | 17.92 |
| 02 | 21.23612 | 23.25528 | 1004 | 10/9 | 10:50 | Mohutsiwa Gabadirwe | BGI | 4.28 |
| 03 | 21.23696 | 23.27134 | 1001 | 10/10 | 13:00 | Oliver Moses | ORI | 10.08 |
| 04* | 21.23880 | 23.28254 | 989 | 10/10 | 14:26 | Thebe Kemosedile | ORI | 13.16 |
| 05 | 21.23959 | 23.29484 | 991 | 10/11 | 07:28 | Mohutsiwa Gabadirwe | BGI | 2.59 |
| 06* | 21.23887 | 23.29740 | 994 | 10/11 | 07:54 | Peter Jenniskens | ORI | 8.55 |
| 07 | 21.23888 | 23.29785 | 994 | 10/11 | 08:00 | Sarah M. Tsenene | DWNP | 0.51 |
| 08 | 21.23862 | 23.29811 | 996 | 10/11 | 08:24 | Kagiso Kgetse | DWNP | 3.95 |
| 09* | 21.23689 | 23.30645 | 996 | 10/11 | 09:10 | Kagiso Kgetse | DWNP | 4.96 |
| 10 | 21.23633 | 23.30866 | 997 | 10/11 | 09:45 | Kagiso Kgetse | DWNP | 0.89 |
| 11 | 21.23968 | 23.30086 | 991 | 10/11 | 12:45 | Sara M. Tsenene | DWNP | 4.51 |
| 12 | 21.23979 | 23.29714 | 990 | 10/11 | 14:10 | Mohutsiwa Gabadirwe | BGI | 3.60 |
| 13* | 21.23708 | 23.27264 | 998 | 10/12 | 07:14 | Kagiso Kgetse | DWNP | 3.76 |
| 14* | 21.23629 | 23.29703 | 996 | 10/12 | 09:06 | Kabelo Dikole | BGI | 4.35 |
| 15 | 21.23626 | 23.29700 | 995 | 10/12 | 09:10 | Oliver Moses | ORI | 2.71 |
| 16 | 21.24087 | 23.29789 | 994 | 10/12 | 09:45 | Peter Jenniskens | ORI | 1.35 |
| 17* | 21.23889 | 23.29038 | 1005 | 10/12 | 12:50 | Kagiso Kgetse | DWNP | 4.93 |
| 18* | 21.23765 | 23.28587 | 997 | 10/12 | 13:20 | Mohutsiwa Gabadirwe | BGI | 0.90 |
| 19* | 21.23812 | 23.28381 | 996 | 10/12 | 13:38 | Tim Cooper | ORI | 6.19 |
| 20 | 21.23758 | 23.28099 | 994 | 10/12 | 13:53 | Peter Jenniskens | ORI | 1.85 |
| 21 | 21.23743 | 23.28025 | 992 | 10/12 | 14:04 | Mohutsiwa Gabadirwe | BGI | 7.60 |
| 22 | 21.23619 | 23.27041 | 997 | 10/12 | 15:07 | Peter Jenniskens | ORI | 4.73 |
| 23 | 21.23646 | 23.26306 | 991 | 10/12 | 15:37 | Mohutsiwa Gabadirwe | BGI | 8.85 |

Notes: *) Meteorites studied here; †) BIUST = Botswana International University of Science and Technology, BGI = Botswana Geoscience Institute, ORI = Okavango Research Institute of the University of Botswana at Maun, DWNP = Department of Wildlife and National Parks.

**Table 8.** Physical description of the meteorites photographed in Fig. 3.

| MP-# | Mass (g) | Physical description |
|---|---|---|
| 01 | 17.92 | black shiny, fully crusted, in partly-brecciated and homogeneous, largest of the pieces |
| 02 | 4.28 | black shiny, fully crusted |
| 03 | 10.1 | black shiny fully crusted, oriented with flow lines, with exposed gray fine grained part |
| 04 | 12.8 | shiny greyish, thin crusted with clear flow lines, homogenous |
| 05 | 2.59 | shiny crusted, angular, big crystals, breccia |
| 06 | 8.4 | black shiny fully crusted, partly broken and exposed, large crystals, brittle, homogeneous |
| 07 | 0.51 | black shiny fully crusted, very small, smallest of pieces |
| 08 | 3.95 | black shiny crusted and angular |
| 09 | 5.42 | shiny black thin crusted, homogenous |
| 10 | 0.89 | black shiny fully crusted with irregular/angular surface, small |
| 11 | 4.51 | black shiny, fully crusted and angular |
| 12 | 3.6 | black shiny fully crusted, clear breccia |
| 13 | 3.76 | black shiny crusted (partly peeling off) with exposed parts |
| 14 | 4.35 | black shiny fully crusted, visible larger crystals |
| 15 | 2.71 | shiny fully crusted and oriented visible larger crystals |
| 16 | 1.35 | black shiny fully crusted with visible large crystal |
| 17 | 5.49 | black shiny, fully crusted with broken/exposed interior, clear breccia |
| 18 | 7.9 | shiny partly crusted (peeling off) with flow lines, clear breccia, crumbling on cutting |
| 19 | 6.19 | shiny thinly crusted, oriented with flow lines |
| 20 | 1.85 | fully crusted, with clear large crystals |
| 21 | 7.6 | black, fully crusted, visible larger crystals |
| 22 | 4.73 | shiny crusted |
| 23 | 8.85 | shiny light greenish gray thinly crusted, large crystals visible. |

**Table 9.** Activity concentration of cosmogenic radionuclides in MP-01, corrected to the time of fall and concentrations of primordial radionuclides.

| radionuclide | Half-life | Fraction left | Activity dpm $kg^{-1}$ | Concentration ng $g^{-1}$ | Notes |
|---|---|---|---|---|---|
| $^{52}Mn$ | 5.6d | 0.005% | -.- | -.- | Now too weak |
| $^{48}V$ | 16.0d | 3.1% | -.- | -.- | Now too weak |
| $^{51}Cr$ | 27.7d | 13.5% | 120 ± 50 | -.- | |
| $^{7}Be$ | 53.1d | 35.2% | 110 ± 20 | -.- | |
| $^{58}Co$ | 70.9d | 45.7% | < 4.0 | -.- | low abundance |
| $^{56}Co$ | 77.3d | 48.8% | < 4.6 | -.- | low abundance |
| $^{46}Sc$ | 83.8d | 51.6% | 8 ± 2 | -.- | |
| $^{57}Co$ | 271.8d | 81.5% | < 1.9 | -.- | low abundance |
| $^{54}Mn$ | 312.3d | 83.7% | 86.7 ± 7.0 | -.- | |
| $^{22}Na$ | 2.60y | 94.3% | 77.3 ± 5.4 | -.- | |
| $^{60}Co$ | 5.27y | 97.2% | 1.8 ± 0.5 | -.- | |
| $^{44}Ti$ | 60y | 99.7% | < 4.1 | -.- | |
| $^{26}Al$ | $7.170\times10^{5}$y | 100% | 96.1 ± 6.6 | -.- | |
| $^{40}K$ | $1.251\times10^{9}$y | 100% | -.- | $(210 \pm 20) \times10^{3}$ | |
| $^{232}Th$ | $14.051\times10^{9}$y | 100% | -.- | 265 ± 12 | |
| $^{238}U$ | $4.470\times10^{9}$y | 100% | -.- | 87.4 ± 4.2 | |

Notes: The combined expanded uncertainties include a 1σ uncertainty of 10% in the detector efficiency, upper limits are given with expansion factor k = 1.645 corresponding to an about 90% confidence level.

**Table 10.** Combined mineral compositional and overall textures data.

| Sample | MP-06 (0.04 g) | MP-09 (0.10 g) | MP-12 (0.22 g) | MP-18 (0.03 g) | MP-19 (0.52 g) |
|---|---|---|---|---|---|
| *Lithology* | diogenite | cumulate eucrite | polymict eucrite breccia | polymict eucrite breccia | cumulate eucrite |
| *Abundances†* | | | | | |
| low-Ca pyroxene | 94 | 40 | 45 | 48 | 44 |
| high-Ca pyroxene | -.- | -.- | 9 | 3 | -.- |
| olivine | tr | -.- | -.- | -.- | -.- |
| feldspar | 5 | 60 | 42 | 45 | 55 |
| silica | -.- | -.- | 3 | 3 | -.- |
| chromite | 1 | tr | Tr | -.- | 1 |
| ilmenite | -.- | -.- | 0.5 | 0.5 | -.- |
| troilite | tr | tr | 0.5 | 0.5 | -.- |
| kamacite | tr | -.- | -.- | -.- | -.- |
| phosphate | -.- | -.- | Tr | -.- | -.- |
| *Textures overall* | "cumulate" | "cumulate" | polymict breccia | polymict breccia | "cumulate" |
| grain shapes | irregular grains | irregular grains | variable | variable | irregular grains |
| grain size | 50 µm - 2 mm | 0.5 - 2 mm | 10 µm - 2 mm | 10 µm - 0.5 mm | 0.5 - 2.5 mm |
| pyroxenes | homoge-neous | homoge-neous | zoned & finely exsolved | zoned & coarsely exsolved | homoge-neous |

Notes: †) Modal abundances visually estimated.

**Table 10.** Compositional data for main minerals.

| Sample | MP-06 (0.04 g) | MP-09 (0.10 g) | MP-12 (0.22 g) | MP-18 (0.03 g) | MP-19 (0.52 g) |
|---|---|---|---|---|---|
| feldspar | | | | | |
| mole % An | 94.69 ± 0.71 | 95.56 ± 0.54 | 84.71 ± 6.02 | 84.62 ± 4.20 | 95.39 ± 0.69 |
| mole % Or | 0.15 ± 0.04 | 0.19 ± 0.10 | 0.78 ± 0.63 | 0.87 ± 0.51 | 0.18 ± 0.10 |
| n | 15 | 18 | 52 | 20 | 37 |
| low-Ca pyroxene | | | | | |
| mole % En | 69.81 ± 1.14 | 62.28 ± 0.75 | 51.00 ± 8.61 | 47.71 ± 8.86 | 61.25 ± 1.21 |
| mole % Wo | 3.42 ± 0.39 | 5.07 ± 1.02 | 4.88 ± 2.22 | 6.22 ± 4.09 | 4.28 ± 1.62 |
| mg# | 72.28 ± 0.91 | 65.60 ± 0.55 | 53.63 ± 8.91 | 50.73 ± 8.23 | 64.01 ± 1.77 |
| n | 7 | 20 | 55 | 76 | 49 |
| high-Ca pyroxene | | | | | |
| mole % En | | | 34.82 ± 2.73 | 34.07 ± 1.71 | |
| mole % Wo | | | 38.98 ± 2.38 | 39.58 ± 2.87 | |
| Mg# | | | 57.09 ± 4.36 | 56.44 ± 2.37 | |
| n | | | 8 | 27 | |
| olivine | | | | | |
| mole % Fo | 60.55 ± 2.78 | | | | |
| Fe/Mn | 40.26 ± 1.23 | | | | |
| n | 13 | | | | |
| chromite | | | | | |
| Cr (apfu)† | 1.32 ± 0.03 | 1.36 ± 0.01 | 1.34 ± 0.06 | 1.29 ± 0.09 | 1.30 ± 0.06 |
| Al (apfu) | 0.67 ± 0.03 | 0.61 ± 0.01 | 0.60 ± 0.06 | 0.59 ± 0.10 | 0.68 ± 0.05 |
| Ti (apfu) | 0.010 ± 0.002 | 0.013 ± 0.002 | 0.032 ± 0.012 | 0.040 ± 0.008 | 0.012 ± 0.003 |
| Mg# | 17.45 ± 0.97 | 9.83 ± 1.00 | 4.73 ± 0.51 | 4.25 ± 0.77 | 11.05 ± 1.74 |
| Cr#* | 66.40 ± 1.70 | 69.04 ± 0.56 | 69.10 ± 2.96 | 68.86 ± 4.90 | 65.61 ± 2.66 |
| n | 9 | 3 | 9 | 5 | 13 |
| ilmenite | | | | | |
| Mg# | | | 2.40 ± 0.27 | 3.13 ± 0.83 | |
| Fe/Mn | | | 47.84 ± 3.04 | 48.35 ± 3.81 | |
| n | | | 10 | 13 | |
| troilite | | | | | |
| Fe/S | 1.74 ± 0.01 | | 1.74 ± 0.02 | | |
| n | 8 | | 9 | | |
| kamacite | | | | | |
| wt. % Ni | 5.07 ± 0.22 | | | | |
| wt. % Co | 1.28 ± 0.05 | | | | |
| n | 14 | | | | |

Notes: *) Cr# = 100* molar Cr/(Cr+Al); †) "apfu" is atoms per formula unit.

**Table 11.** Description of the MP-18 breccia. Pyx: Pyroxene; Pl: Plagioclase.

| Feature | Breccia A | Breccia B |
|---|---|---|
| Large (> 0.3 mm) clasts | Pyx:Pl ~ 50:50 | Pyx:Pl ~ 70:30 |
| Clast shape | Rounded to elongate-rounded | Angular-irregular, curved edges common |
| Large clasts | >50% of Pyx clasts display coarse exsolution lamellae; Rounded; Fractures displace exsolution lamellae | <20% of Pyx clasts display fine exsolution lamellae; no coarse lamellae; Highly angular |
| Lithic clasts | Unclear owing to indistinct matrix-clast boundaries | ~10 vol%, largest (1 mm) displays m/g to f/g euhedral/subhedral Pl laths + zoned Pyx |
| Distinguishing features | Larger average clast size; | Smaller clast size; |
| | No Pyx Fe-Mg zoning; | Strong Pyx Fe-Mg zoning, including in lithic clast; |
| | Silica more abundant; Finely disseminated sulphide common in clasts and matrix; | Silica less abundant; found in 0.5 mm polymineralic clast (Pyx-Pl-troilite) that is partially fused (fusion crust); Apart from the silica-bearing clast mentioned above, |
| | Ilmenite clasts (< 0.08 mm) | only a few Pyx clasts show limited sulphide inclusions; Chromite + ilmenite clasts (av. 0.1-0.15 mm) |
| Matrix | Pyx:Pl ~ 60:40; Annealed (sintered/ recrystallized?) Pyx-Pl-troilite (≤ 0.05 mm; troilite < 0.02 mm) | Pyx:Pl ~ 30:70; Angular fragments < 0.02 mm |

**Table 12.** Most likely classifications based on similarities in outward appearance with petrographically studied samples.

| MP-# | Classification | MP-# | Classification | MP-# | Classification |
|---|---|---|---|---|---|
| 01¶ | Howardite? | 09*† | Cumulate Eucrite | 17* | Howardite |
| 02 | Howardite? | 10 | Diogenite? | 18*† | Howardite |
| 03 | Howardite? | 11 | Eucrite? | 19*† | Cumulate Eucrite |
| 04* | Eucrite? | 12*† | Howardite | 20 | Howardite? |
| 05 | Diogenite? | 13* | Diogenite? | 21 | Howardite? |
| 06*† | Diogenite | 14 | Howardite? | 22 | Howardite? |
| 07 | Eucrite? | 15 | Eucrite? | 23 | Eucrite? |
| 08 | Diogenite? | 16 | Howardite? | | |

Notes: *) Sampled; †) Petrographic analysis; ¶) CT scan.

**Table 13.** Natural (top) and induced (bottom) Thermoluminescence of MP-09.

| Fragment# | Glow curves | Value (cps) | Average ± 1σ (cps) |
|---|---|---|---|
| *Natural TL:* | | *LT/HT* | |
| 1 | 2019-3-12/1 | 1.407 | |
| 2 | 2019-3-19/1 | 1.234 | 1.32 ± 0.10 |
| *Induced TL:* | | *TL** | |
| 1 | 2019-3-12/2,3,4 | 241,000 | |
| 2 | 2019-3-19/2,3,4 | 243,000 | 242,000 ± 1,000 |

Notes: *) Dhajala = 40,000 cps.

**Table 14.** Magnetic Susceptibility, in descending order.

| MP-# | Type | Logχ (in $10^{-9}$ $m^3$/kg) | N | Petrography |
|---|---|---|---|---|
| 06 | dio | 3.26 ± 0.01 | 10 | diogenite |
| 13 | dio | 3.13 ± 0.01 | 10 | -.- |
| 17 | how | 2.94 ± 0.01 | 10 | -.- |
| 18 | how | 2.94 ± 0.01 | 10 | howardite |
| 12 | how | 2.85 ± 0.01 | 10 | howardite |
| 09 | euc | 2.64 ± 0.01 | 10 | cumulate euc |
| 01 | how | 2.60 ± 0.01 | 10 | -.- |
| 19 | euc | 2.53 ± 0.01 | 10 | cumulate euc |
| 04 | euc | 2.42 ± 0.01 | 10 | -.- |

**Table 15.** Elemental abundances. Elements used to spike solution: Re, In, Bi. "bdl" = below detection limit. Results are compared to Sariçiçek samples SC12 and SC14 (Unsalan *et al.* 2019), and to the sample of diogenites, cumulate eucrites, and polymict breccias in (Mittlefehldt 2015).

| | Z | Unit | MP-06 (dio) | Diogenite† | MP-09 (euc) | Cumulate Eucrite† | MP-17 (how) | SC12 (how) | SC14 (how) | Polymict Breccia† |
|---|---|---|---|---|---|---|---|---|---|---|
| Li | 3 | µg/g | 10.8 | -.- | 8.06 | -.- | 4.55 | 5.68 | 5.70 | -.- |
| Be | 4 | µg/g | 0.752 | -.- | 0.198 | -.- | 0.268 | 0.174 | -.- | -.- |
| Na | 11 | wt. % | 0.689 | 0.01-0.02-0.16* | 0.287 | 0.13-0.20-0.32* | 0.483 | 0.255 | 0.200 | 0.05-0.33-0.48* |
| Mg | 12 | wt. % | 8.72 | 12.0-16.0-21.5 | 10.6 | 4.5-6.5-8.3 | 10.5 | 9.94 | 9.90 | 3.5-7.5-15 |
| Al | 13 | wt. % | 4.33 | 0.1-0.5-2.4 | 12.1 | 3.2-7.8-9.2 | 11.4 | 3.77 | 5.23 | 1.2-6.0-7.1 |
| Si | 14 | wt. % | -.- | 17.2-24.4-26.0 | -.- | 22.0-22.4-23.4 | -.- | -.- | -.- | 21.1-23.0-24.5 |
| P | 15 | wt. % | -.- | -.- | -.- | -.- | -.- | -.- | -.- | -.- |
| Cl | 17 | wt. % | -.- | -.- | -.- | -.- | -.- | -.- | -.- | -.- |
| K | 19 | wt. % | 0.393 | 0.0004-0.007-0.03 | 0.250 | 0.007-0.012-0.04 | 0.068 | 0.0248 | 0.022 | 0.002-0.03-0.15 |
| Ca | 20 | wt. % | 6.10 | 0.2-0.8-3.1 | 8.81 | 4.3-7-7.8 | 8.92 | 5.34 | 6.28 | 1.4-6-8.4 |
| Sc | 21 | µg/g | 77.4 | -.- | 96.8 | -.- | 42.0 | 24.3 | 25.1 | -.- |
| Ti | 22 | µg/g | 2682 | 100-600-2000 | 2293 | 200-1600-2500 | 6128 | 2630 | -.- | 800-4000-6200 |
| V | 23 | µg/g | 610 | -.- | 583 | -.- | 132 | 102 | 88.7 | -.- |
| Cr | 24 | µg/g | 2966 | -.- | 1713 | -.- | 5196 | 7560 | -.- | -.- |
| Mn | 25 | µg/g | 1560 | -.- | 1758 | -.- | 6542 | 4638 | 4560 | -.- |
| Fe | 26 | wt. % | 7.98 | 7.6-13.2-19.6 | 13.9 | 9.8-12.0-16.1 | 29.2 | 14.4 | 14.6 | 12.2-14.2-18.2 |
| Co | 27 | µg/g | 16.1 | 10-22-60 | 7.98 | 6-9-13 | 2.20 | 42.5 | 25.0 | 3.5-14-340 |
| Ni | 28 | µg/g | 344 | 1-60-150 | 173 | 0.5-4-8 | 46.6 | 530 | 150 | 6-100-7000 |
| Cu | 29 | µg/g | 2.58 | -.- | 1.25 | -.- | 0.350 | 3.32 | 3.4 | -.- |
| Zn | 30 | µg/g | 33.7 | -.- | 13.6 | -.- | 4.02 | 2.09 | 2.1 | -.- |
| Ga | 31 | µg/g | -.- | -.- | -.- | -.- | -.- | 0.887 | 1.7 | -.- |
| Ge | 32 | µg/g | -.- | -.- | -.- | -.- | -.- | 11.6 | -.- | -.- |
| As | 33 | µg/g | 11.7 | -.- | 12.8 | -.- | 3.30 | 13.6 | -.- | -.- |
| Se | 34 | µg/g | 56.9 | -.- | 63.0 | -.- | 16.2 | 114 | -.- | -.- |
| Rb | 37 | µg/g | 0.586 | 0.01-0.05-0.13 | 0.431 | 0.05-0.07-0.08 | 0.214 | 0.209 | 0.200 | 0.15-0.3-1.0 |
| Sr | 38 | µg/g | 41.7 | -.- | 63.5 | -.- | 98.3 | 39.3 | 43.0 | -.- |
| Y | 39 | µg/g | 8.06 | -.- | 7.85 | -.- | 26.9 | 9.96 | 13.0 | -.- |
| Zr | 40 | µg/g | 16.5 | -.- | 13.4 | -.- | 67.1 | 40.9 | 34.0 | -.- |
| Nb | 41 | µg/g | 2.09 | -.- | 1.32 | -.- | 4.50 | 1.48 | -.- | -.- |
| Mo | 42 | µg/g | 5.13 | -.- | 3.20 | -.- | 0.754 | 0.977 | -.- | -.- |
| Ru | 44 | µg/g | 0.261 | -.- | 0.129 | -.- | 0.035 | 0.036 | 0.020 | -.- |
| Rh | 45 | µg/g | 0.424 | -.- | 0.217 | -.- | 0.060 | -.- | -.- | -.- |
| Pd | 46 | µg/g | 0.164 | -.- | 0.100 | -.- | 0.114 | -.- | -.- | -.- |
| Ag | 47 | µg/g | -.- | -.- | -.- | -.- | -.- | 0.338 | -.- | -.- |
| Cd | 48 | µg/g | 0.097 | -.- | 0.048 | -.- | 0.016 | 0.119 | -.- | -.- |
| | 50 | µg/g | 0.459 | -.- | 0.133 | -.- | 0.036 | 0.052 | -.- | -.- |
| Sb | 51 | µg/g | 0.296 | -.- | 0.148 | -.- | 0.043 | -.- | -.- | -.- |
| Te | 52 | µg/g | -.- | -.- | -.- | -.- | -.- | 0.007 | -.- | -.- |

**Table 15** (cont.)

| | Z | Units | MP-06 (dio) | Diogenite† | MP-09 (euc) | Cumulate Eucrite† | MP-17 (how) | SC12 (how) | SC14 (how) | Polymict Breccia |
|---|---|---|---|---|---|---|---|---|---|---|
| Cs | 55 | µg/g | 0.058 | 0.0008-0.003-0.01 | 0.066 | 0.016-0.002-0.01 | 0.036 | -.- | 0.007 | 0.004-0.008-0.09 |
| Ba | 56 | µg/g | 28.2 | -.- | 31.7 | -.- | 52.6 | 10.1 | 12.0 | -.- |
| La | 57 | µg/g | 1.30 | -.- | 0.892 | -.- | 4.18 | 1.37 | 1.79 | -.- |
| Ce | 58 | µg/g | 3.17 | -.- | 2.22 | -.- | 10.7 | 4.06 | 4.91 | -.- |
| Pr | 59 | µg/g | 0.528 | -.- | 0.364 | -.- | 1.62 | 0.623 | 0.75 | -.- |
| Nd | 60 | µg/g | 2.95 | -.- | 2.12 | -.- | 7.77 | 2.95 | 3.56 | -.- |
| Sm | 62 | µg/g | 1.39 | 0.002-0.1-0.2 | 0.963 | 0.09-0.3-0.8 | 2.17 | 1.02 | 1.23 | 0.13-1.5-2.7 |
| Eu | 63 | µg/g | 0.329 | -.- | 0.925 | -.- | 0.884 | 0.331 | 0.410 | -.- |
| Gd | 64 | µg/g | 1.06 | -.- | 0.869 | -.- | 3.70 | 1.34 | 1.62 | -.- |
| Tb | 65 | µg/g | 0.196 | -.- | 0.170 | -.- | 0.674 | 0.260 | 0.32 | -.- |
| Dy | 66 | µg/g | 1.20 | -.- | 1.20 | -.- | 4.65 | 1.43 | 1.68 | -.- |
| Ho | 67 | µg/g | 0.278 | -.- | 0.292 | -.- | 1.02 | 0.349 | 0.42 | -.- |
| Er | 68 | µg/g | 0.837 | -.- | 0.848 | -.- | 2.96 | 0.998 | 1.21 | -.- |
| Tm | 69 | µg/g | 0.154 | -.- | 0.140 | -.- | 0.42 | 0.158 | 0.19 | -.- |
| Yb | 70 | µg/g | 0.914 | -.- | 0.934 | -.- | 2.81 | 0.939 | 1.14 | -.- |
| Lu | 71 | µg/g | 0.129 | -.- | 0.142 | -.- | 0.401 | 0.171 | 0.20 | -.- |
| Hf | 72 | µg/g | 0.519 | 0.003-0.08-0.6 | 0.421 | 0.05-0.3-0.5 | 2.11 | 0.848 | 0.80 | 0.03-1.1-3.7 |
| Ta | 73 | µg/g | 0.090 | -.- | 0.052 | -.- | 0.298 | 0.074 | 0.06 | -.- |
| W | 74 | µg/g | 0.142 | -.- | 0.188 | -.- | 0.039 | 0.197 | -.- | -.- |
| Os | 76 | µg/g | 0.047 | -.- | 0.024 | -.- | 0.006 | 0.010 | -.- | -.- |
| Ir | 77 | µg/g | 0.014 | -.- | 0.007 | -.- | 0.002 | 0.008 | 0.010 | -.- |
| Pt | 78 | µg/g | 0.033 | -.- | 0.017 | -.- | 0.005 | 0.011 | 0.018 | -.- |
| Au | 79 | µg/g | 0.034 | -.- | 0.018 | -.- | 0.005 | 0.033 | -.- | -.- |
| Tl | 81 | µg/g | bdl | -.- | bdl | -.- | bdl | 0.002 | bdl | -.- |
| Pb | 82 | µg/g | 0.416 | -.- | 0.170 | -.- | 0.065 | 0.196 | -.- | -.- |
| Th | 90 | µg/g | 0.138 | -.- | 0.013 | -.- | 0.582 | 0.181 | 0.230 | -.- |
| U | 92 | µg/g | 0.047 | -.- | 0.025 | -.- | 0.155 | 0.034 | 0.050 | -.- |

Notes: †) Mittlefehldt (2015); *) Three values give the range of measurements (lowest - most common - and highest reported values).

**Table 16.** Oxygen isotopes for Motopi Pan (MP) and Sariçiçek (SC).

| MP-# | mg | date | $\delta^{17}O'$ | $\delta^{18}O'$ | $\Delta^{17}O'$ | N | SC | mg | date | $\delta^{17}O'$ | $\delta^{18}O'$ | $\Delta^{17}O'$ | N |
|---|---|---|---|---|---|---|---|---|---|---|---|---|---|
| 04 euc | 3.6 | 2-Apr-19 | 1.764 | 3.858 | -0.272 | 1 | 12 how | 1.8 | 19-Feb-16 | 1.627 | 3.666 | -0.309 | 1 |
| 06 dio | 2.4 | 2-Apr-19 | 1.668 | 3.703 | -0.288 | 1 | | 2.1 | 19-Feb-16 | 1.519 | 3.450 | -0.303 | 1 |
| 09 euc | 2.2 | 2-Apr-19 | 1.719 | 3.791 | -0.283 | 1 | | 2.1 | 19-Feb-16 | 1.437 | 3.453 | -0.386 | 1 |
| 12 how | 3.9 | 3-Apr-19 | 1.821 | 4.020 | -0.301 | 1 | | 1.1 | 19-Feb-16 | 1.632 | 3.638 | -0.289 | 1 |
| 13 how | 1.5 | 2-Apr-19 | 1.747 | 3.860 | -0.290 | 1 | | 1.5 | 19-Feb-16 | 1.571 | 3.502 | -0.278 | 1 |
| | 1.7 | 3-Apr-19 | 1.815 | 3.995 | -0.294 | 1 | 14 how | 1.1 | 11-Dec-15 | 1.589 | 3.554 | -0.288 | 1 |
| 17 how | 2.3 | 2-Apr-19 | 1.625 | 3.620 | -0.287 | 1 | | 1.1 | 11-Dec-15 | 1.542 | 3.505 | -0.309 | 1 |
| | 1.3 | 3-Apr-19 | 1.675 | 3.725 | -0.292 | 1 | | 2.2 | 11-Dec-15 | 1.602 | 3.627 | -0.313 | 1 |
| 18 how | 3.3 | 3-Apr-19 | 1.667 | 3.709 | -0.292 | 1 | | 1.9 | 11-Dec-15 | 1.543 | 3.543 | -0.328 | 1 |
| | 3.9 | 3-Apr-19 | 1.289 | 2.980 | -0.284 | 1 | | 1.7 | 19-Feb-16 | 1.487 | 3.498 | -0.360 | 1 |
| 19 euc | 3.2 | 3-Apr-19 | 1.775 | 3.894 | -0.281 | 1 | | 1.4 | 19-Feb-16 | 1.706 | 3.700 | -0.248 | 1 |
| -.- | -.- | -.- | -.- | -.- | -.- | - | | 1.5 | 19-Feb-16 | 1.631 | 3.662 | -0.303 | 1 |
| -.- | -.- | -.- | -.- | -.- | -.- | - | | 1.6 | 11-Dec-15 | 1.621 | 3.633 | -0.297 | 1 |

**Table 17.** Helium and Ne concentrations (in $10^{-8}$ $cm^3$ STP/g) and isotopic ratios in Motopi Pan.

| MP-# | Mass (mg) | $^{4}He$ | $^{3}He/^{4}He$ × 10000 | $^{20}Ne$ | $^{20}Ne/^{22}Ne$ | $^{21}Ne/^{22}Ne$ | $^{3}He_{(=cos)}$ | $^{21}Ne_{(=cos)}$* |
|---|---|---|---|---|---|---|---|---|
| 06 diog | 81.01±0.06 | 232.4±2.1 | 1583±18 | 8.260±0.049 | 0.8415±0.0035 | 0.9205±0.0038 | 36.77±0.23 | 9.036±0.053 |
| 09 eucr | 95.65±0.04 | 129.1±1.4 | 1078±13 | 4.443±0.019 | 1.0307±0.0030 | 0.8703±0.0020 | 13.92±0.09 | 3.749±0.016 |
| 12 how | 63.43±0.03 | 5703±140 | 40.0±1.0 | 4.298±0.019 | 0.8339±0.0029 | 0.8771±0.0025 | 22.82±0.14 | 4.521±0.018 |
| 18L how | 37.58±0.15 | 4932±137 | 48.6±1.4 | 4.645±0.030 | 0.8342±0.0029 | 0.8702±0.0031 | 23.99±0.18 | 4.846±0.031 |
| 18S how | 12.88±0.11 | 5565±55 | 48.5±0.7 | 4.712±0.048 | 0.8303±0.0038 | 0.8761±0.0037 | 26.96±0.28 | 4.973±0.050 |

Notes: *) Only MP-09 contains a small amount of $^{20}Ne_{tr}$ of ~0.9 × $10^{-8}$ $cm^3$ STP/g (see Fig. 27). Here, the cosmogenic $^{21}Ne/^{22}Ne$ ~0.895-0.910 endmember was constrained by extrapolation from $Ne_{tr}$ (air or Q-Ne – Busemann *et al.* 2000) through the measured data point to a typical $(^{20}Ne/^{22}Ne)_{cos}$ range of 0.704–0.933 (Wieler 2002). Typical blanks are for $^{3,4}He$, $^{20,21,22}Ne$ (in $10^{-12}$ $cm^3$ STP): 0.2, 500, 20, 0.2, 2, respectively. Blank corrections for all He and Ne isotope measurements were <1 % except for $^{20}Ne$ in MP-18L/S (1.7 % / 3.7 %).

**Table 18.** Argon concentrations (in $10^{-8}$ cm$^3$ STP/g) and isotopic ratios in Motopi Pan.

| MP-# | $^{36}Ar$ | $^{36}Ar/^{38}Ar$ | $^{40}Ar/^{36}Ar$ | $^{36}Ar_{tr}$* | $^{38}Ar_{cos}$* |
|---|---|---|---|---|---|
| 06 diog | 0.854±0.026 | 0.831±0.016 | 97±6 | 0.211±0.024 | 0.988±0.030 |
| 09 eucr | 10.27±0.31 | 2.055±0.026 | 254±8 | 7.99±0.25 | 3.49±0.11 |
| 12 how | 3.451±0.084 | 0.8779±0.0045 | 467±12 | 1.020±0.089 | 3.739±0.093 |
| 18L how | 3.97±0.10 | 1.006±0.011 | 352±12 | 1.601±0.094 | 3.649±0.097 |
| 18S how | 5.02±0.13 | 1.140±0.006 | 347±10 | 2.46±0.11 | 3.94±0.10 |

Notes: *) Determined by decomposition of $^{36}Ar/^{38}Ar$ into trapped air or Q with $(^{36}Ar/^{38}Ar)_{tr}$ in the range 5.30–5.34 and $(^{36}Ar/^{38}Ar)_{cos} = 0.65 \pm 0.02$. Typical blanks are for $^{36,40}Ar$ (in $10^{-10}$ cm$^3$ STP): 0.5, 130, respectively. Blank corrections for most Ar isotope measurements were <2.5 % except for $^{36,40}Ar$ in MP-18L/S (<8 %) and MP-06 (~13 and 30 %, respectively).

**Table 19.** Krypton concentrations (in $10^{-10}$ $cm^3$ STP/g) and isotopic ratios in Motopi Pan.

| | $^{84}Kr$ | $^{78}Kr/^{84}Kr$ | $^{80}Kr/^{84}Kr$ | $^{82}Kr/^{84}Kr$ | $^{83}Kr/^{84}Kr$ | $^{86}Kr/^{84}Kr$ |
|---|---|---|---|---|---|---|
| MP-# | | | | $^{84}Kr$ = 100 | | |
| 06 diog | 0.388±0.041 | 0.72±0.13 | 4.64±0.73 | 22.3±3.5 | 22.3±3.5 | 30.9±5.5 |
| 09 eucr | 33.62±0.16 | 0.6495±0.0064 | 4.078±0.026 | 20.48±0.11 | 20.45±0.13 | 30.56±0.18 |
| 12 how | 1.4014±0.0087 | 2.95±0.13 | 10.38±0.19 | 29.46±0.64 | 31.45±0.33 | 28.76±0.60 |
| 18L how | 1.589±0.088 | 2.33±0.15 | 9.45±0.62 | 27.8±2.1 | 29.6±2.1 | 29.2±2.8 |
| 18S how | 2.088±0.039 | 2.20±0.12 | 9.45±0.39 | 27.0±1.1 | 28.83±0.83 | 29.62±0.91 |

Notes: The blank for $^{84}Kr$ is typically 1.5 × $10^{-12}$ $cm^3$ STP. Blank corrections for Kr isotope measurements were <1 % in MP-09 and between 3 % (for some light isotopes $^{78,80}Kr$) and 50 % in all other samples.

**Table 20.** Xenon concentrations ($10^{-10}$ $cm^3$ STP/g) and isotopic ratios in Motopi Pan.

| MP-# | $^{132}Xe$ | $^{124}Xe/^{132}Xe$ | $^{126}Xe/^{132}Xe$ | $^{128}Xe/^{132}Xe$ | $^{129}Xe/^{132}Xe$ | $^{130}Xe/^{132}Xe$ | $^{131}Xe/^{132}Xe$ | $^{134}Xe/^{132}Xe$ | $^{136}Xe/^{132}Xe$ |
|---|---|---|---|---|---|---|---|---|---|
| | | $^{132}Xe$ = 100 | | | | | | | |
| 06 diog | 0.132±0.010 | 0.485±0.092 | 0.240±0.069 | 7.34±0.84 | 98±11 | 14.9±1.8 | 78.9±8.6 | 37.8±4.2 | 33.7±3.3 |
| 09 eucr | 5.797±0.090 | 0.3855±0.0065 | 0.350±0.006 | 7.037±0.058 | 98.72±0.54 | 15.12±0.11 | 79.18±0.55 | 38.61±0.27 | 32.99±0.22 |
| 12 how | 0.738±0.012 | 1.858±0.067 | 3.378±0.078 | 11.17±0.25 | 98.1±1.3 | 17.21±0.30 | 88.4±1.5 | 42.16±0.63 | 37.34±0.44 |
| 18L how | 0.560±0.023 | 2.07±0.14 | 3.35±0.18 | 11.34±0.66 | 98.1±5.4 | 17.2±1.0 | 92.1±4.8 | 42.3±2.3 | 39.1±1.9 |
| 18S how | 0.560±0.023 | 1.72±0.18 | 2.61±0.13 | 10.44±0.46 | 100.3±4.0 | 16.92±0.77 | 90.6±3.6 | 41.9±1.7 | 37.4±1.7 |

Notes: The blank for $^{132}Xe$ is typically $1.8 \times 10^{-13}$ $cm^3$ STP. Blank corrections for Xe isotope measurements were <1 % in MP-09 and between <2 % (for some light isotopes $^{124,126,128}Xe$) and 21 % in all other samples.

**Table 21.** Production rate ($P_x$) ranges determined for radii 40–80 cm† (radii range from radionuclides, see main text, normalised to a density of 2.85 g/cm$^3$, the Leya & Masarik (2009) model uses a density of 3.5 g/cm$^3$), with the chemistry given in Table 22 (and Table 15 for Na only), ($^{22}Ne/^{21}Ne)_{cos}$¶ and cosmogenic $^3He$, $^{21}Ne$ and $^{38}Ar$ (Tables 17–18). In brackets we also give production rates and CRE ages determined with the model by Eugster & Michel (1995).

| MP-# | shielding depth cm | $P_3$ | $P_{21}$ | $P_{38}$ | $T_3$ | $T_{21}$ | $T_{38}$ | $T_{preferred}$ |
|---|---|---|---|---|---|---|---|---|
| | | 10$^{-8}$ cm$^3$/g/Ma | | | Ma | | | |
| 06 diog | 23–49 | 1.920–2.125 (1.674) | 0.398–0.435 (0.362) | 0.050–0.055 (0.060) | 17–19 (22) | 21–23 (25) | 18–20 (16) | **18–22** |
| 09 eucr | 17–66 | 1.817–2.065 (1.680) | 0.301–0.329 (0.239) | 0.180–0.208 (0.149) | 7–8* (8*) | 11–13* (16*) | 17–20 (23) | **17–20** |
| 12 how | 17–80 | 1.766–1.995 (1.643) | 0.279–0.307 (0.206) | 0.157–0.178 (0.144) | 11–13* (14*) | 15–16 (22) | 21–24 (26) | **16–22** |
| 18L how | 16–80 | 1.769–1.996 (1.646) | 0.270–0.300 (0.195) | 0.166–0.192 (0.151) | 12–14* (15*) | 16–18 (25) | 19–22 (24) | **17–21** |
| 18S how | 26–66 | 2.265–3.544 (1.649) | 0.274–0.299 (0.199) | 0.179–0.193 (0.151) | 13–16 (16) | 16–18 (25) | 20–22 (26) | **17–21** |

Notes: †) We restricted the production rate determination to the range ~40 – 80 cm given by radionuclides; ¶) ($^{21}Ne/^{22}Ne)_{cos}$ as measured (Table 18, extrapolated in the case of MP-09 due to the presence of a small $Ne_{tr}$ component, see Fig. 3.4); *) potentially affected by diffusive loss.

**Table 22.** Chemical composition of Motopi Pan ("MP") and Sariçiçek ("SC") samples.

| Sample: | | MP-06 | MP-09 | MP-12 | MP-18 | SC12 |
|---|---|---|---|---|---|---|
| Type: | | dio | euc | how | how | how |
| Mass: | (mg) | 48.8 | 75.0 | 28.0 | 49.7 | [1] |
| Si † | (wt%) | 24.4 | 23.7 | 23.3 | 23.7 | -.- |
| Mg | (wt%) | 14.0 | 5.9 | 6.1 | 5.4 | 9.9 |
| Al | (wt%) | 1.3 | 9.0 | 5.5 | 6.0 | 3.8 |
| Ca | (wt%) | 1.7 | 7.2 | 6.1 | 6.6 | 5.3 |
| Mn | (wt%) | 0.40 | 0.26 | 0.43 | 0.39 | 0.46 |
| Fe | (wt%) | 13.6 | 8.1 | 14.5 | 13.4 | 14.4 |
| K | (ppm) | 25 | 100 | 360 | 360 | 248 |
| Ti | (ppm) | 400 | 230 | 3350 | 3800 | 2630 |
| Co | (ppm) | 130 | 410 | 9 | 10 | 43 |
| Ni | (ppm) | 69 | 23 | 25 | 31 | 530 |
| POEM | (wt%) | 12 | 97 | 85 | 83 | 60 |

Notes: [1] Unsalan *et al.* (2019); †) *Si was estimated from other elements assuming* $SiO_2$ = *100% - [sum of all other oxides]*. POEM = Percentage of Eucrite Material, based on Ca and Al concentrations (Mittlefehldt *et al.* 2013) for MP-09 only POEM Ca is reported as Al content is outside the normal range for HED meteorites.

**Table 23.** Cosmogenic radionuclide concentrations.

| Sample: | | MP-06 | MP-09 | MP-12 | MP-18 |
|---|---|---|---|---|---|
| Type: | | dio | euc | how | how |
| Mass: | (mg) | 48.8 | 75.0 | 28.0 | 49.7 |
| $^{10}$Be | (dpm/kg) | 22.3 ± 0.2 | 22.1 ± 0.2 | 23.3 ± 0.3 | 21.7 ± 0.2 |
| $^{26}$Al | (dpm/kg) | 77.6 ± 1.1 | 105.9 ± 1.5 | -.- | -.- |
| $^{36}$Cl | (dpm/kg) | 7.4 ± 0.1 | 20.8 ± 0.2 | 19.0 ± 0.3 | 19.0 ± 0.2 |
| $^{36}$Cl* | (dpm/kg[Fe*]) | 24.1 ± 0.3 | 25.8 ± 0.3 | 24.4 ± 0.3 | 23.4 ± 0.3 |

Notes: *) In dpm/kg[Fe*], with Fe* = Fe + 10Ca + 50K.

**Table 24**. CRE ages based on cosmogenic $^{38}$Ar concentration and relationship between calculated $^{38}$Ar and $^{36}$Cl production rates from Leya & Masarik (2009).

| MP-# | $^{36}$Cl [dpm/kg] | P($^{38}$Ar) [E-8 cc/g/Ma] | $^{38}$Ar [E-8 cc/g] | T($^{38}$Ar) [Ma] |
|---|---|---|---|---|
| 06 dio | 7.4 | 0.058 | 0.988 | 17.1 |
| 09 euc | 20.8 | 0.220 | 3.49 | 15.7 |
| 12 how | 19.0 | 0.185 | 3.74 | 20.2 |
| 18 how | 19.0 | 0.185 | 3.75 | 20.2 |

**Table 25.** SIMS U-Pb isotopic data of zircon from MP-17.

| Spot | U (ppm) | Th (ppm) | Th/U | $^{207}Pb^*/^{206}Pb^*$ | ±1σ (%) | $^{207}Pb^*/^{235}U$ | ±1σ (%) | $^{206}Pb^*/^{238}U$ | ±1σ (%) | $t_{207/206}$ (Ma) | ±1σ | $t_{207/235}$ (Ma) | ±1σ | $t_{206/238}$ (Ma) | ±1σ |
|---|---|---|---|---|---|---|---|---|---|---|---|---|---|---|---|
| 5μm | | | | | | | | | | | | | | | |
| 1 | 344 | 139 | 0.40 | 0.6287 | 0.86 | 85.5 | 1.7 | 0.987 | 1.50 | 4569 | 12 | 4525 | 18 | 4427 | 48 |
| 2 | 178 | 111 | 0.58 | 0.6248 | 0.77 | 88.1 | 1.8 | 1.024 | 1.58 | 4560 | 11 | 4555 | 18 | 4544 | 52 |
| 3 | 631 | 295 | 0.46 | 0.6234 | 0.33 | 85.1 | 1.6 | 0.991 | 1.52 | 4557 | 5 | 4521 | 16 | 4440 | 49 |
| 4 | 396 | 230 | 0.55 | 0.6324 | 0.67 | 85.8 | 1.7 | 0.985 | 1.53 | 4578 | 10 | 4529 | 17 | 4420 | 49 |
| 2μm | | | | | | | | | | | | | | | |
| 5 | 325 | 126 | 0.47 | 0.6276 | 0.52 | 106.8 | 2.1 | 1.235 | 2.06 | 4567 | 7 | 4749 | 22 | 5186 | 74 |
| 6 | 68 | 41 | 0.54 | 0.6303 | 1.49 | 86.7 | 3.1 | 0.998 | 2.76 | 4573 | 21 | 4538 | 32 | 4461 | 90 |
| 7 | 186 | 62 | 0.37 | 0.6259 | 0.85 | 105.4 | 2.2 | 1.222 | 2.01 | 4563 | 12 | 4735 | 22 | 5147 | 72 |
| 8 | 276 | 189 | 0.60 | 0.6246 | 2.01 | 89.5 | 3.1 | 1.039 | 2.34 | 4560 | 29 | 4570 | 31 | 4594 | 77 |

Notes: * denotes radiogenic, using the CDT Pb as common-lead compositions $^{206}Pb/^{204}Pb$ = 9.307, $^{207}Pb/^{206}Pb$ = 1.09861 from (Tatsumoto *et al.* 1973).

**Table 26.** SIMS U-Pb isotopic data of phosphate from MP-17.

| Spot | Mineral | U (ppm) | Th (ppm) | Th/U | $^{204}Pb$ (cps) | $^{204}Pb/^{206}Pb$ | $f_{206}$ (%) | $^{207}Pb/^{206}Pb$ | ±1σ (%) | $^{206}Pb/^{238}U$ | ±1σ (%) |
|---|---|---|---|---|---|---|---|---|---|---|---|
| 4 | Merrillite | 11. | 451 | 40.9 | n.d. | ---- | ---- | 0.4718 | 0.89 | 1.285 | 2.56 |
| 6 | Apatite | 67 | 126 | 1.87 | n.d. | ---- | ---- | 0.4839 | 0.39 | 1.002 | 2.29 |
| 7 | Apatite | 17 | 25 | 1.48 | n.d. | ---- | ---- | 0.5065 | 0.82 | 0.975 | 2.36 |
| 12 | Apatite | 28 | 78 | 2.76 | 0.033 | 0.0001 | 0.07 | 0.5292 | 0.54 | 1.151 | 2.51 |
| 13 | Apatite | 47 | 150 | 3.19 | n.d. | ---- | ---- | 0.5302 | 0.86 | 1.145 | 2.53 |
| 14 | Apatite | 58 | 112 | 1.92 | 0.019 | 0.0001 | 0.06 | 0.4994 | 0.37 | 1.065 | 2.76 |
| 15 | Apatite | 127 | 208 | 1.64 | n.d. | ---- | ---- | 0.5068 | 0.88 | 1.120 | 2.43 |
| 16 | Apatite | 35 | 37 | 1.06 | n.d. | ---- | ---- | 0.5097 | 0.49 | 0.972 | 2.08 |
| 18 | Merrillite | 9 | 1935 | 217 | 0.033 | 0.0000 | 0.00 | 0.4661 | 0.87 | 1.402 | 2.81 |
| 19 | Apatite | 26 | 68 | 2.62 | 0.067 | 0.0003 | 0.29 | 0.5351 | 1.13 | 1.138 | 2.08 |

Note: "n.d." is not determined.

**Table 26 (cont.).**

| Spot | Mineral | $^{207}Pb^*/^{206}Pb^*$ | ±1σ (%) | $^{207}Pb^*/^{235}U$ | ±1σ (%) | $^{206}Pb^*/^{238}U$ | ±1σ (%) | $t_{207/206}$ (Ma) | ±1σ | $t_{207/235}$ (Ma) | ±1σ | $t_{206/238}$ (Ma) | ±1σ |
|---|---|---|---|---|---|---|---|---|---|---|---|---|---|
| 4 | Merrillite | 0.4717 | 0.89 | 83.5 | 2.7 | 1.285 | 2.56 | 4149 | 13 | 4501 | 28 | 5327 | 94 |
| 6 | Apatite | 0.4838 | 0.39 | 66.8 | 2.3 | 1.002 | 2.29 | 4187 | 6 | 4278 | 23 | 4476 | 74 |
| 7 | Apatite | 0.5065 | 0.82 | 68.0 | 2.5 | 0.974 | 2.36 | 4254 | 12 | 4296 | 25 | 4385 | 75 |
| 12 | Apatite | 0.5288 | 0.55 | 83.8 | 2.6 | 1.150 | 2.51 | 4317 | 8 | 4505 | 26 | 4935 | 87 |
| 13 | Apatite | 0.5301 | 0.86 | 83.6 | 2.7 | 1.144 | 2.53 | 4321 | 13 | 4502 | 27 | 4918 | 88 |
| 14 | Apatite | 0.4993 | 0.37 | 73.3 | 2.8 | 1.065 | 2.76 | 4233 | 5 | 4370 | 28 | 4675 | 92 |
| 15 | Apatite | 0.5067 | 0.88 | 78.2 | 2.6 | 1.120 | 2.43 | 4255 | 13 | 4436 | 26 | 4844 | 83 |
| 16 | Apatite | 0.5096 | 0.49 | 68.3 | 2.1 | 0.972 | 2.08 | 4263 | 7 | 4299 | 22 | 4378 | 66 |
| 18 | Merrillite | 0.4649 | 0.90 | 89.7 | 2.9 | 1.400 | 2.81 | 4128 | 13 | 4573 | 30 | 5643 | 106 |
| 19 | Apatite | 0.5339 | 1.14 | 83.6 | 2.4 | 1.136 | 2.08 | 4331 | 17 | 4502 | 24 | 4893 | 72 |

Notes: *) denotes radiogenic, using the CDT Pb as common-lead compositions $^{206}Pb/^{204}Pb = 9.307$, $^{207}Pb/^{206}Pb = 1.09861$ from (Tatsumoto *et al.* 1973). The uncertainties for individual U-Pb isotopic data analyses are reported as 1σ. The intercept age and Pb-Pb ages, quoted at the 95% confidence level, were calculated using ISOPLOT 3.0 (Ludwig 2003). Correction of the common Pb was made by measuring the amount of $^{204}Pb$ and the CDT Pb isotopic compositions $^{206}Pb/^{204}Pb = 9.307$, $^{207}Pb/^{206}Pb = 1.09861$ (Tatsumoto *et al.* 1973).

**Table 27a.** Amino acid abundances in parts-per-billion (ppb; ng/g) in the free (nonhydrolyzed) and total (6M HCl-hydrolyzed) hot-water extracts of four meteorites and assedociated sand from the recovery sites.

| Amino Acid | **Eucrite MP-04 77.4 mg** | | **Sand MP-04 138.4 mg** | | **Eucrite MP-19 62.2 mg** | | **Sand MP-19 103.0 mg** | |
|---|---|---|---|---|---|---|---|---|
| | Free | Total | Free | Total | Free | Total | Free | Total |
| acidic amino acid | | | | | | | | |
| D-aspartic acid | <0.1 | 540±10 | 550±40 | 24600±1300 | <0.1 | 480±40 | 290±20 | 7754±389 |
| L-aspartic acid | 190±10 | 870±20 | 3400±180 | 6500±4700 | 160±10 | 820±30 | 1300±80 | 24801±1181 |
| D-glutamic acid | <0.1 | 390±4 | 280±20 | 11800±500 | 5±4 | 460±10 | 90±20 | 3302±113 |
| L-glutamic acid | 57±5 | 1700±100 | 7700±160 | 37900±900 | 60±10 | 2500±200 | 3340±60 | 22685±3773 |
| hydroxy amino acid | | | | | | | | |
| D-serine | 214±5 | 440±10 | 170±7 | 2472±30 | 133±4 | 1110±40 | 109±6 | 1679±44 |
| L-serine | 721±24 | 650±20 | 2070±70 | 16500±100 | 490±20 | 1860±60 | 920±40 | 8623±131 |
| D-threonine | <0.1 | 5±1 | <0.1 | 87±1 | <0.1 | 10±1 | <0.1 | 45±1 |
| L-threonine | 274±6 | 610±20 | <0.1 | 40000±300 | 203±7 | 1160±50 | <0.1 | 24946±321 |
| C2 amino acid | | | | | | | | |
| glycine | 2720±40 | 6600±200 | 5200±160 | 44400±300 | 1570±80 | 13000±400 | 3140±70 | 33773±282 |
| C3 amino acid | | | | | | | | |
| β-alanine | 145±6 | 496±5 | 5040±60 | 6720±80 | 188±2 | 510±10 | 3120±60 | 3426±58 |
| D-alanine | 95±4 | 236±3 | 486±8 | 8400±100 | 55±5 | 351±7 | 248±6 | 2983±47 |
| L-alanine | 597±6 | 686±5 | 2690±30 | 51500±500 | 193±8 | 1790±20 | 1240±20 | 31872±349 |
| C4 amino acid | | | | | | | | |
| D,L-α-amino-*n*-butyric acid | 2±2 | 71±2 | 191±6 | 4600±200 | <0.1 | 98±6 | 72±4 | 1357±27 |
| D-β-amino-*n*-butyric acid | <0.1 | 25±2 | 35±2 | 105±4 | 2±2 | 15±3 | 23±1 | 47±1 |
| L-β-amino-*n*-butyric acid | <0.1 | 17±1 | 34±1 | 124±2 | <0.1 | 10±1 | 25.6±0.3 | 52±1 |
| γ-amino-*n*-butyric acid | 43±5 | 628±8 | 1160±10 | 5830±70 | 370±10 | 660±10 | 850±10 | 2709±46 |
| α-aminoisobutyric acid | 6±3 | 17±2 | 6±1 | 30±1 | 11±2 | 15±1 | 6±1 | 16±3 |
| C5 amino acid | | | | | | | | |
| D-valine | <0.1 | 32±1 | <0.1 | 4800±100 | <0.1 | 72±4 | 80±50 | 499±40 |
| L-valine | 388±9 | 566±7 | 1700±30 | 32800±700 | 91±3 | 1550±40 | 737±10 | 14294±125 |
| D-isovaline | <0.1 | <0.1 | <0.1 | <0.1 | <0.1 | <0.1 | <0.1 | <0.1 |
| L-isovaline | <0.1 | <0.1 | <0.1 | <0.1 | <0.1 | <0.1 | <0.1 | <0.1 |
| C6 amino acid | | | | | | | | |
| ε-amino-*n*-caproic acid | 160±2 | 3100±70 | <0.1 | <0.1 | 271±6 | 9000±300 | 1±1 | < 1 |

Notes: tr = trace amounts detected but could not be quantified.

**Table 27b.** As Table 27a, for other MP meteorites.

| Amino Acid | Howardite MP-17 75.2 mg | | Sand MP-12 120.4 mg | | Diogenite MP-06 91.3 mg | | Sand MP-06 118.9 mg | |
|---|---|---|---|---|---|---|---|---|
| | Free | Total | Free | Total | Free | Total | Free | Total |
| acidic amino acid | | | | | | | | |
| D-aspartic acid | <0.1 | 92±3 | 580±20 | 14100±600 | 33±1 | 125±10 | 310±20 | 28000±3000 |
| L-aspartic acid | <0.1 | 143±5 | 2800±60 | 33000±2000 | 100±1 | 209±8 | 1790±80 | 61000±5000 |
| D-glutamic acid | <0.1 | 54±4 | 92±3 | 6900±300 | 10±2 | 44±5 | <0.1 | 10500±200 |
| L-glutamic acid | <0.1 | 260±20 | 2190±50 | 32100±200 | 36±5 | 260±10 | 2700±300 | 44000±2000 |
| hydroxy amino acid | | | | | | | | |
| D-serine | 0.9±0.5 | 80±2 | 76±2 | 810±70 | 62±2 | 52±1 | 73±1 | 2270±50 |
| L-serine | 4.1±0.2 | 112±3 | 1820±50 | 9400±500 | 261±4 | 73±1 | 660±10 | 12900±400 |
| D-threonine | <0.1 | <0.1 | <0.1 | 28±1 | <0.1 | 0.5±0.3 | <0.1 | 64±1 |
| L-threonine | 2.8±0.2 | 121±2 | 1050±30 | 26000±1000 | 143±2 | 157±5 | 800±20 | 37000±1000 |
| C2 amino acid | | | | | | | | |
| glycine | 16±1 | 1780±30 | 4600±100 | 23000±1000 | 1750±80 | 955±9 | 1850±10 | 36000±2000 |
| C3 amino acid | | | | | | | | |
| β-alanine | 1.7±0.1 | 88±1 | 400±200 | 1600±100 | 124±3 | 77±1 | 411±7 | 3880±30 |
| D-alanine | 1.2±0.1 | 32.3±0.3 | 239±7 | 2600±200 | 60±2 | 57.9±0.3 | 178±3 | 4870±30 |
| L-alanine | 8.3±0.1 | 113±1 | 1590±20 | 1900±1000 | 199±3 | 293±2 | 970±10 | 34590±50 |
| C4 amino acid | | | | | | | | |
| D,L-α-amino-n-butyric acid | <0.1 | 10.1±0.2 | 72±4 | 1110±90 | <0.1 | 17±1 | 46±3 | 1250±90 |
| D-β-amino-n-butyric acid | <0.1 | 2.2±0.1 | 35±3 | 55±4 | <0.1 | tr | 14±1 | 31±2 |
| L-β-amino-n-butyric acid | <0.1 | <0.1 | 35±2 | 62±3 | <0.1 | tr | 14±1 | 42±3 |
| ɣ-amino-n-butyric acid | <0.1 | 230±4 | 2140±50 | 3400±300 | 37±1 | 93±1 | 650±5 | 3570±20 |
| α-amino-isobutyric acid | <0.1 | 1.8±0.1 | <0.1 | 13.1±0.4 | 1±1 | tr | 11±2 | 68±6 |
| C5 amino acid | | | | | | | | |
| D-valine | <0.1 | 5.9±0.3 | 6±3 | 710±60 | <0.1 | 9±5 | 10±2 | 1190±80 |
| L-valine | 3.0±0.3 | 132±3 | 610±20 | 12400±400 | 107±3 | 164±3 | 367±3 | 18000±1000 |
| D-isovaline | <0.1 | <0.1 | <0.1 | <0.1 | <0.1 | <0.1 | <0.1 | <0.1 |
| L-isovaline | <0.1 | <0.1 | <0.1 | <0.1 | <0.1 | <0.1 | <0.1 | <0.1 |
| C6 amino acid | | | | | | | | |
| ε-amino-n-caproic acid | <0.1 | 850±20 | 2±1 | <0.1 | 34±3 | 398±8 | <0.1 | <0.1 |

**Fig. 1.** Images of asteroid 2018 LA in the SkyMapper Southern Survey g,r,i, and z color bands (Wolf *et al.* 2018) from top to bottom, respectively.

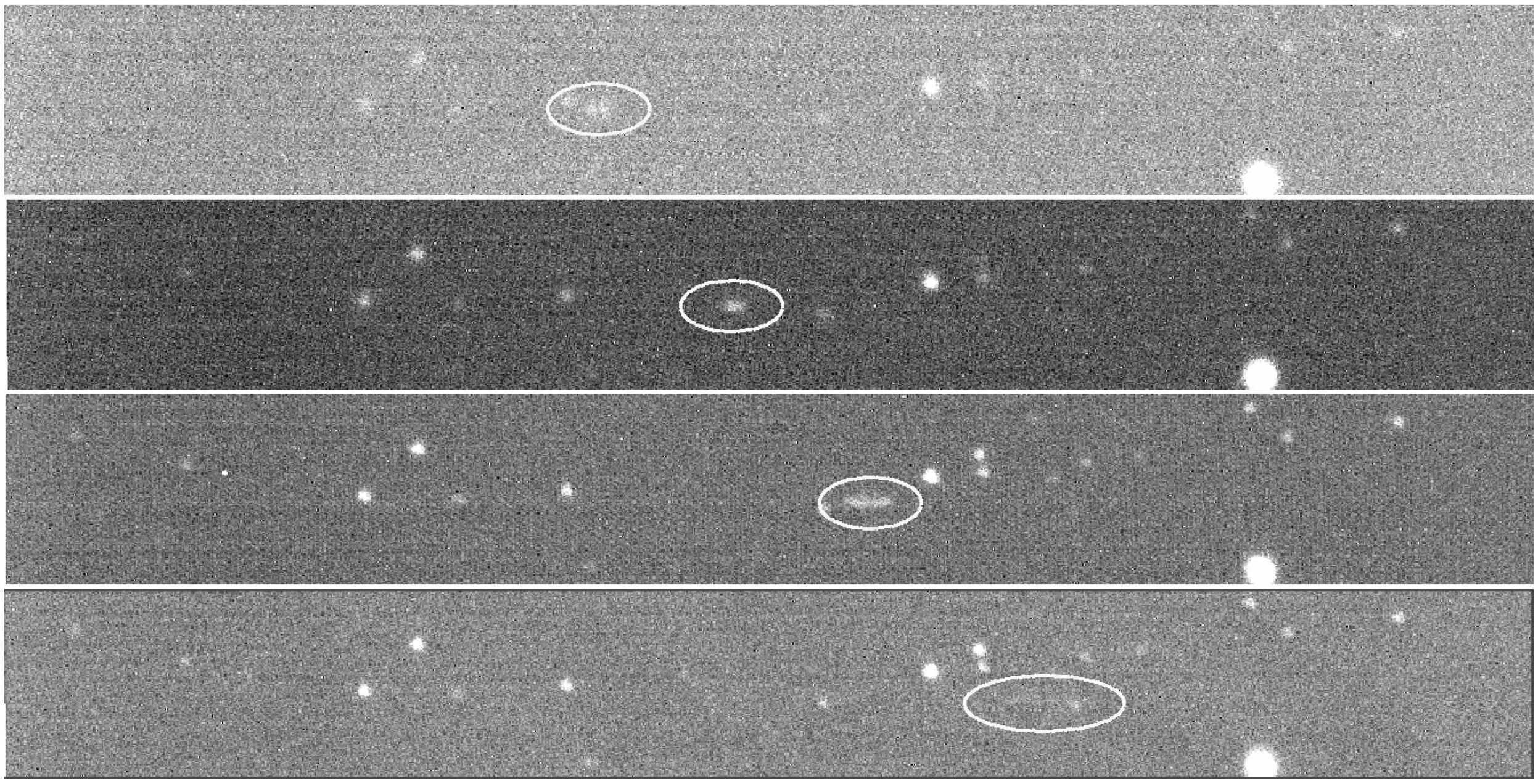

**Fig. 2.** The recovery of 2018 LA. **(a)** The team that found first meteorite (left) and **(b)** the team that found the additional 22 meteorites.

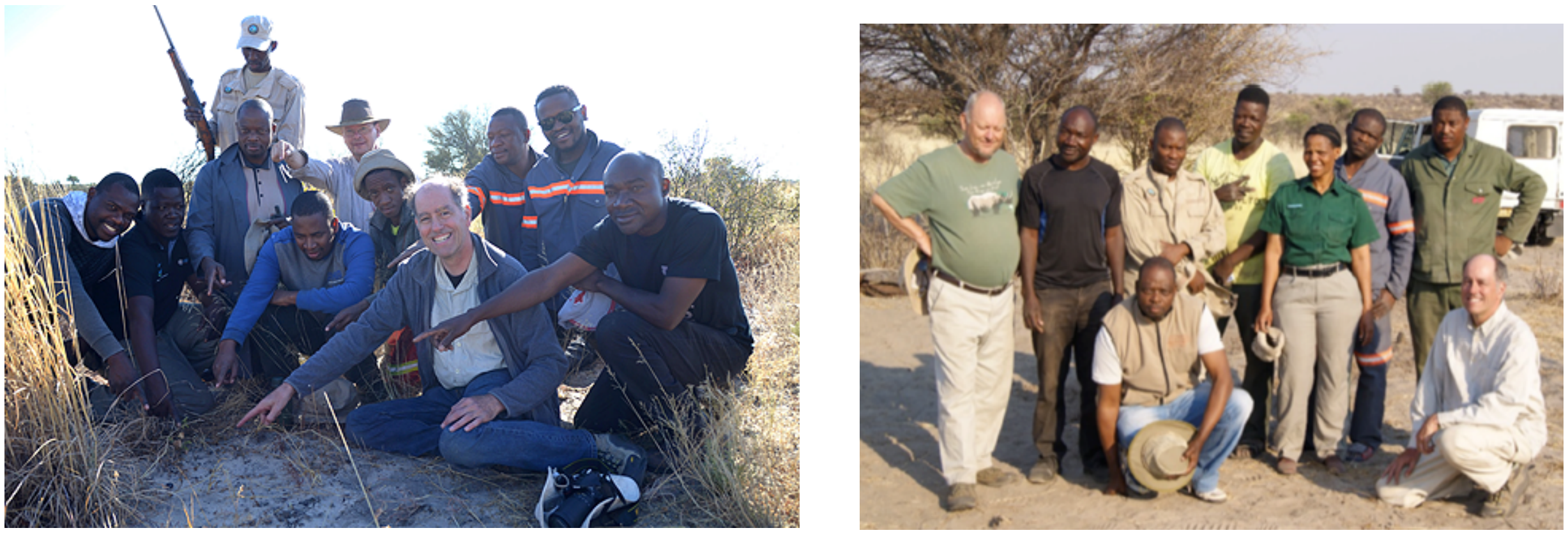

Notes: **(a)** From left: Thembiso Basupi, Thebe Kemosedile, Mohutsiwa Gabadirwe, Ontebogile Mbebane (with rifle), Lesedi Seitshiro (finder), Alexander Proyer, Reginald G. Gababaloke, Peter Jenniskens, Tom Kenny Tom, Phemo Moleje, Oliver Moses, and Jarious Kaekane (taking picture). **(b)** From left: Sitting: Mohutsiwa Gabadirwe and Peter Jenniskens; From left standing: Tim Cooper, Oliver Moses, Kagiso Kgetse, Thebe Kemosedile, Sarah Tsenene, Kabelo Dikole, Babutsi Mosarwa and Odirile Sempho (taking picture).

**Fig. 3.** Asteroid 2018 LA in space (top left image by the Catalina Sky Survey) and the first 23 meteorites recovered on the ground as photographed in-situ (scale of each figure is about 4 × 4 cm). Meteorites are shown in order of find (MP-01 top left, to MP-23 bottom right).

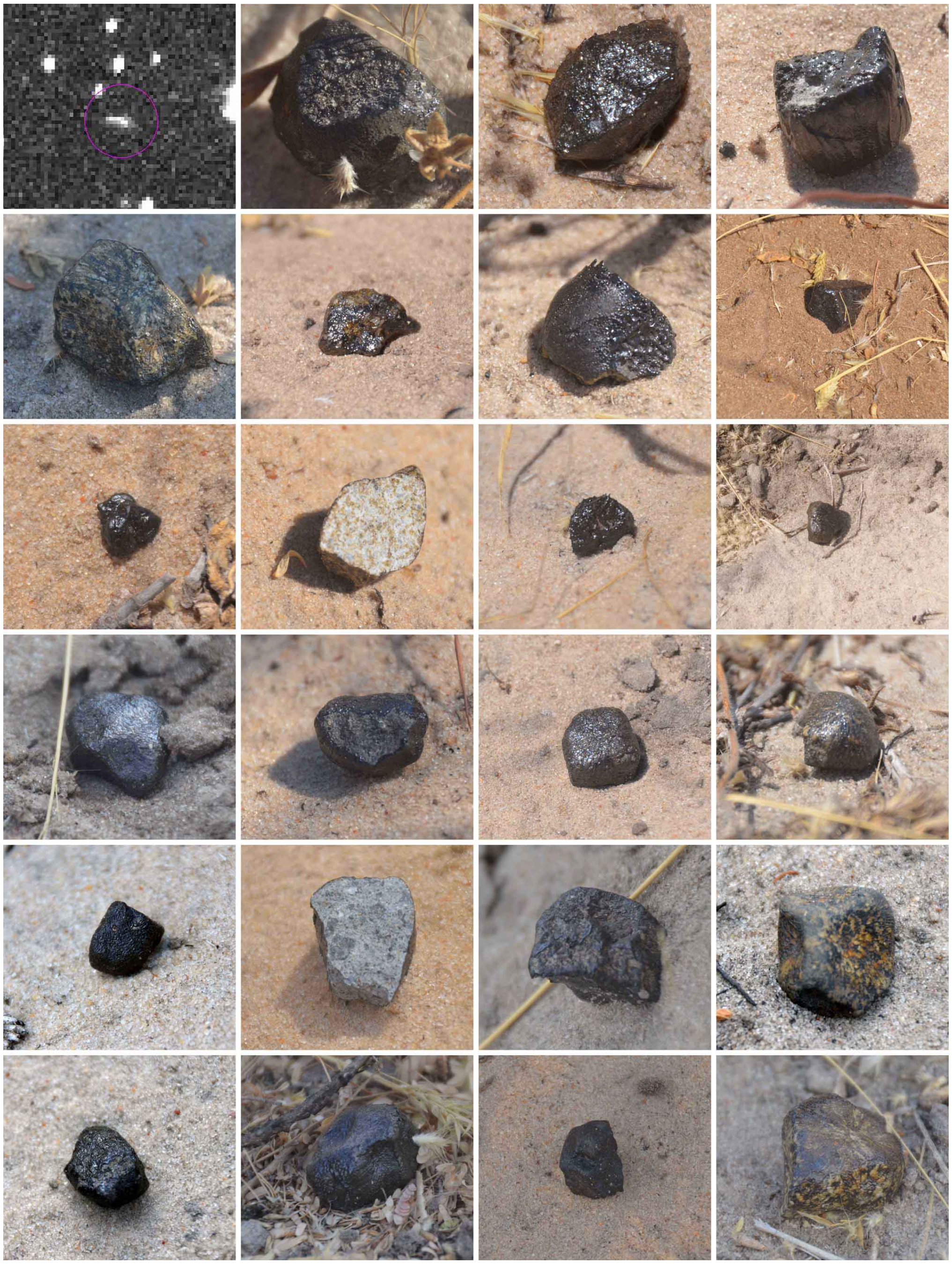

**Fig. 4.** Asteroid spin period. Best sinusoidal fit to the brightness for an ellipsoid c = b (left) and for a triaxial ellipsoid c/b = 0.58 (right). Light blue points = Catalina data near first observation (t=0) and dark blue = near t = 80 min; red = ATLAS data near 200 min; green = SkyMapper data near 330 min.

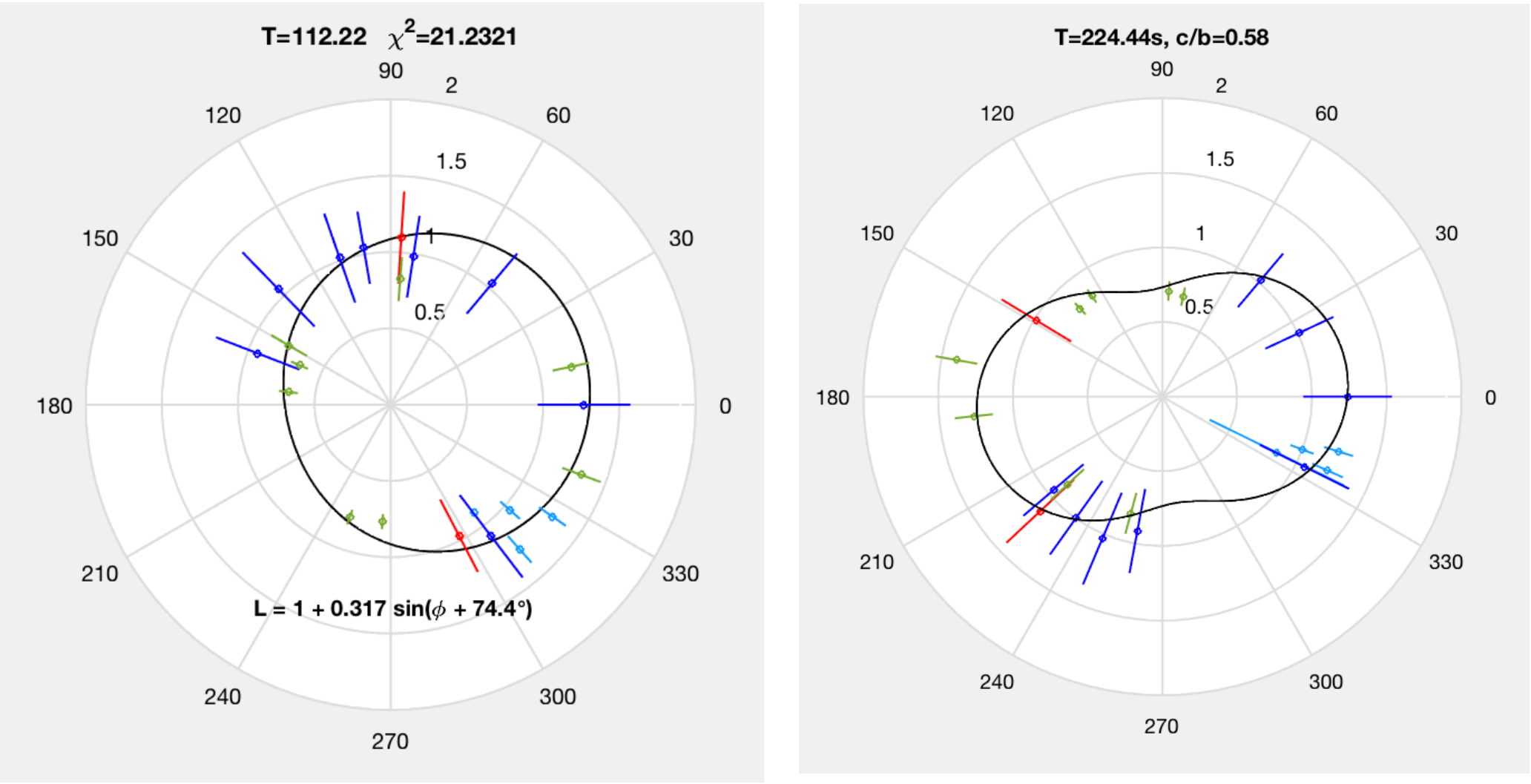

**Fig. 5.** Asteroid shape. Measure of best fit (Log of $\chi^2$) for different axis ratios c/a and b/a.

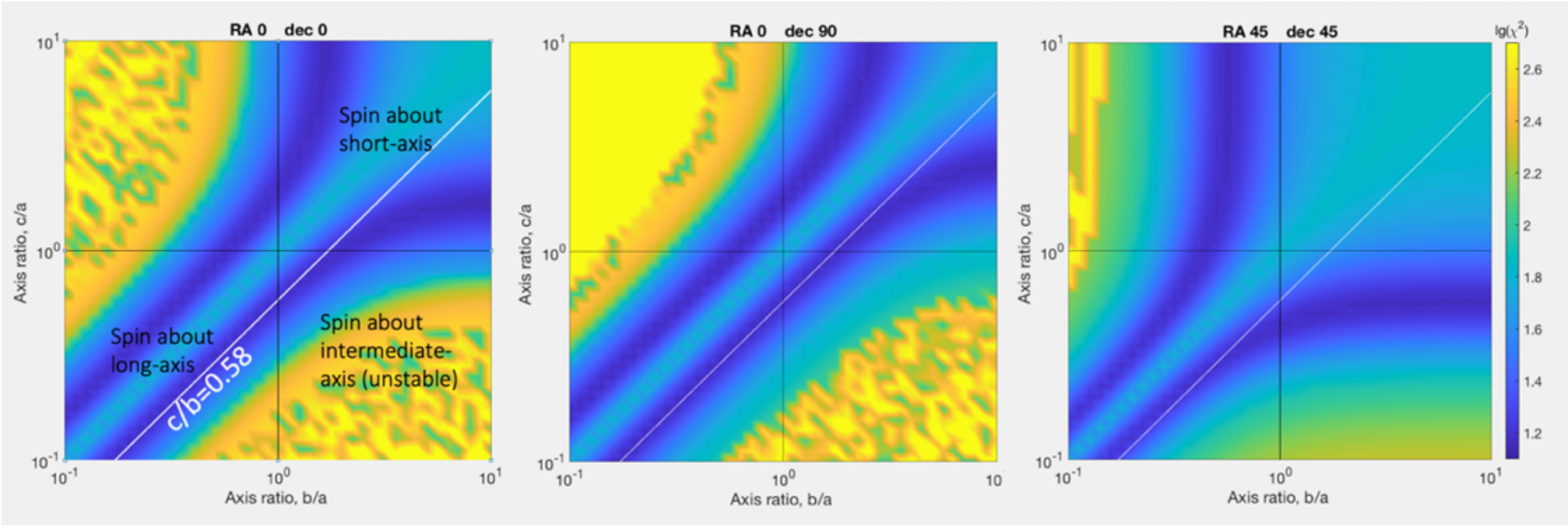

**Fig. 6.** Video observations and calibrations. **(a)** Small tree at Maun Lodge; **(b)** Shadow from tree on kitchen roof just prior to peak brightness. Corresponding features are marked; **(c)** Overview of camera locations and direction to the flare; **(d)** Two-frame difference image shows light from meteor in restaurant door opening at Rakops; **(e)** Door opening showing two poles that cast shadow; **(f)** Sign in Ghanzi; **(g)** Shadow of sign on tiled floor (left); **(h)** Meteor in video frame from Ottosdal; **(i)** Meteor in video frame from Gaborone; **(j)** Calibration of Gaborone video with path of meteor marked.

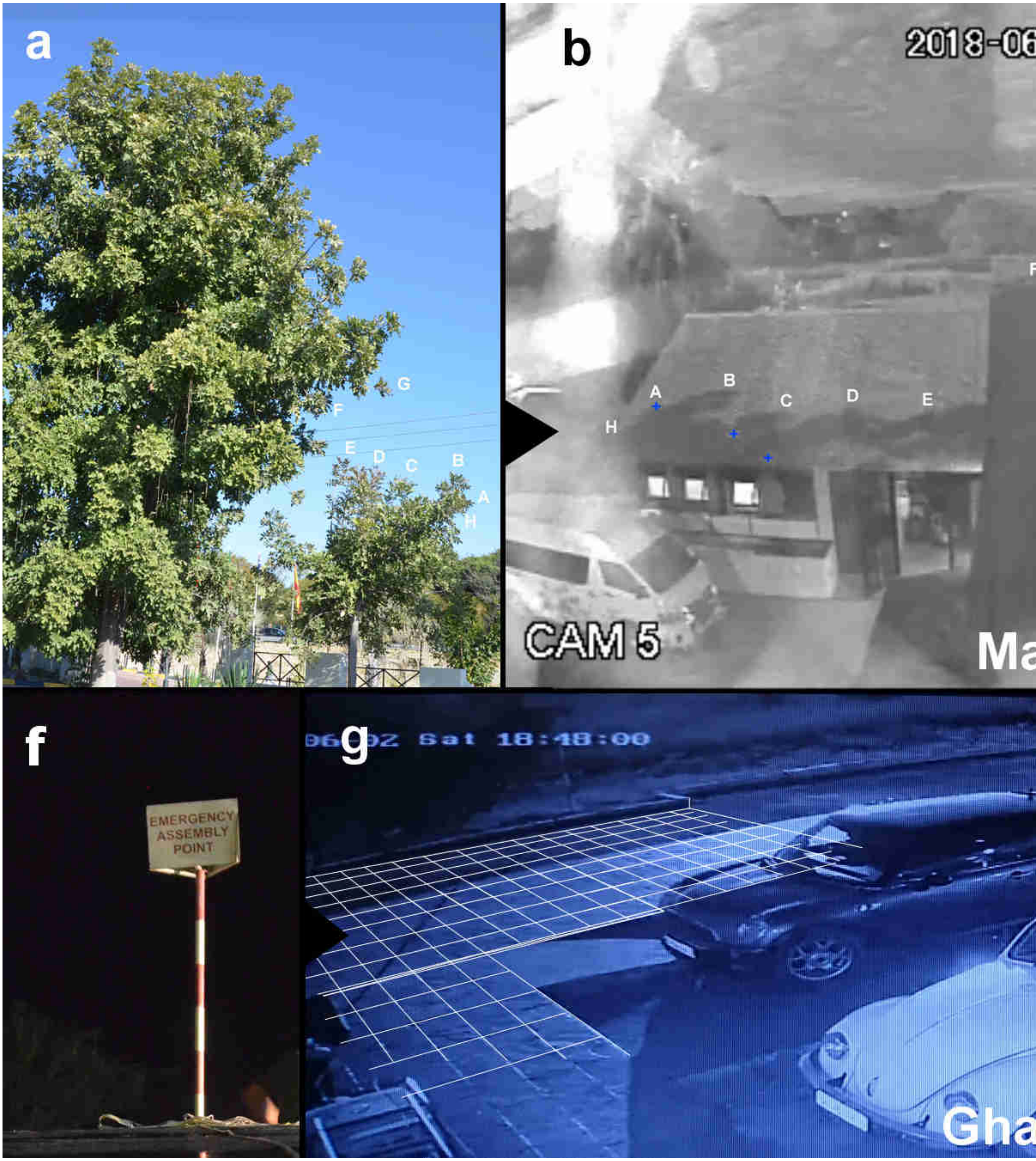

**Fig. 7.** Ground-projected trajectory of asteroid 2018 LA, with orange circles marking the position in 1-km intervals, from 31 km right to 25 km left. Black cross and ellipse show the asteroid position at 27.8 km altitude, while red cross and ellipse show the video-derived position of disruption at 27.8 km.

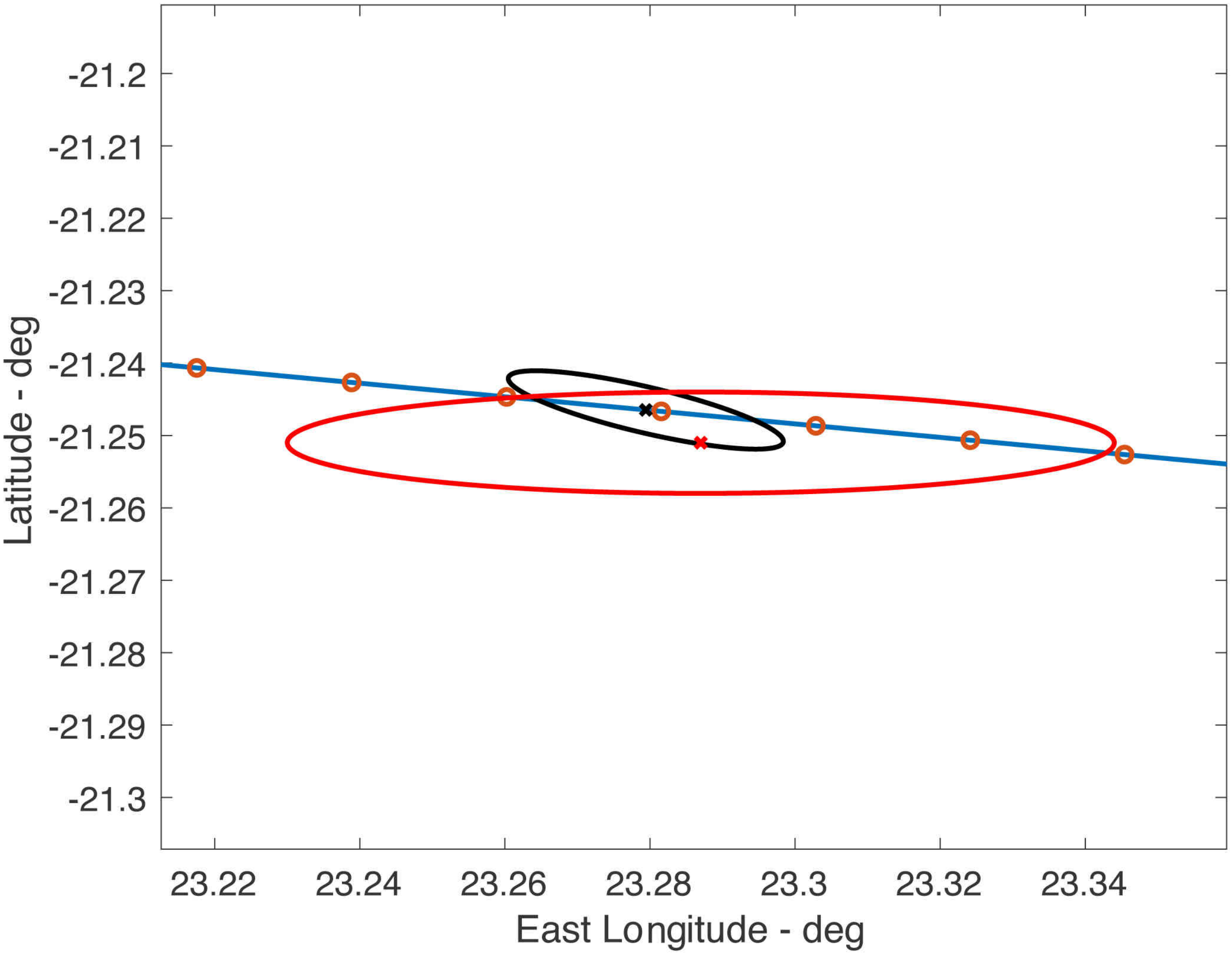

**Fig. 8.** Gaborone meteor video frame (gray) on top of calibration image (color). Letters refer to Table 3.

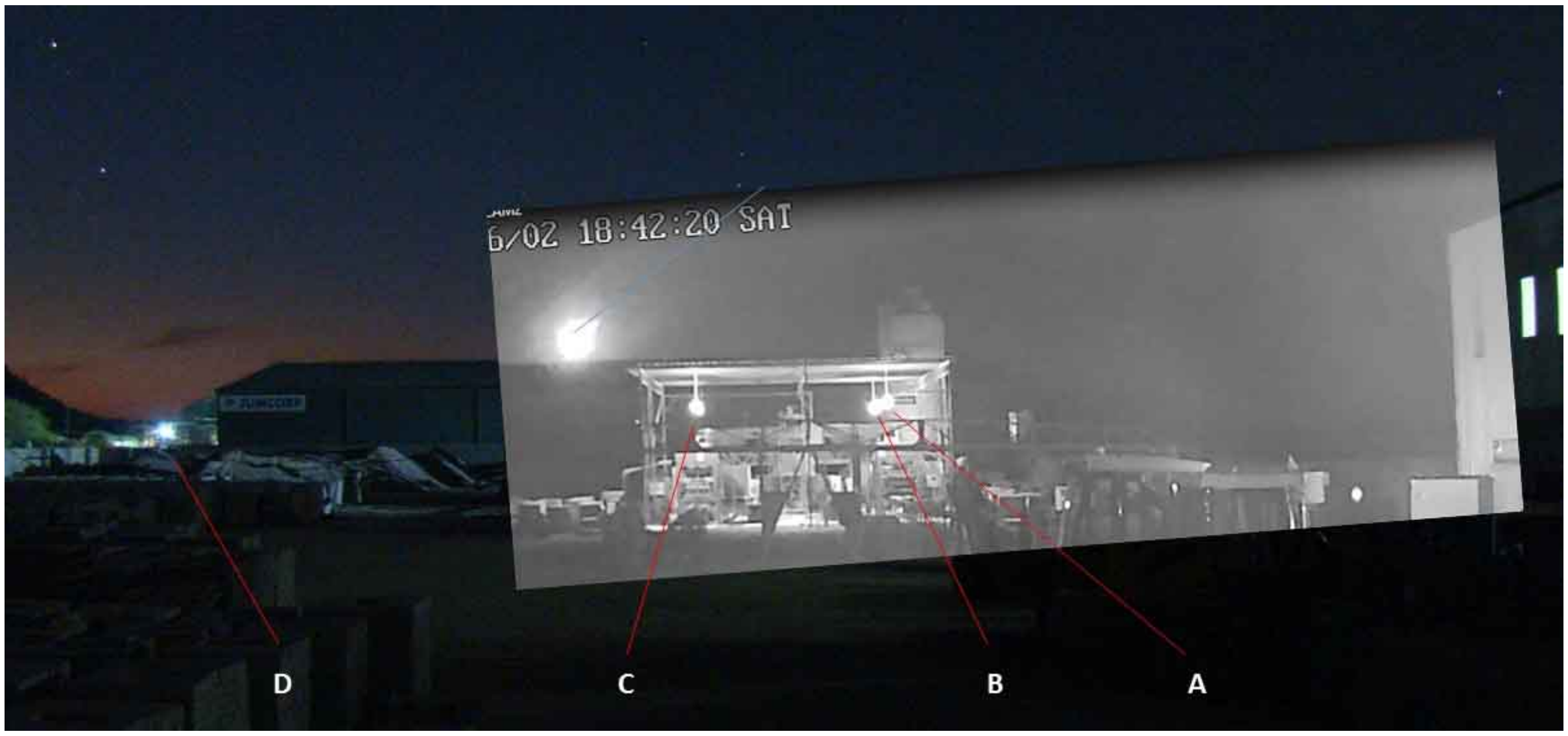

**Fig. 9.** Meteor light curve as a function of altitude. Left: Motopi Pan, with symbols: • = Gaborone video meteor, calibrated to distant lamps; gray circles = Kuruman Radiators South Africa security camera video (Location: 27.47178°S, 23.43287°E, +21m, courtesy of Christian Matthys Grobler), scaled to Gaborone video intensities; small squares = Maun video shadows, scaled to Gaborone video intensities. Right: Sariçiçek, with "*" marking the point of breakup based on the location of meteorites on the ground (Unsalan *et al.* 2019).

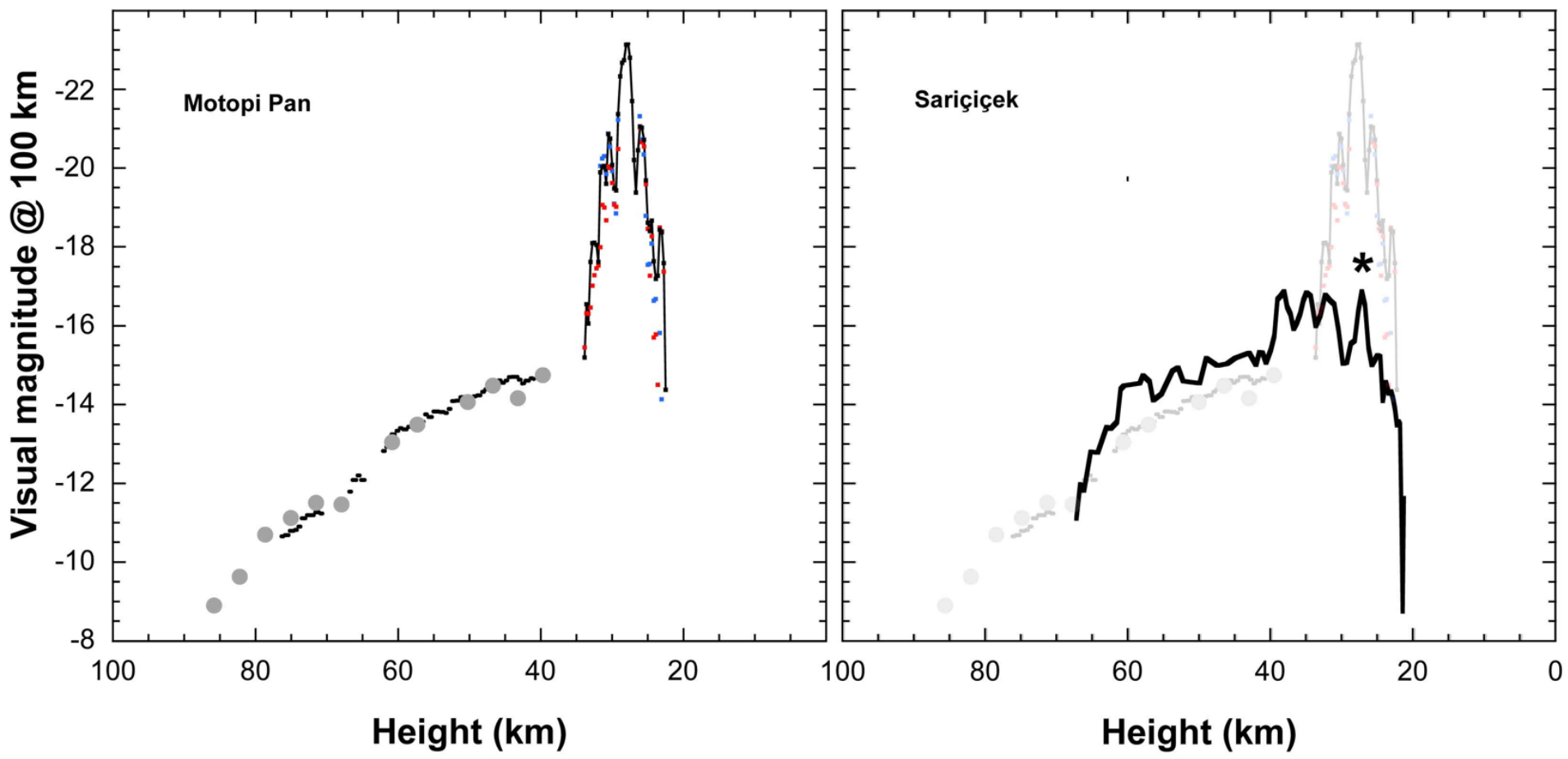

**Fig. 10.** Infrasound signal. Period at maximum amplitude for the bolide signal detected at I47ZA. The dominant period is 3–4 s.

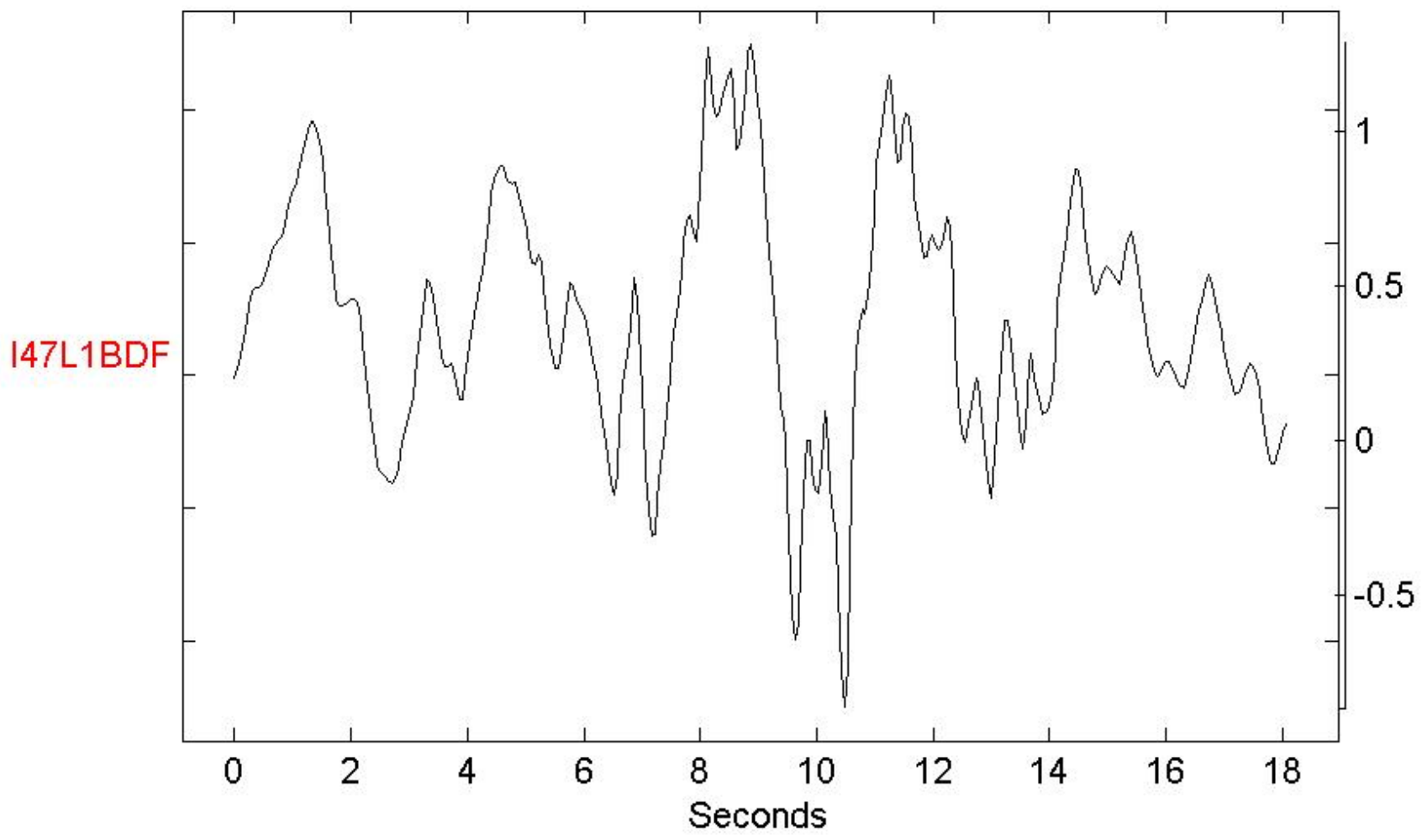

**Fig. 11.** Location of impact orbit in a semi-major axis vs. inclination diagram. Background pattern is the expected distribution of impact orbits for a 20-cm sized meteoroid originating from an inner-belt source region with inclination = 1° (left) and 10° (right) (Jenniskens *et al.* 2020). The large ellipse represents the 3σ measurement of semi-major axis and inclination of the orbit for Sariçiçek, while the single pixel dot that for Motopi Pan (marked by arrows).

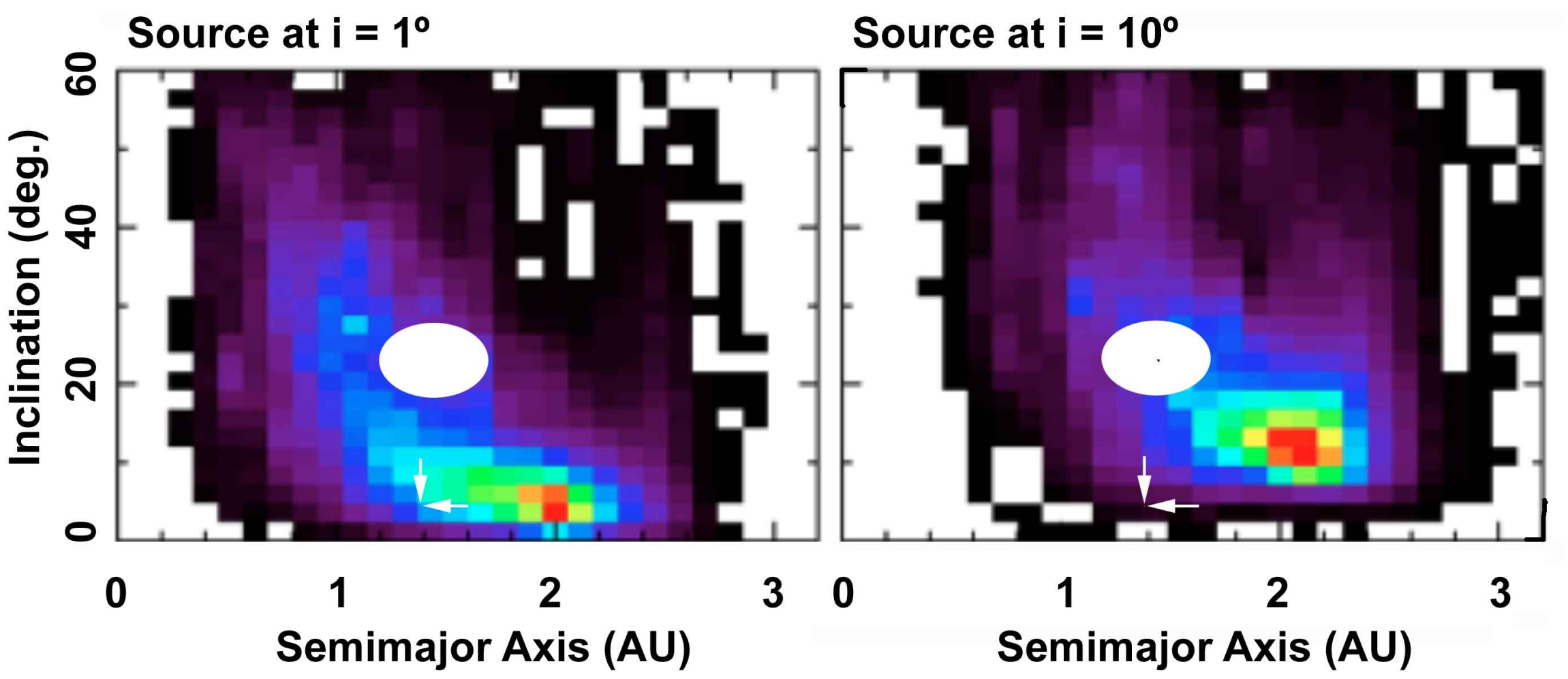

**Fig. 12.** Wind direction along the track of the falling meteorites.

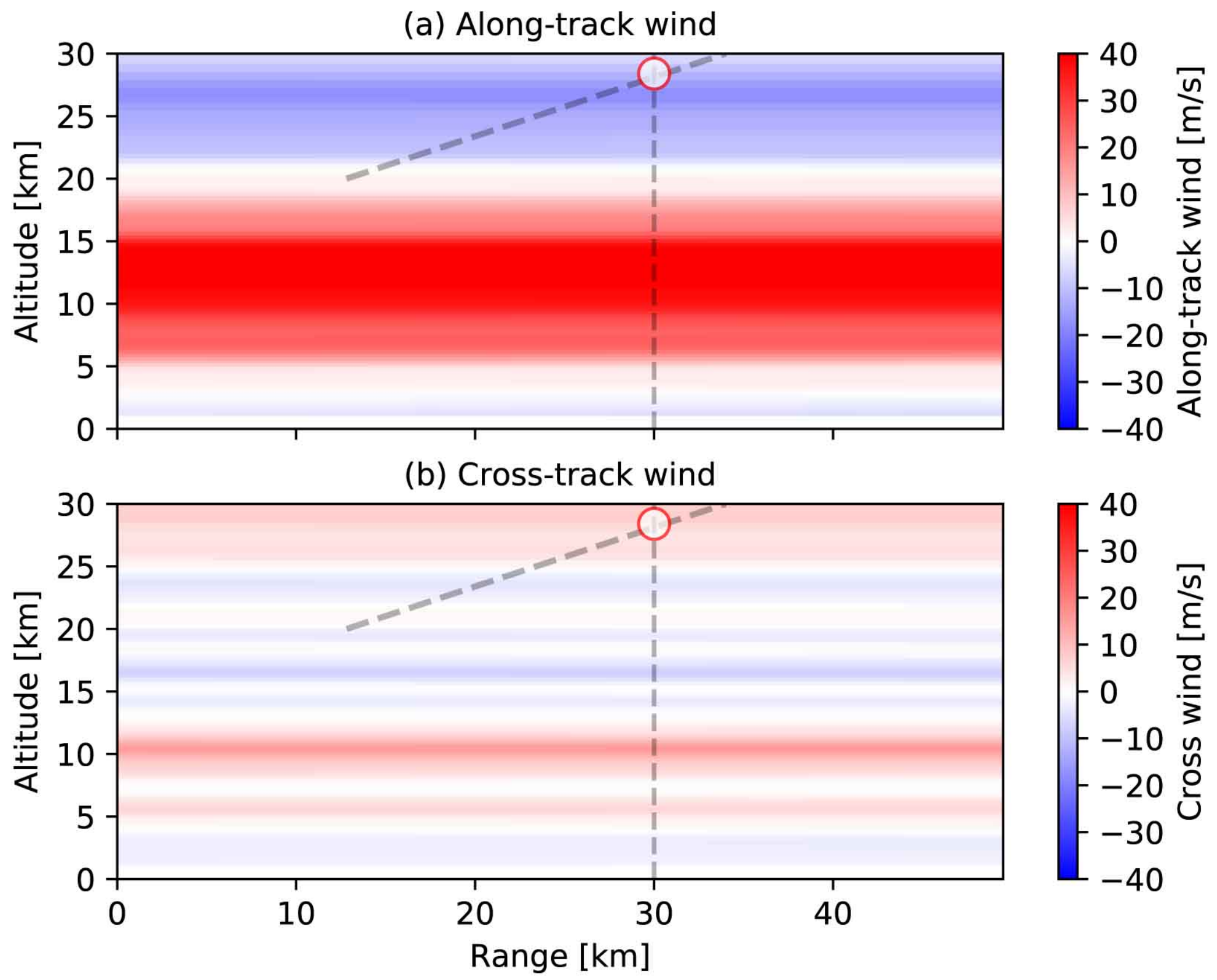

**Fig. 13.** Predicted location of fallen meteorites prior to find. Monte-Carlo derived meteorite strewn field based on preliminary position of disruption (labeled "Start", based on asteroid trajectory and Ottosdal video) for masses of >10 kg (red), 3.0–10 kg (orange), 1.0–3.0 kg (yellow), 0.3–1.0 kg (green), 0.1–0.3 kg (cyan), < 0.1 kg (blue).

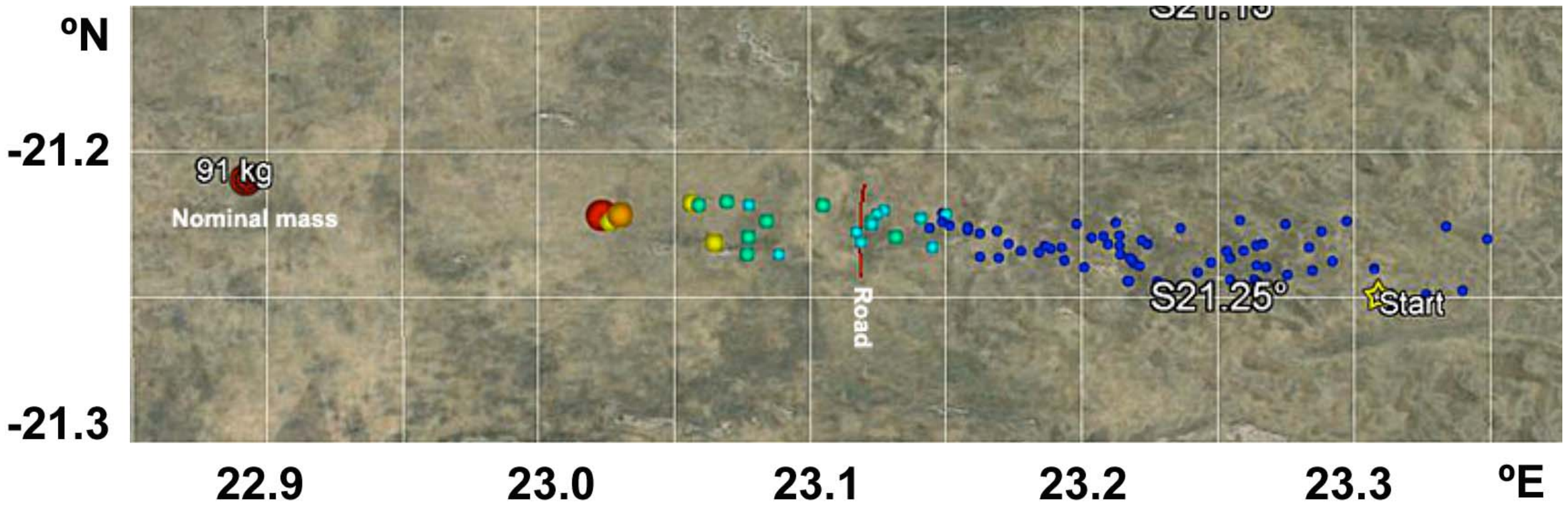

**Fig. 14.** Final solution to the 2018 LA disruption location and that of recovered meteorites. Ground-projected trajectory (blue line, marked every 1 km in altitude, descending right to left) with asteroid position at time of disruption (black ellipse, 1-σ uncertainty), and disruption location at 27.8 km altitude derived from distant video observations (red ellipse, 2-σ). Black dots show the calculated location of meteorite fall positions due to winds and friction, while yellow dots are the recovered meteorites and orange marks the area searched. MP-01 was found ~30-m from camp.

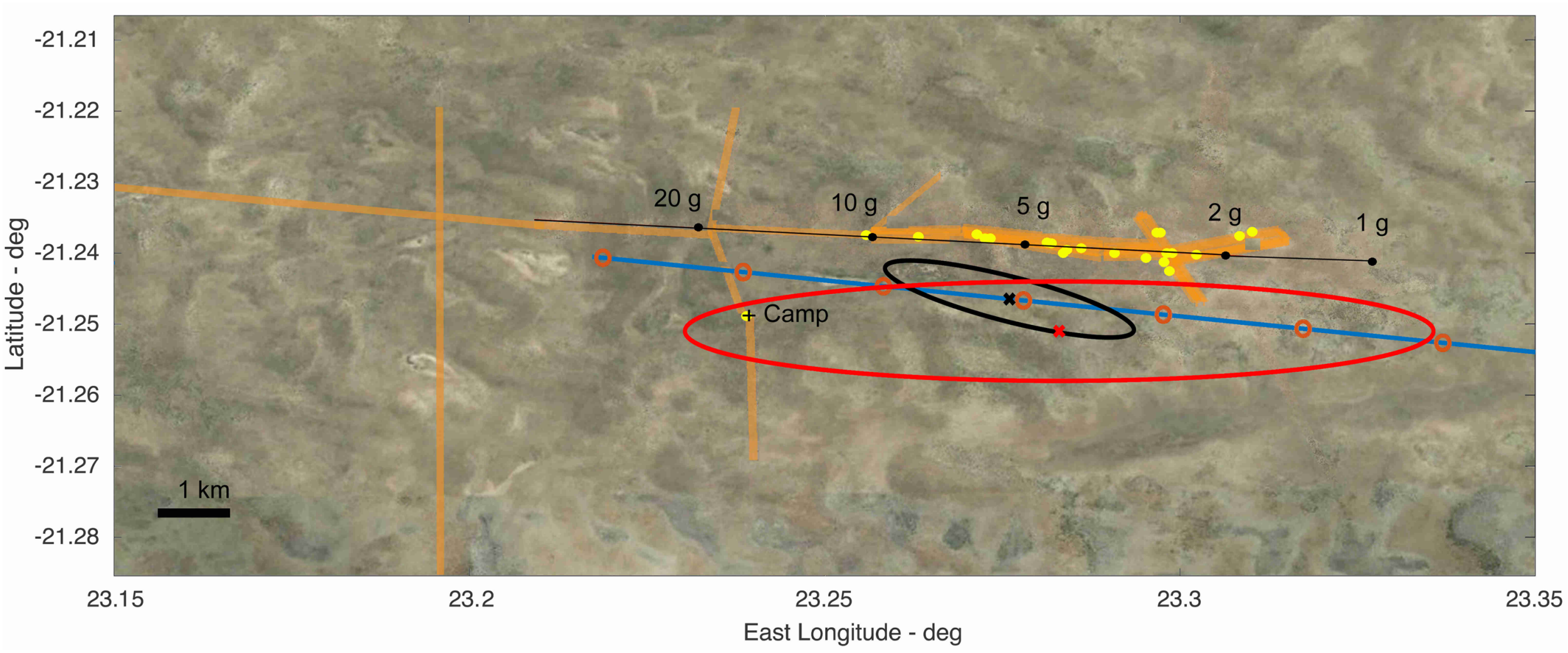

**Fig. 15.** X-ray Computed Tomography of meteorite MP-01. Shown is one cross section at 7 μm/voxel. Light shades are metals and iron sulfides, dark are predominantly silicates.

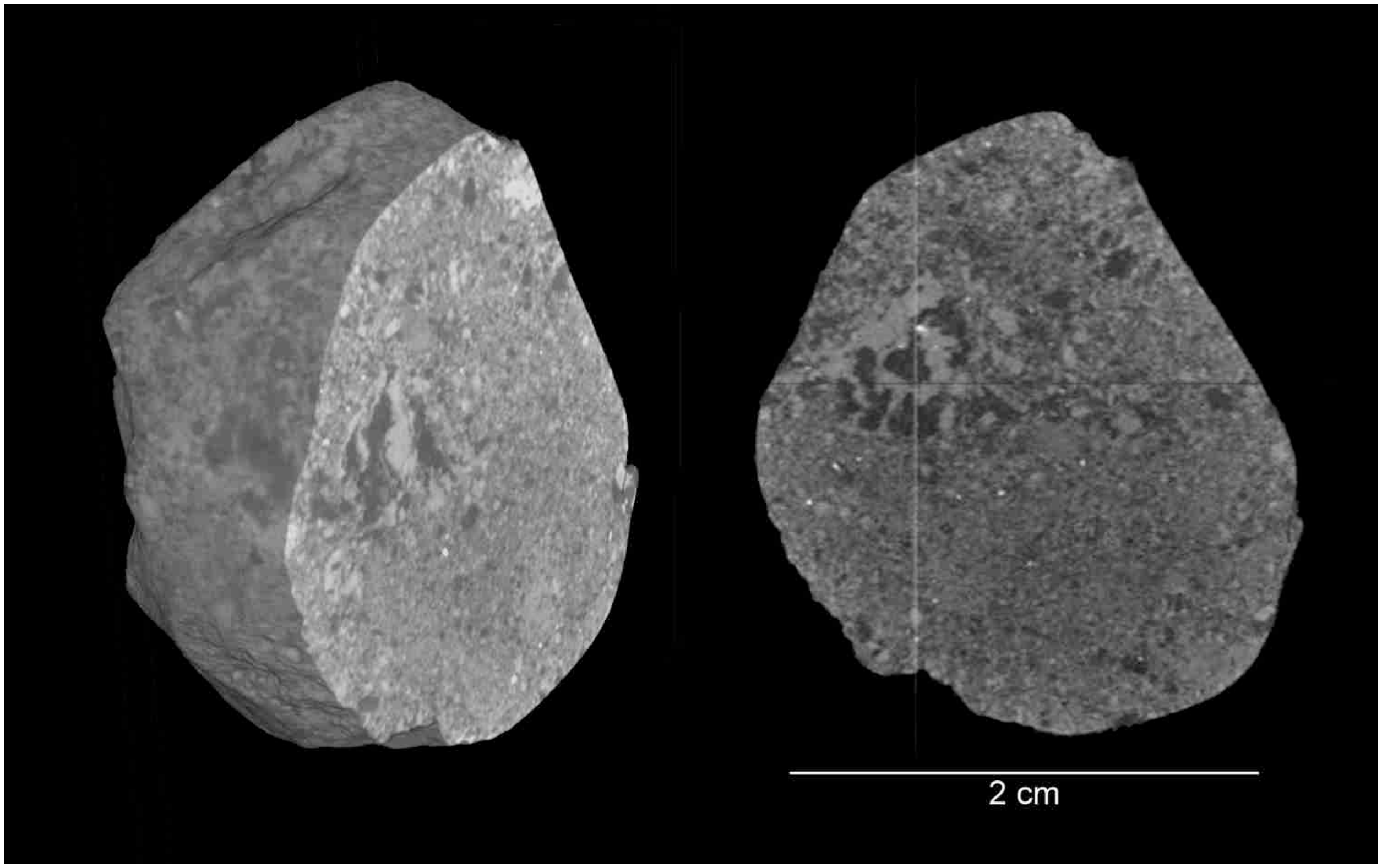

**Fig. 16.** Mosaic of MP-06 and MP-09 from EPMA. Scale bars are 1 mm long.

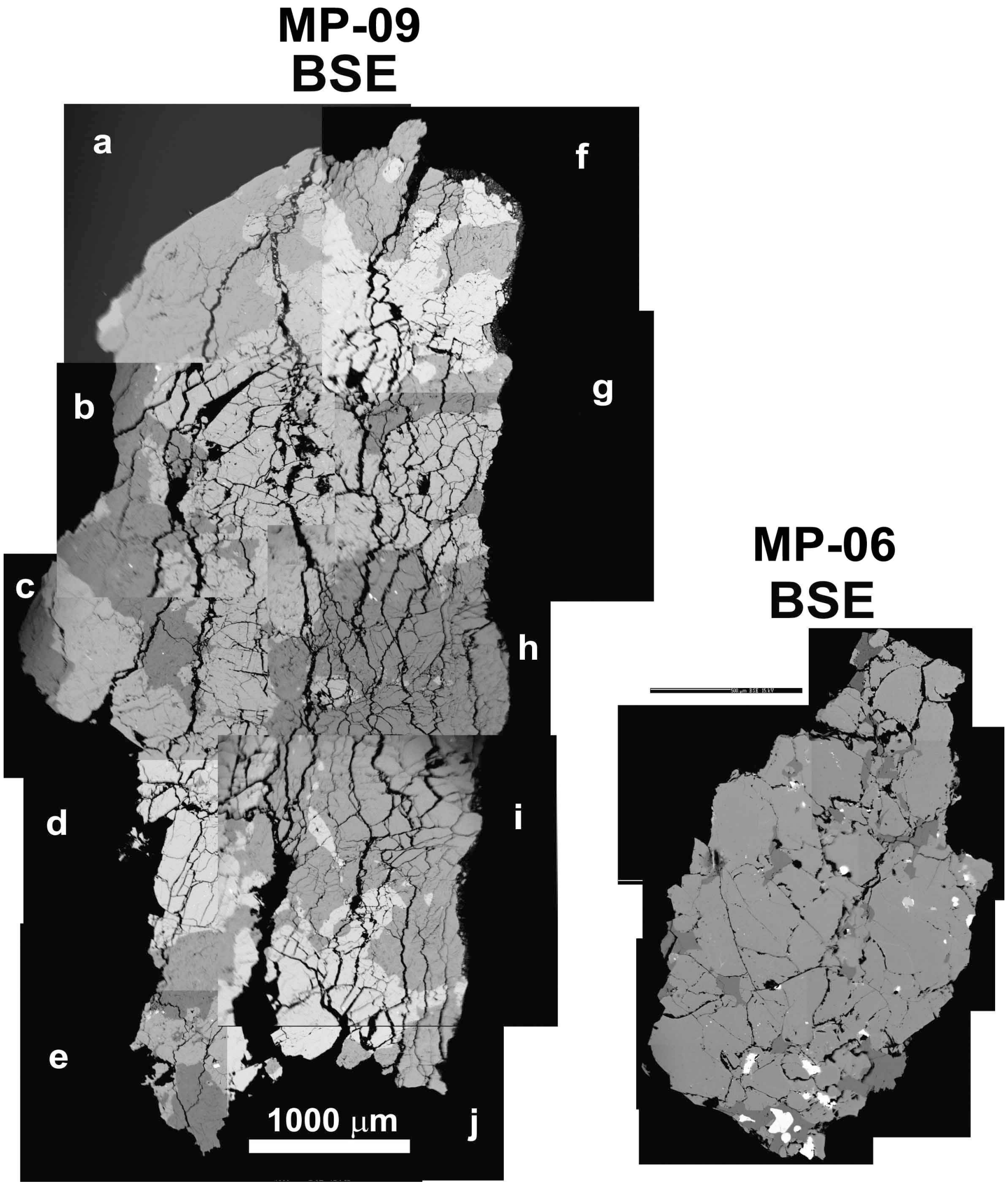

**Fig. 17.** BSE panoramas for the TIMA-X. Scalebars are 2 mm long.

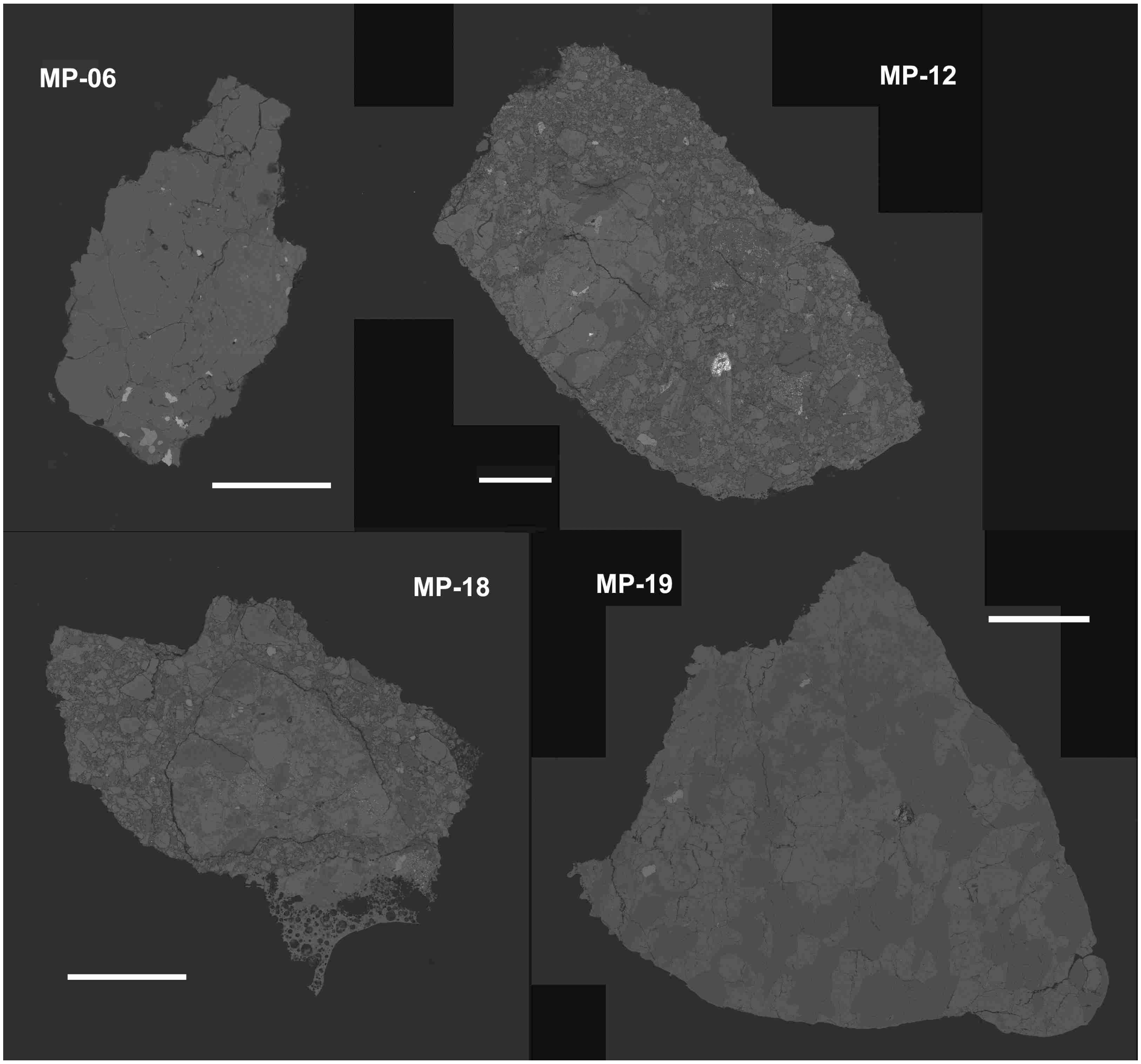

**Fig. 18.** Mg-K X-ray maps of fragments from MP-12 and MP-18.

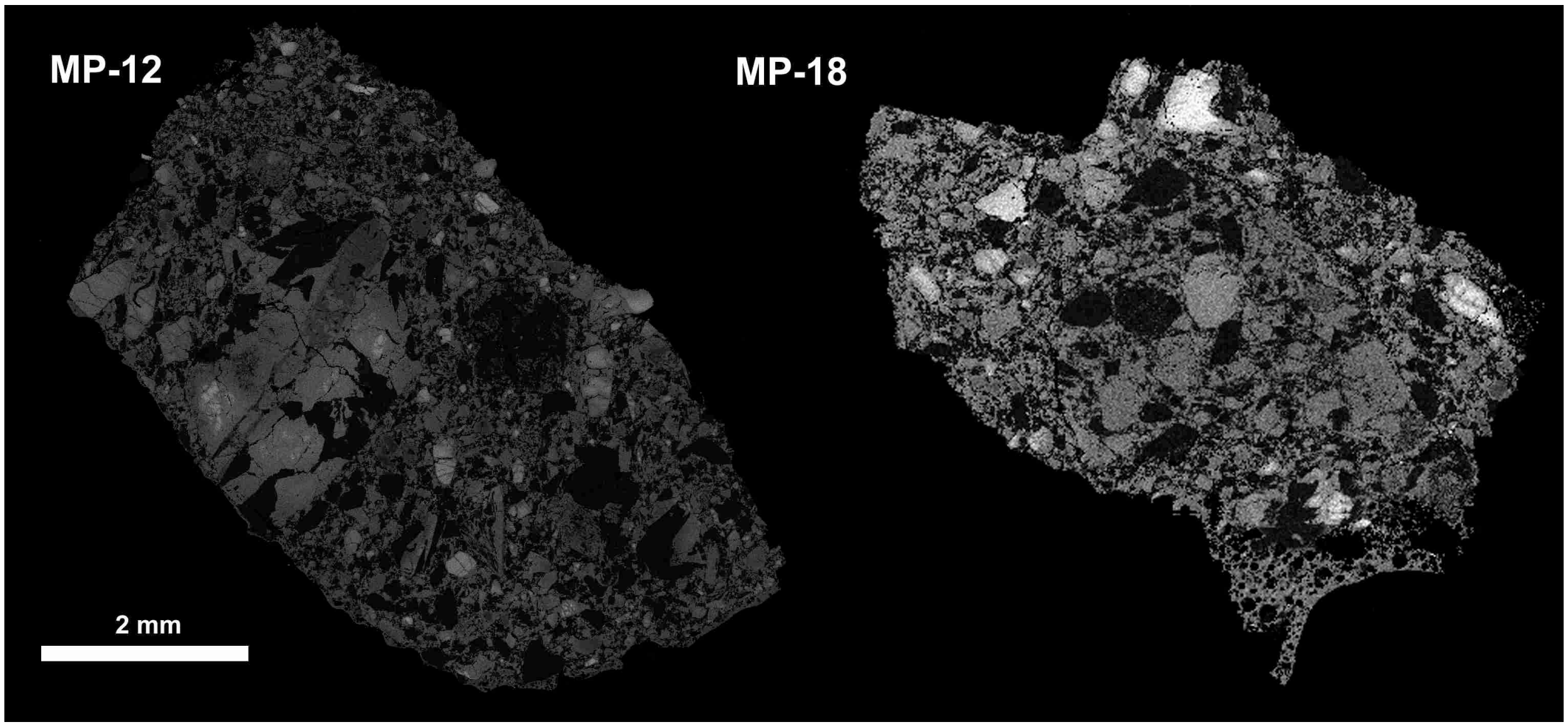

**Fig. 19.** Fragments and powders in white light.

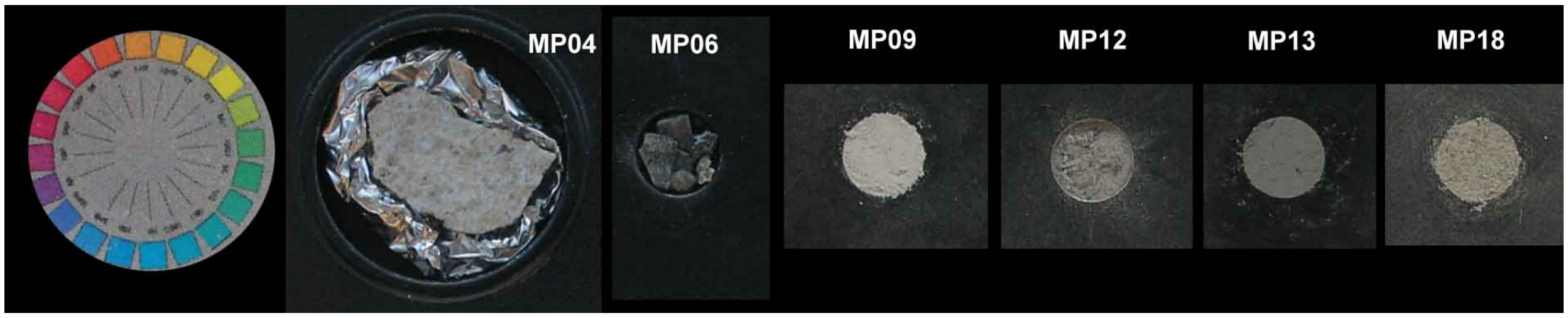

**Fig. 20.** Reflectance spectra over the 0.3–9 μm wavelength range of fragments MP-04 and MP-06 (top) and in powdered form with sizes <45 μm (middle). Bottom panel shows unsorted powders from the cutting of MP-09, -12, -13, and -18.

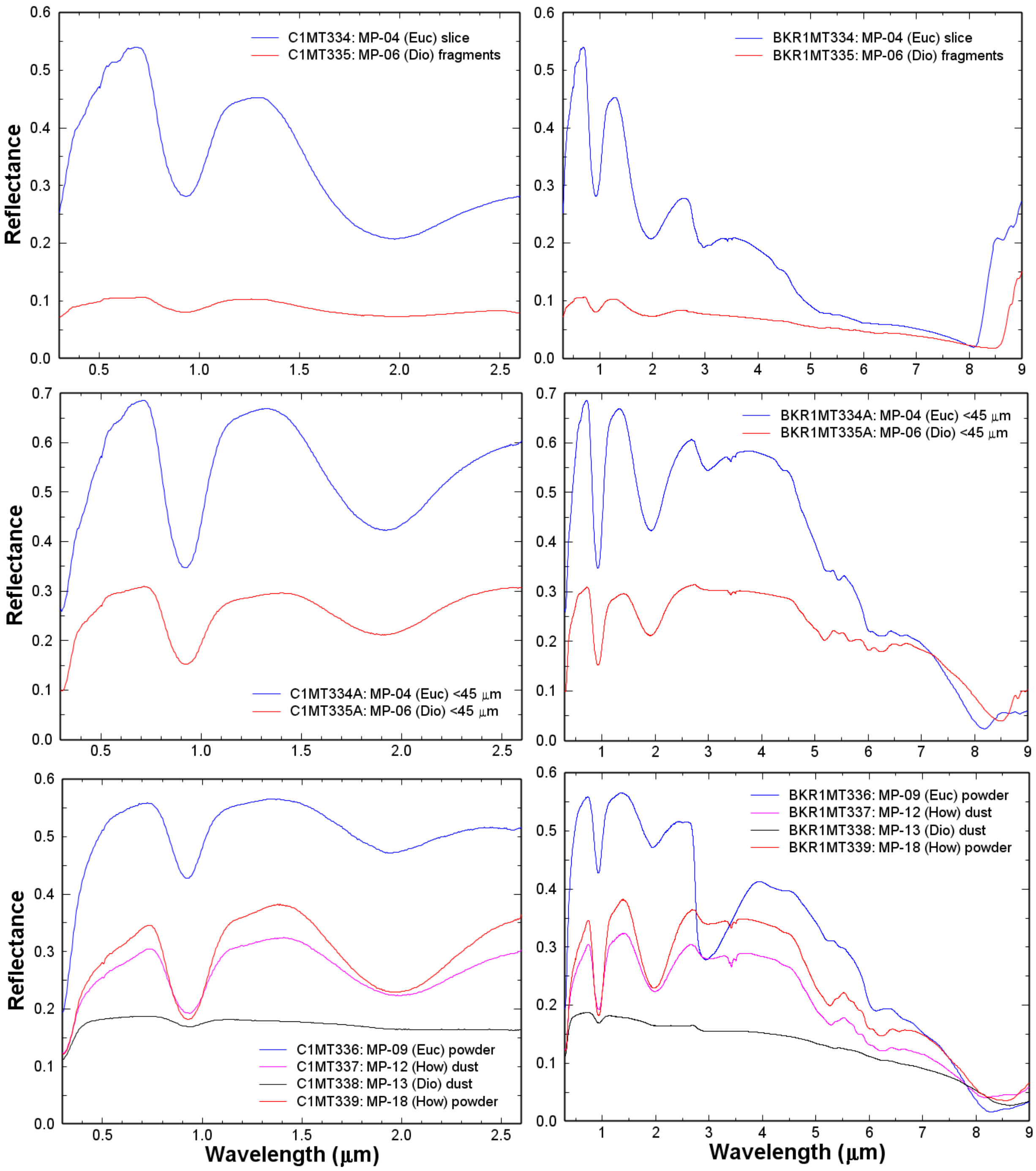

**Fig. 21.** Reflectance spectra over the 10–50 μm wavelength range of powdered fragments MP-04 and MP-06 with sizes < 45 μm (left). Right panel shows unsorted powders from the cutting of MP-09, -12, -13, and -18.

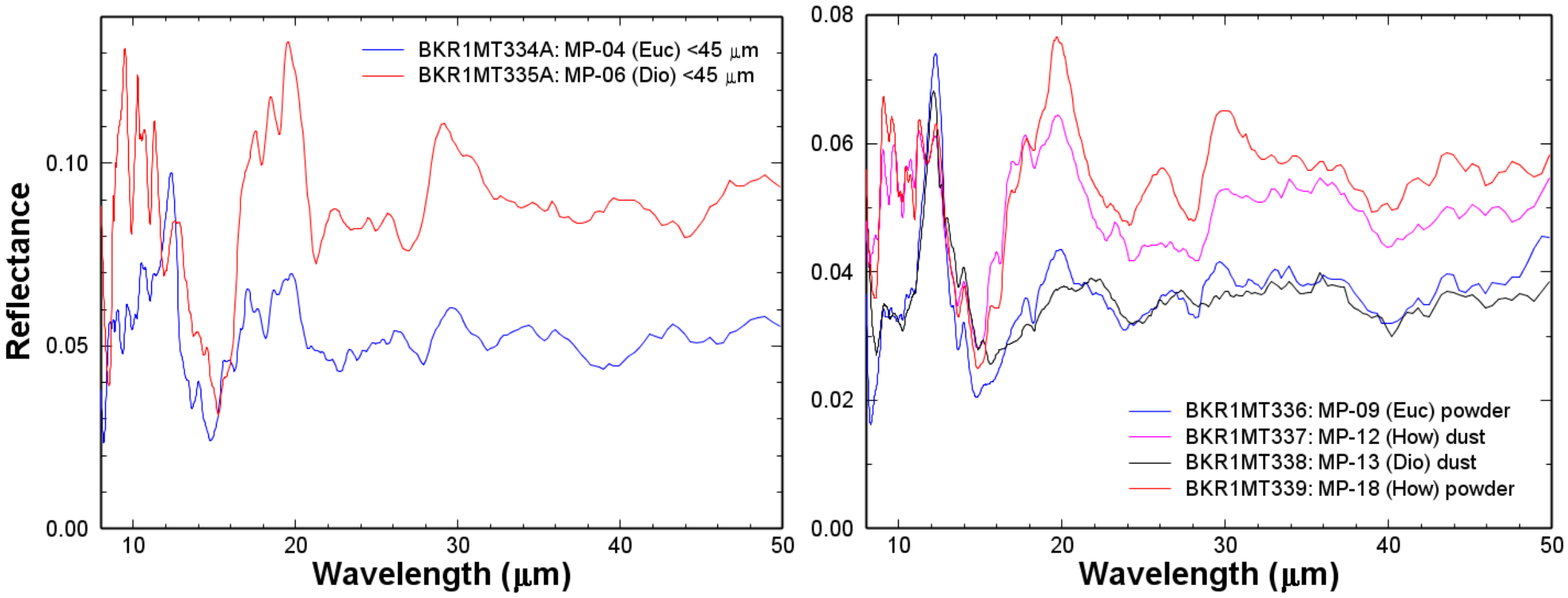

**Fig. 22.** Comparing reflectance spectra of HED meteorites in the RELAB spectral database (thin blue, green and red lines) with those measured for Motopi Pan (black solid and dashed lines).

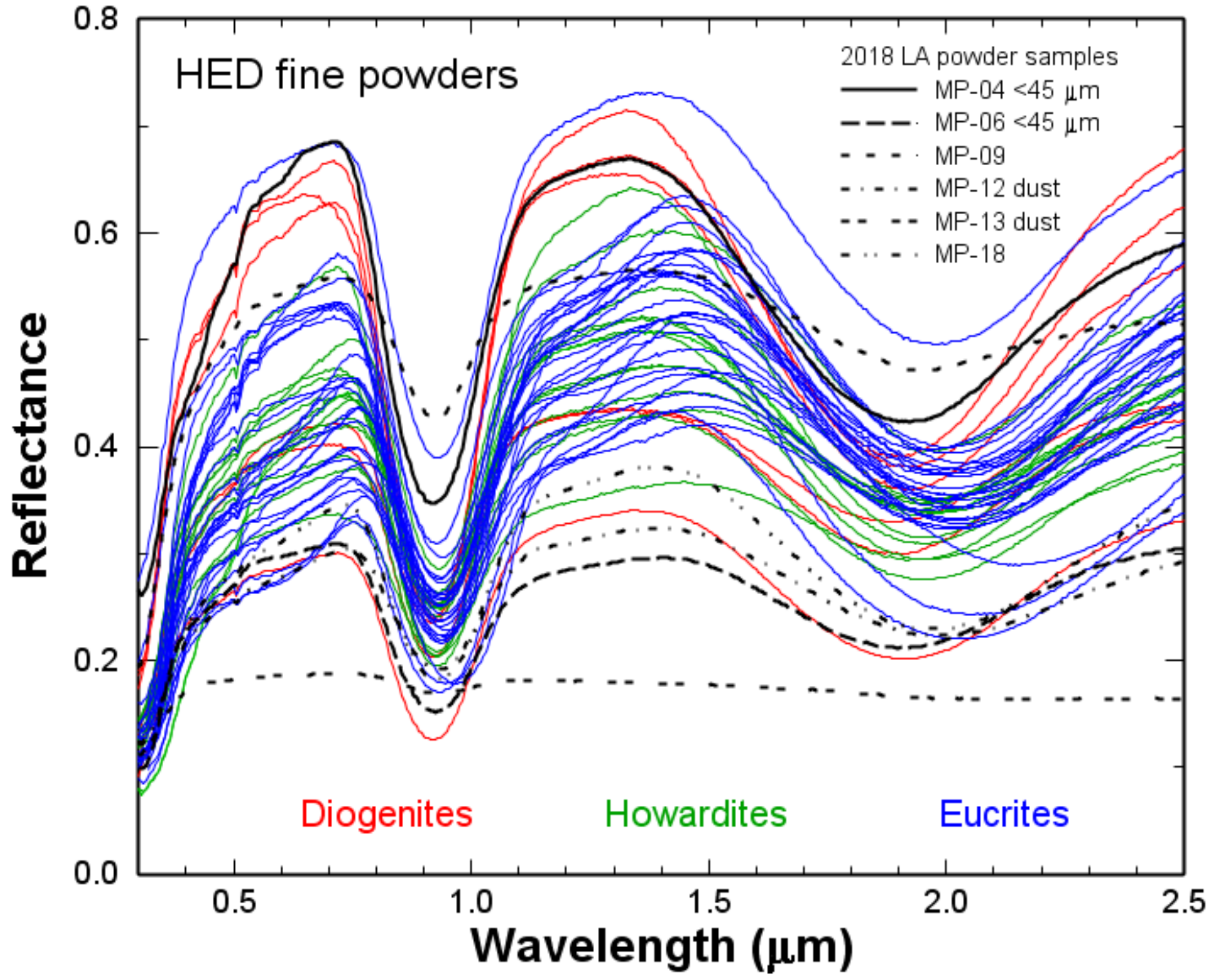

**Fig. 23.** Emissivity spectrum of the samples MP-06 (diogenite), MP-09 (eucrite), and MP-18 (howardite).

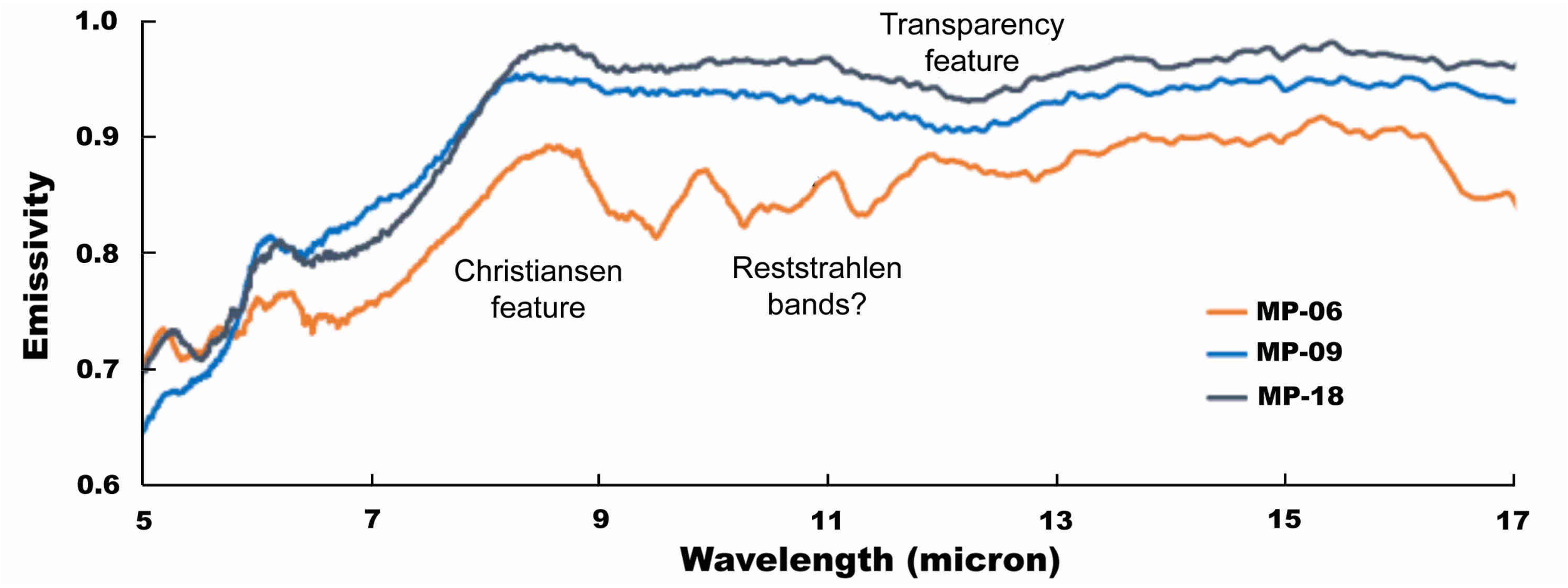

**Fig. 24.** Induced Thermoluminescence of 2018 LA and Sariçiçek compared to other HED meteorites, after Sears *et al.* (2013).

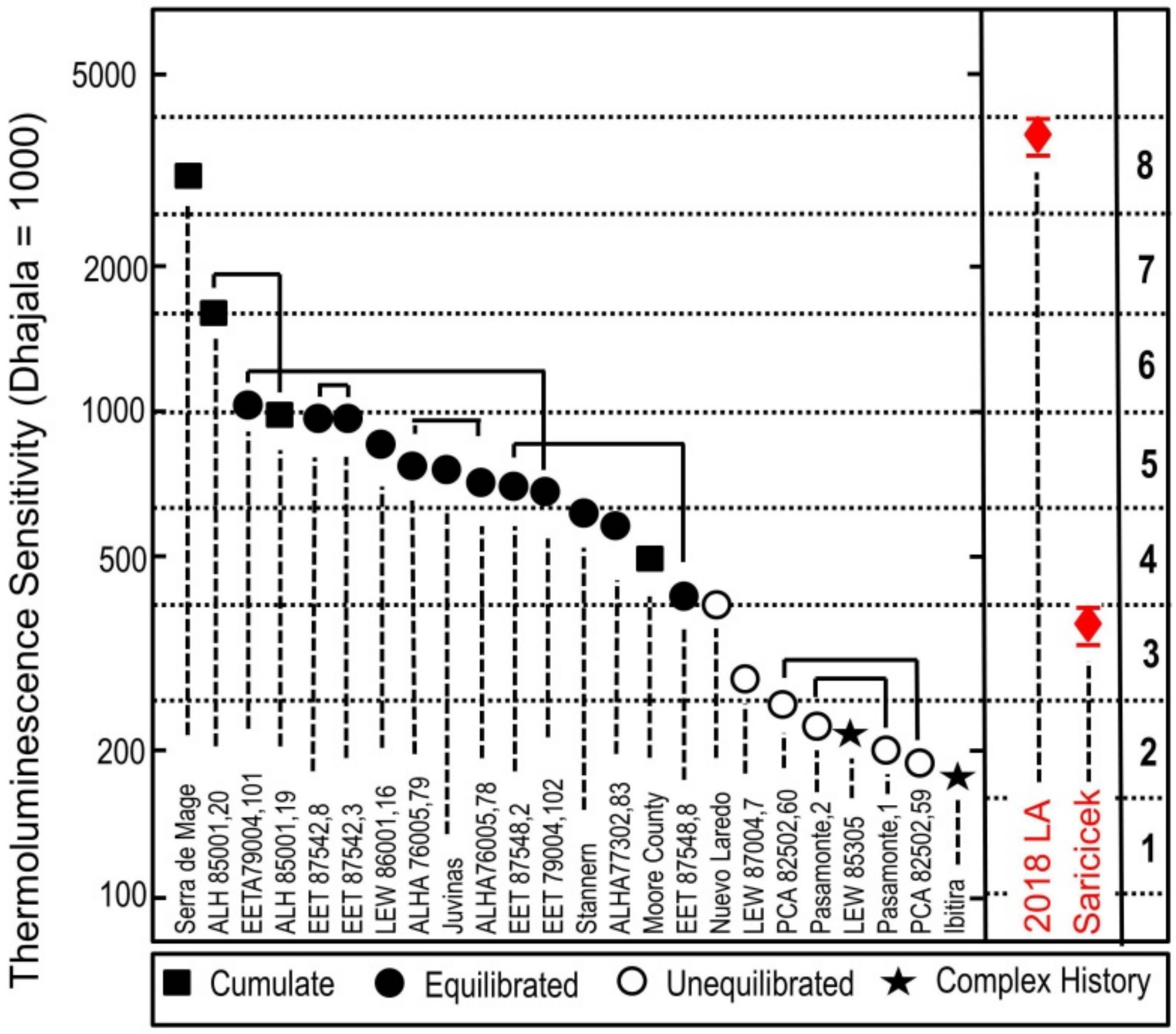

**Fig. 25.** Oxygen isotope diagram. TFL = Terrestrial Fractionation Line.

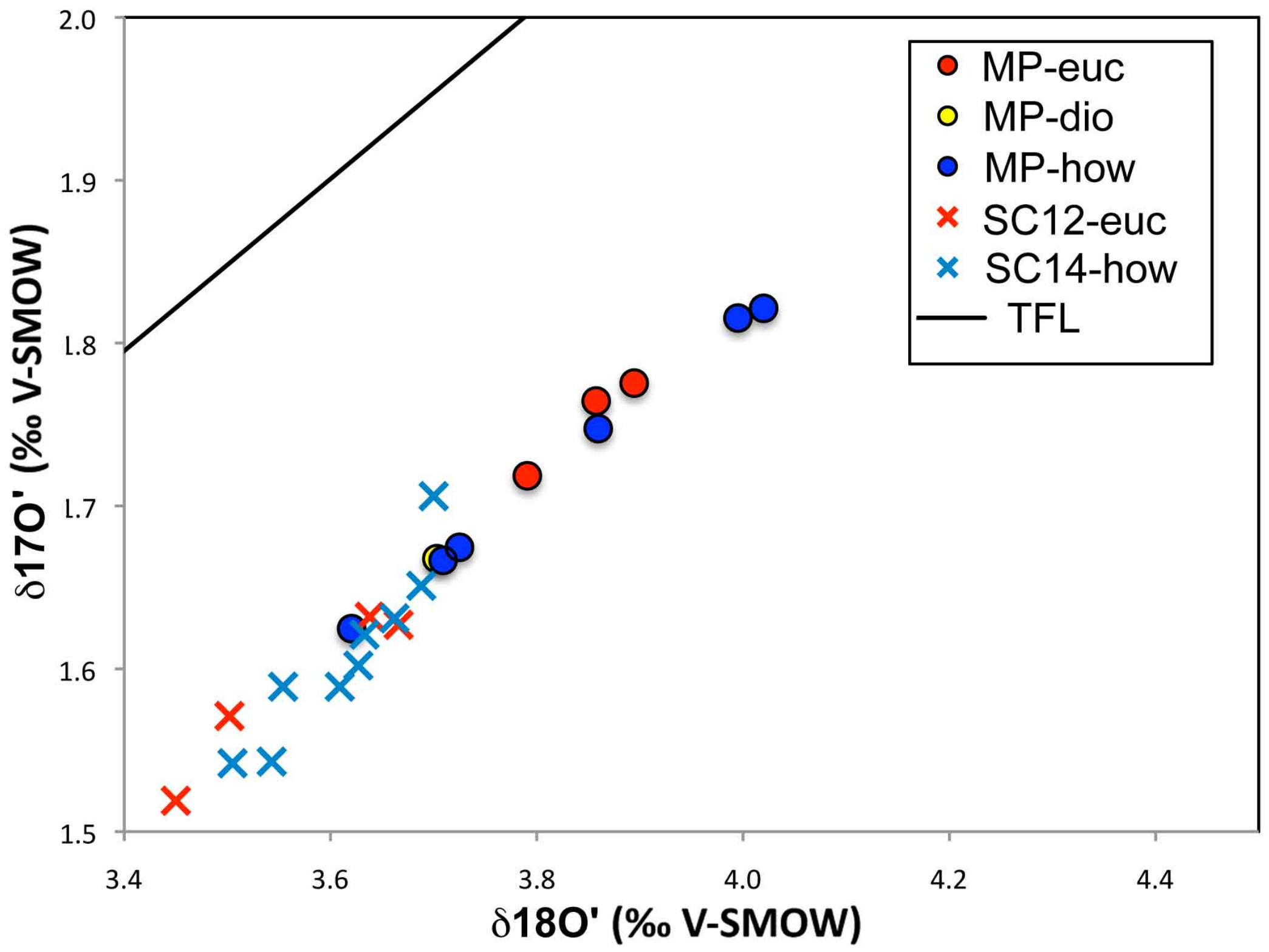

**Fig. 26.** Meteorite cosmochemistry. **(a)** Pb-Pb age from zircons, with the results from individual zircons pointing to an intercept age of 4562.7 ± 6.4 Ma; **(b)** Pb-Pb age from phosphates, with the results from individual apatite grains pointing to an intercept at 4234 ± 41 Ma; **(c)** Oxygen-Chromium isotope diagram with mixing lines for different amounts of certain carbonaceous chondrites. Best-fit CR2 is specifically an admixture with meteorite LAP 02342; **(d)** Effects of exogenous material admixture (Ir) and solar wind implantation ($^{20}$Ne), figure after (Unsalan *et al.* 2019; Warren *et al.* 2009).

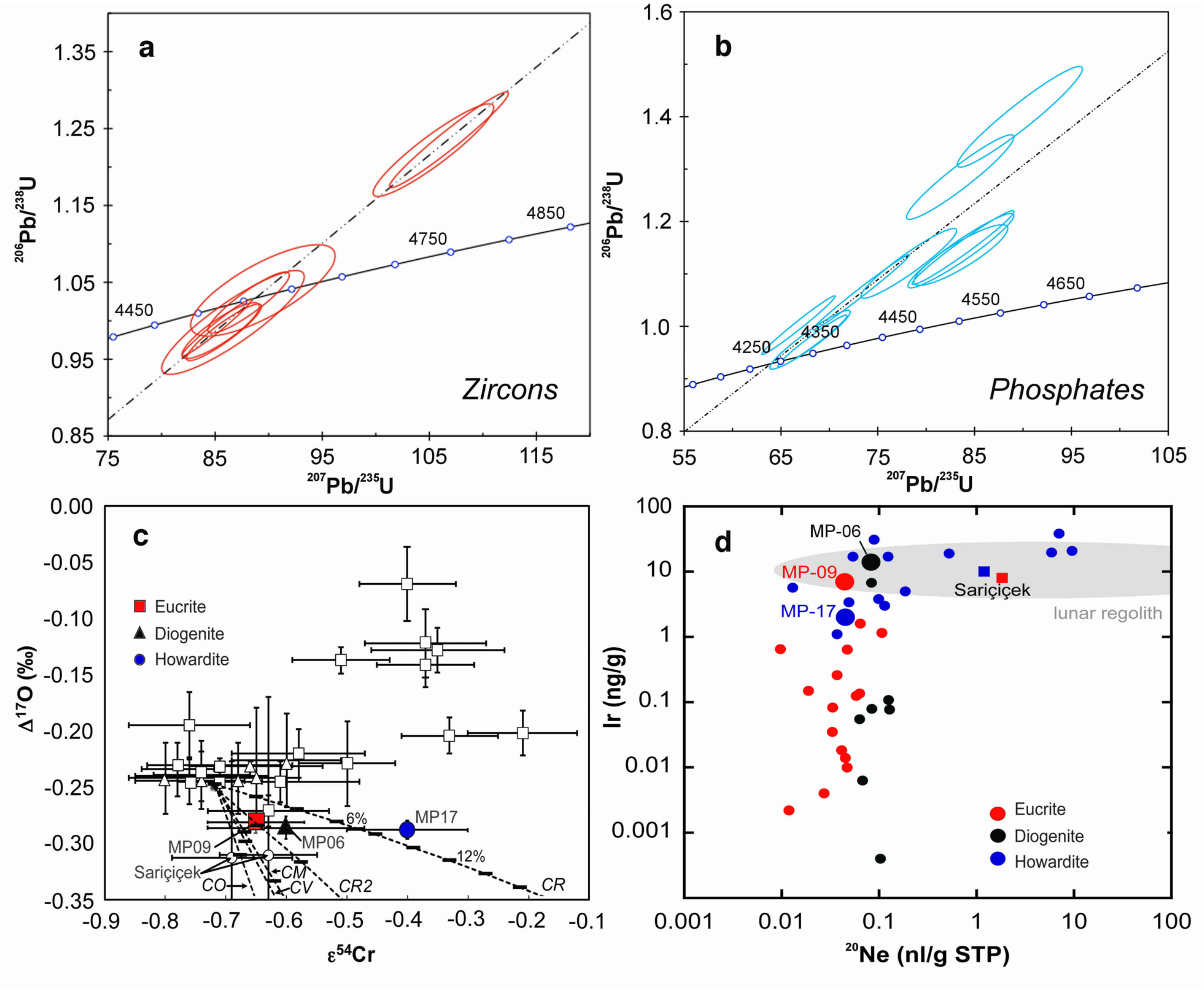

**Fig. 27. (a)** Ne three-isotope plot. All Motopi Pan samples plot close to the cosmogenic Ne endmember, enlarged in **(b)**. Only the eucrite MP-09 contains some trapped Ne ($^{20}Ne_{tr}$ ~0.9 × $10^{-8}$ $cm^3$ STP/g). The cosmogenic $^{21}Ne/^{22}Ne$ ~ 0.895–0.910 endmember was constrained by extrapolation from $Ne_{tr}$ (air or Q-Ne) through the measured data point to a typical $(^{20}Ne/^{22}Ne)_{cos}$ range of 0.704–0.933. The shift of the diogenite MP-06 towards higher $(^{21}Ne/^{22}Ne)_{cos}$ ratios is consistent with the higher Mg content compared to that in eucrites and howardites (see Table 22).

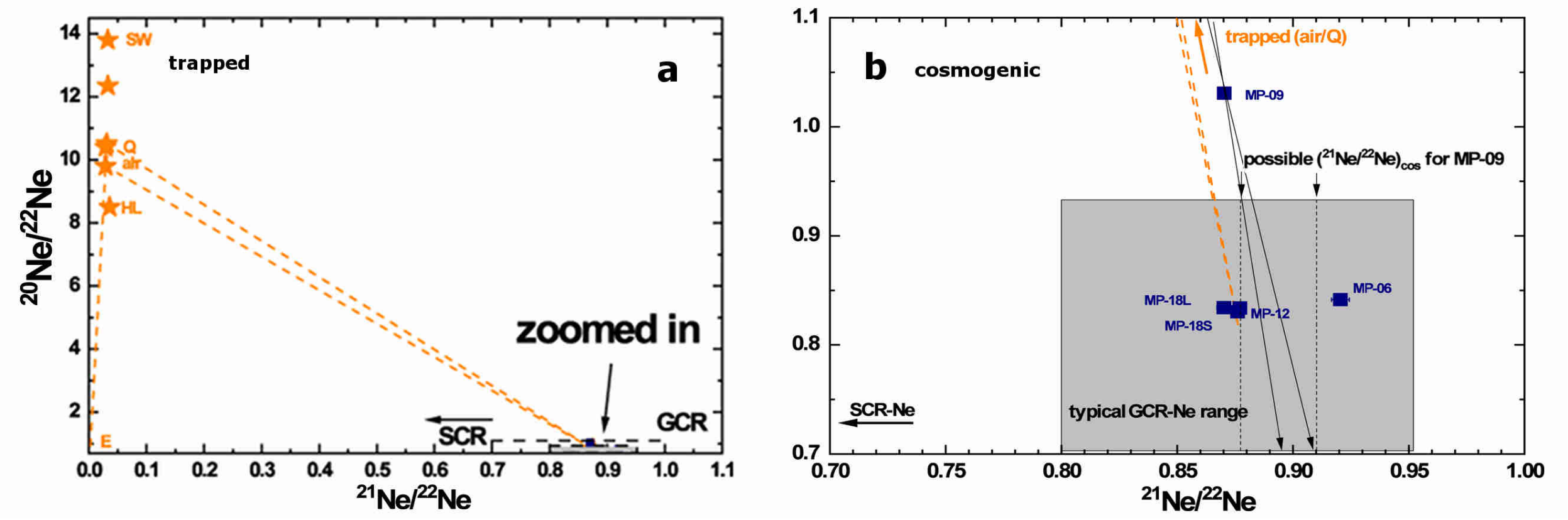

**Fig. 28**. Measured Al and Ca concentrations in four Motopi Pan meteorite samples compared to values in other howardites, eucrites, and diogenite falls and finds from Beck *et al.* (2012).

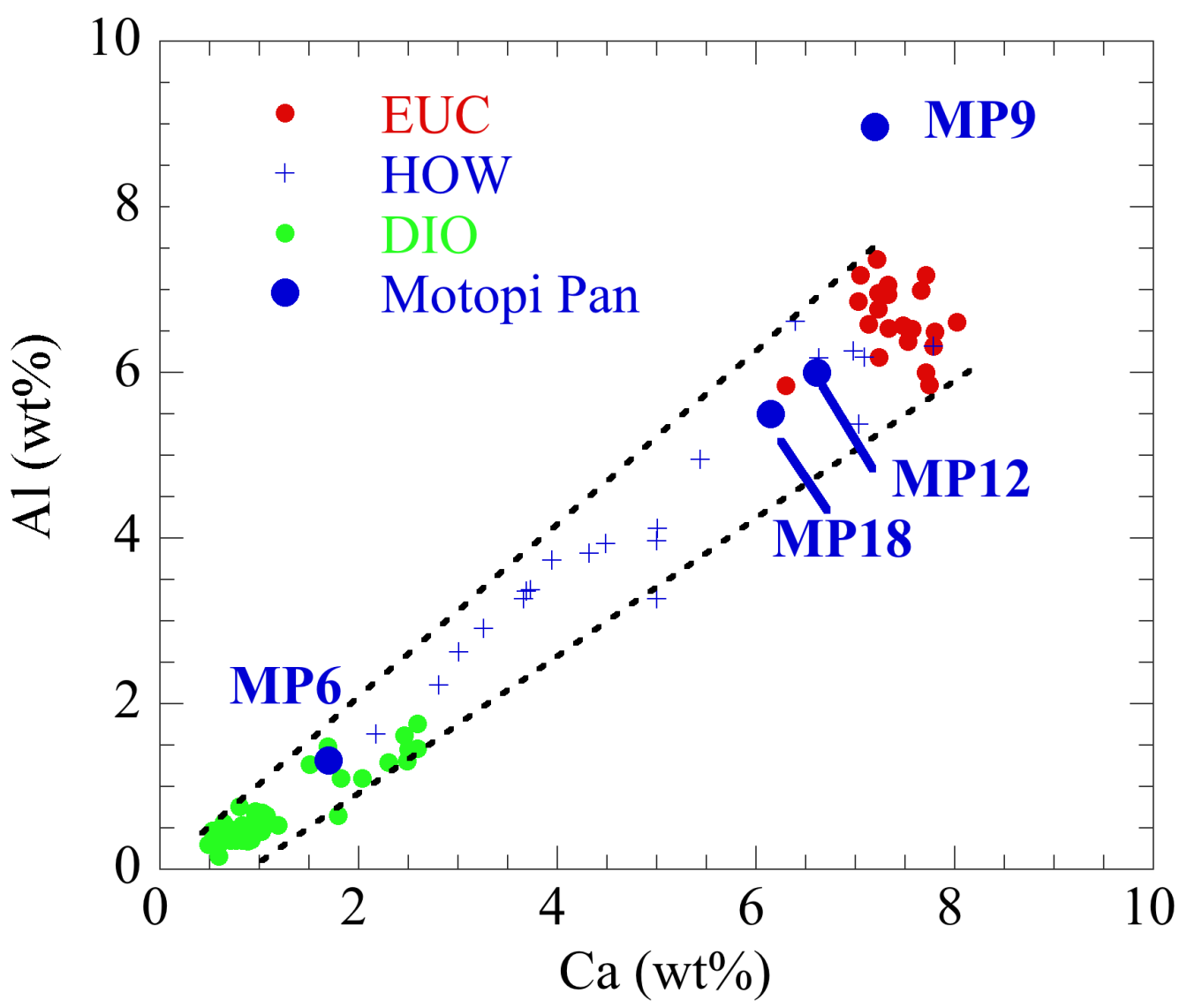

**Fig. 29. (a)** Measured $^{36}Cl$ concentrations as a function of the chemical composition. **(b)** Normalized $^{36}Cl$ concentrations in the MP samples (red symbols and grey bar) with production rate calculations from Leya & Masarik (2009), which were increased by 10%.

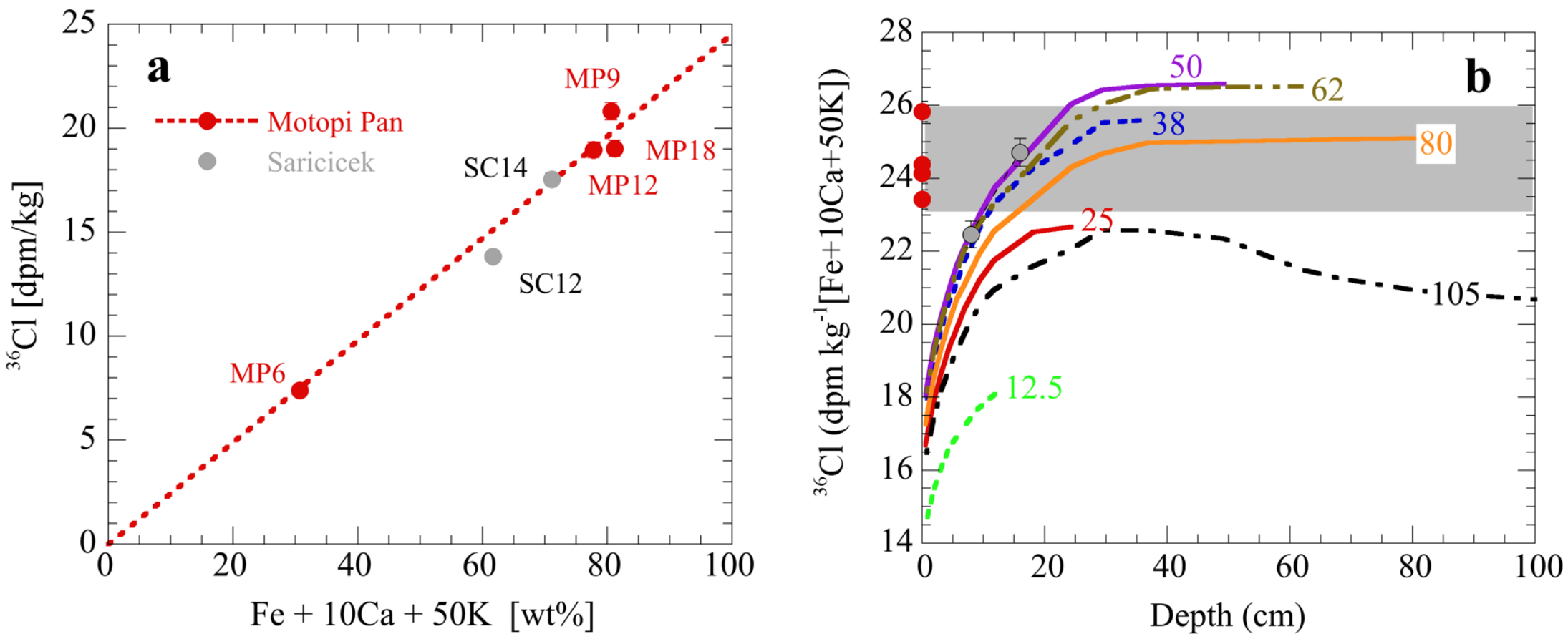

**Fig. 30.** Comparison of the measured concentrations of cosmogenic $^{10}Be$ in Motopi Pan samples (red symbols and grey bar) with calculated production rates from Leya & Masarik (2009) for HED objects with radii from 12.5 cm to 105 cm.

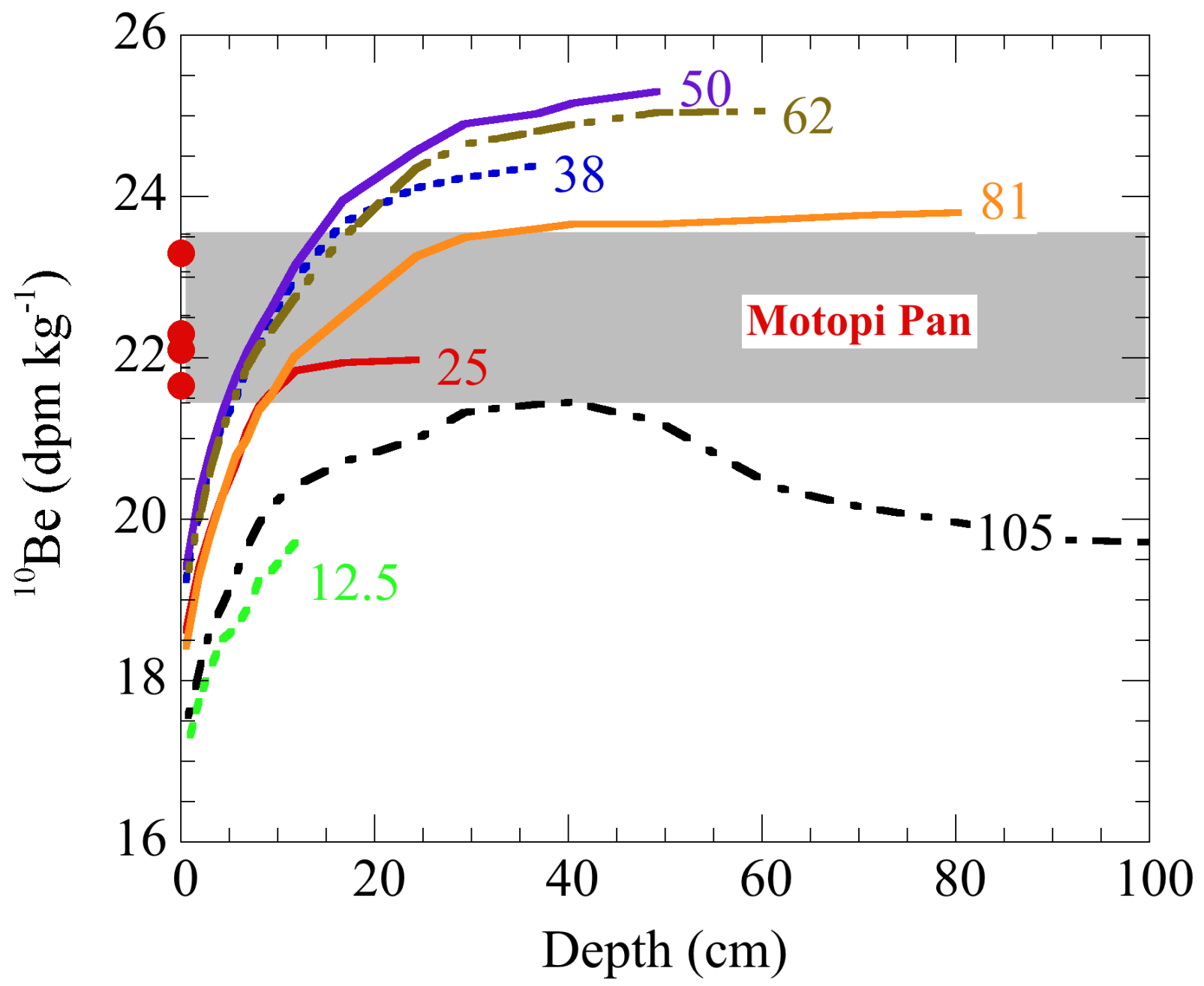

**Fig. 31.** Relationship between calculated production rates of $^{38}Ar$ and $^{36}Cl$ in howardites with different radii and chemical composition of MP-12, using elemental production rates from Leya & Masarik (2009).

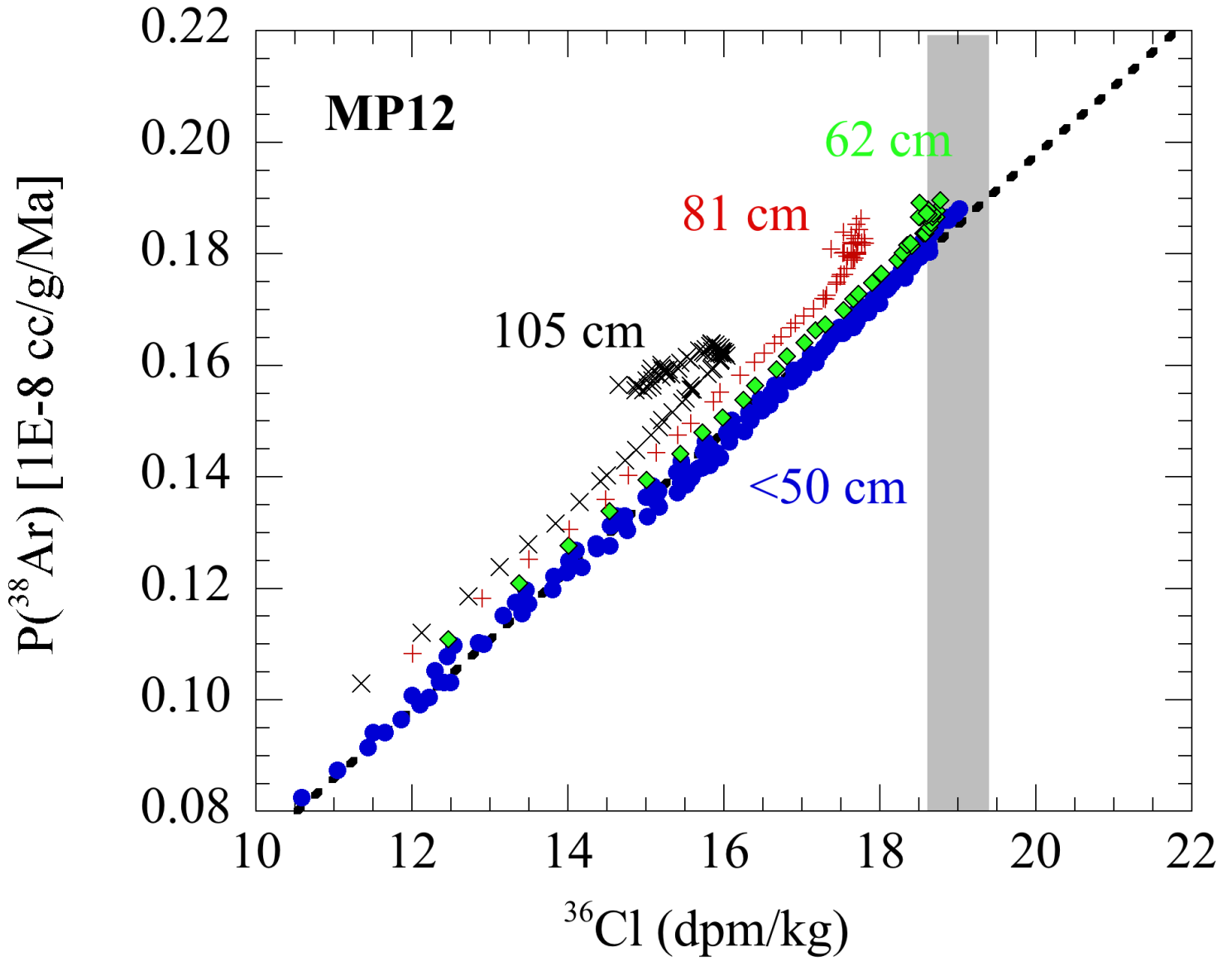

**Fig. 32.** Backscattered electron (BSE) images showing zircon and phosphate occurrences in Motopi Pan (MP-17), which consists of high-Ca pyroxene (cpx), low-Ca pyroxene (opx) and plagioclase (Pl), with minor of ilmenite (Ilm), chromite (Chrm), troilite (FeS), silica ($SiO_2$), zircon (Zr), apatite (Ap) and merrillite (Merr). Phosphate grain number corresponds to the U-Pb data reported in Table 26.

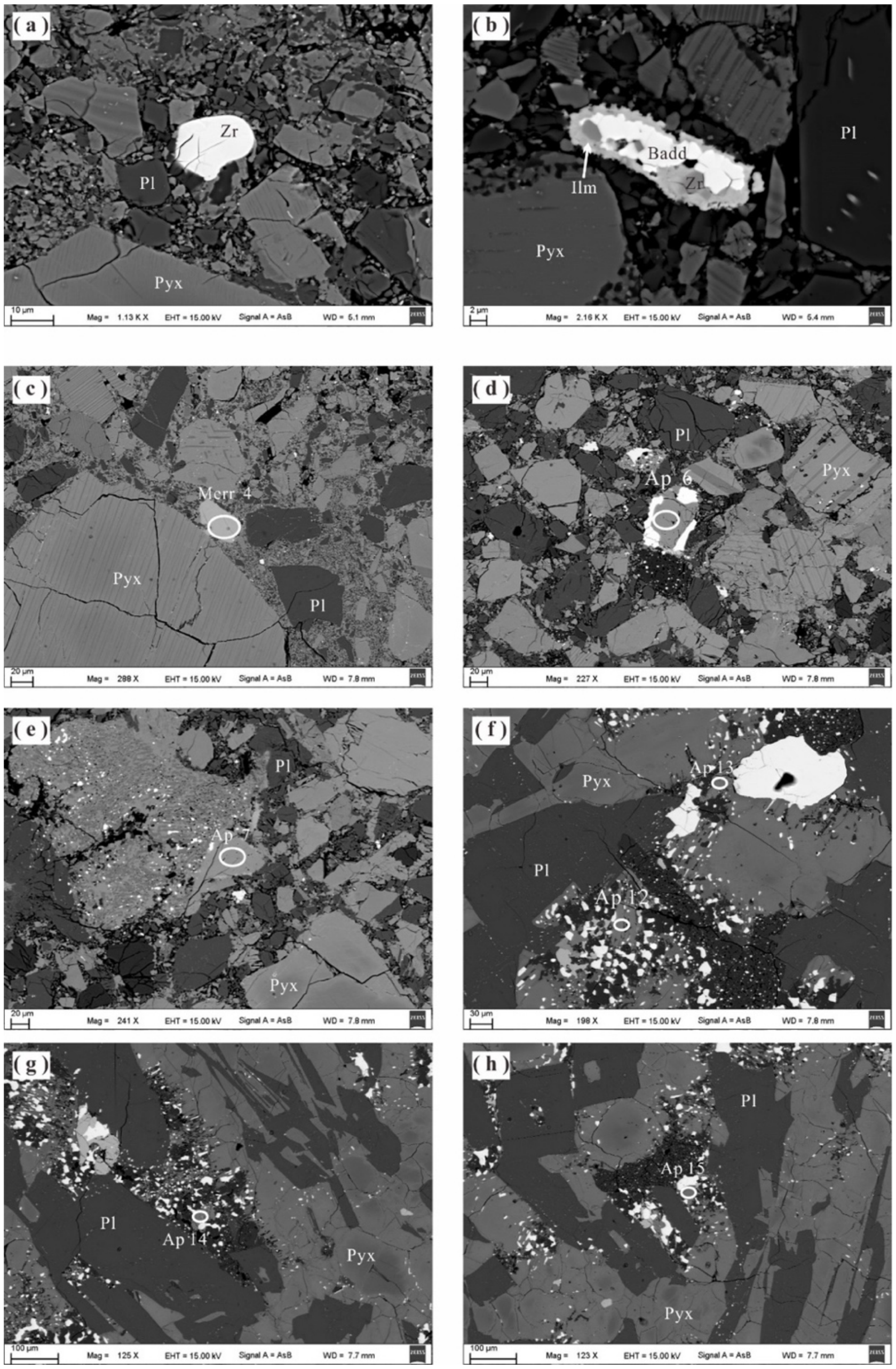

**Fig. 32 (cont.).**

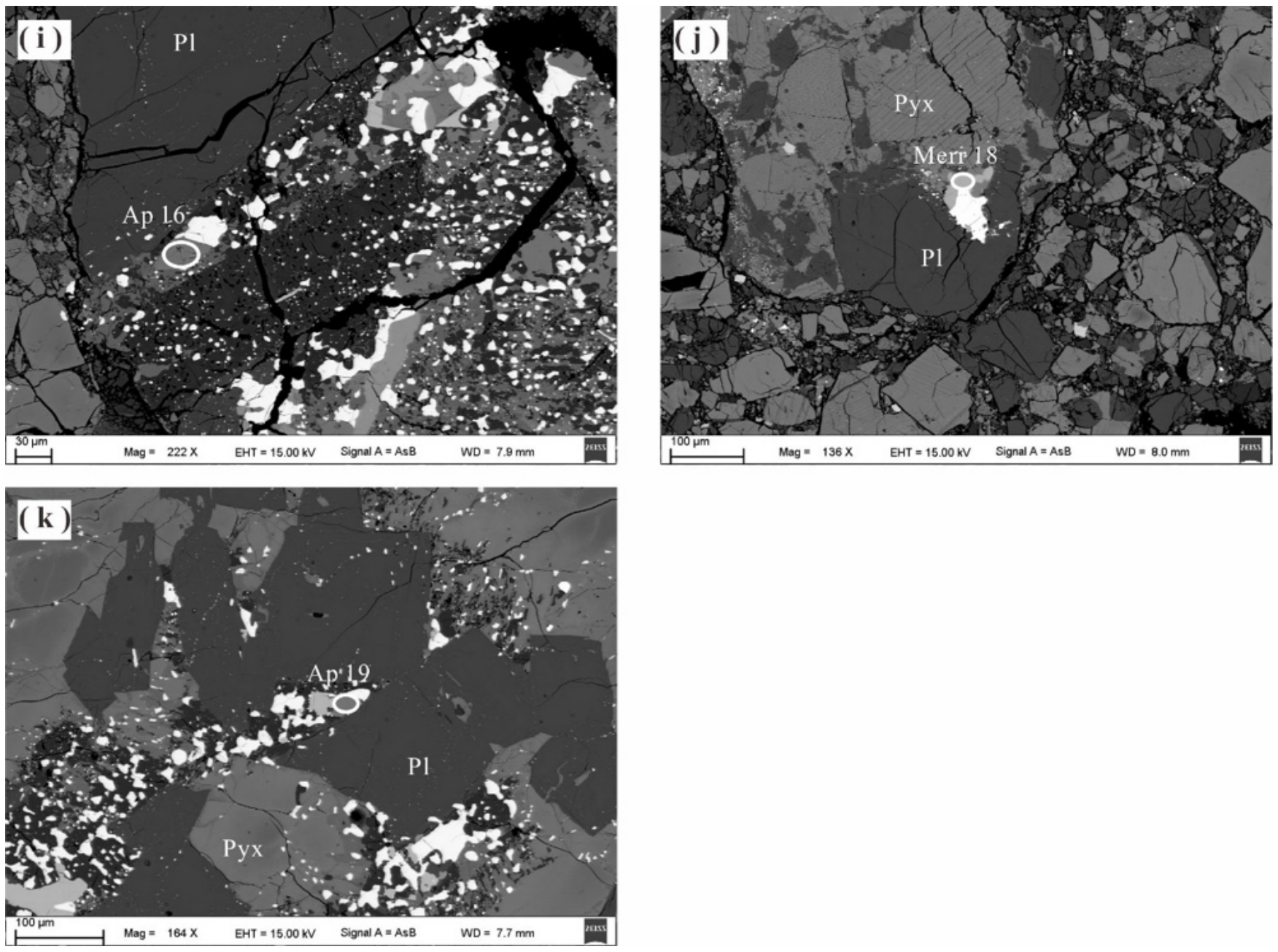
(i)
Pl
Ap 16
30 µm
Mag = 222 X
EHT = 15.00 kV
Signal A = AsB
WD = 7.9 mm
(j)
Pyx
Merr 18
Pl
100 µm
Mag = 136 X
EHT = 15.00 kV
Signal A = AsB
WD = 8.0 mm
(k)
Ap 19
Pl
Pyx
100 µm
Mag = 164 X
EHT = 15.00 kV
Signal A = AsB
WD = 7.7 mm

**Fig. 33.** Methanol-soluble polar organic matter: **(a)** FTICR-MS mass spectra detail on m/z = 319, showing the high chemical diversity of MP-04, similar to howardite Sarcicek and eucrite Tirhert. MP-06 and MP-18 have higher signature in lower mass defects (oxygenated, sulfur rich domain); **(b)** Elementary distributions in CHO, CHNO, CHOS, CHNOS and CHOMg showing the compositional similarities between the analyzed samples; **(c)** Selected double bond equivalent (DBE) analysis of the polar CHO chemical space, complementing the AROMA analysis of the hydrocarbons (blue lines) in Fig. 36.

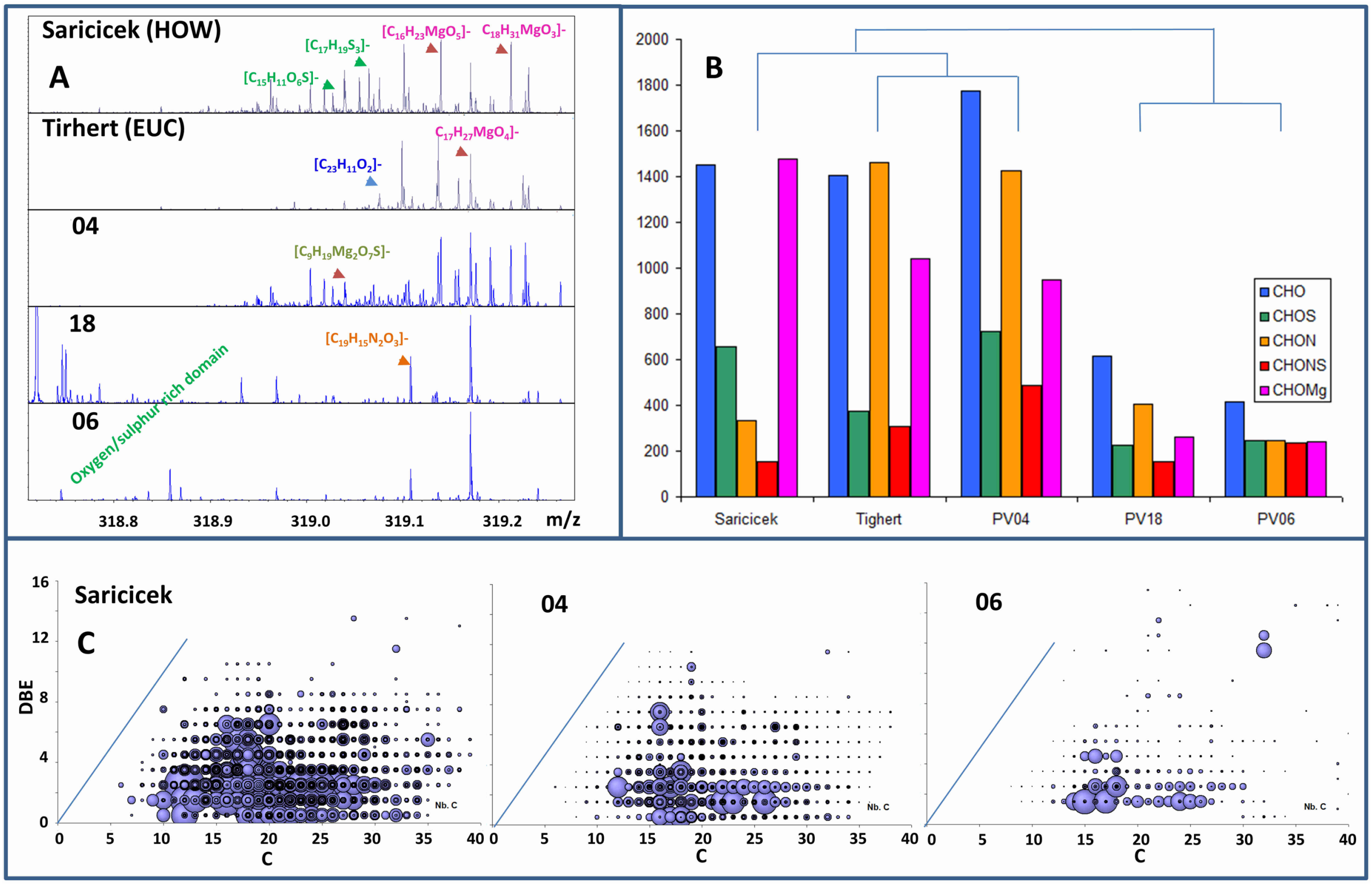

**Fig. 34.** FTICR-MS analysis of the methanol-soluble organic matter of eucrite MP-04, diogenite MP-06 and howardite MP-18 compared with two fresh observed falls (howardite Sariçiçek and eucrite Tirhert). Results are presented as van Krevelen diagrams with bubble size expressing the signal intensity in the mass spectra and CHO molecules shown in blue, CHNO orange, CHOS green and CHOMg pink.

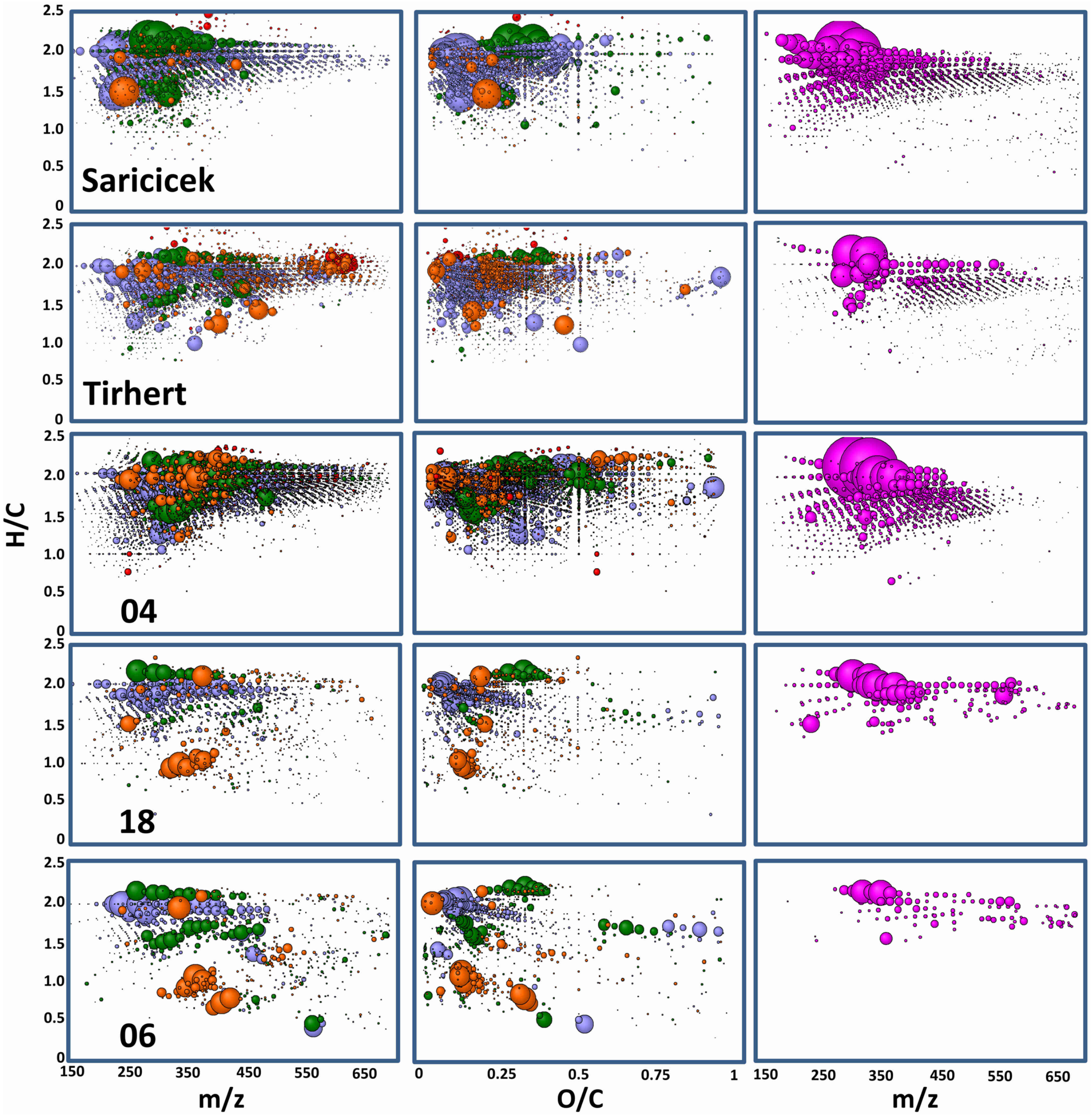

**Fig. 35.** Ion mass spectra for eucrite-looking MP-04, the diogenite-looking MP-06 and MP-13, and the howardite-looking MP-12, -17 and -18.

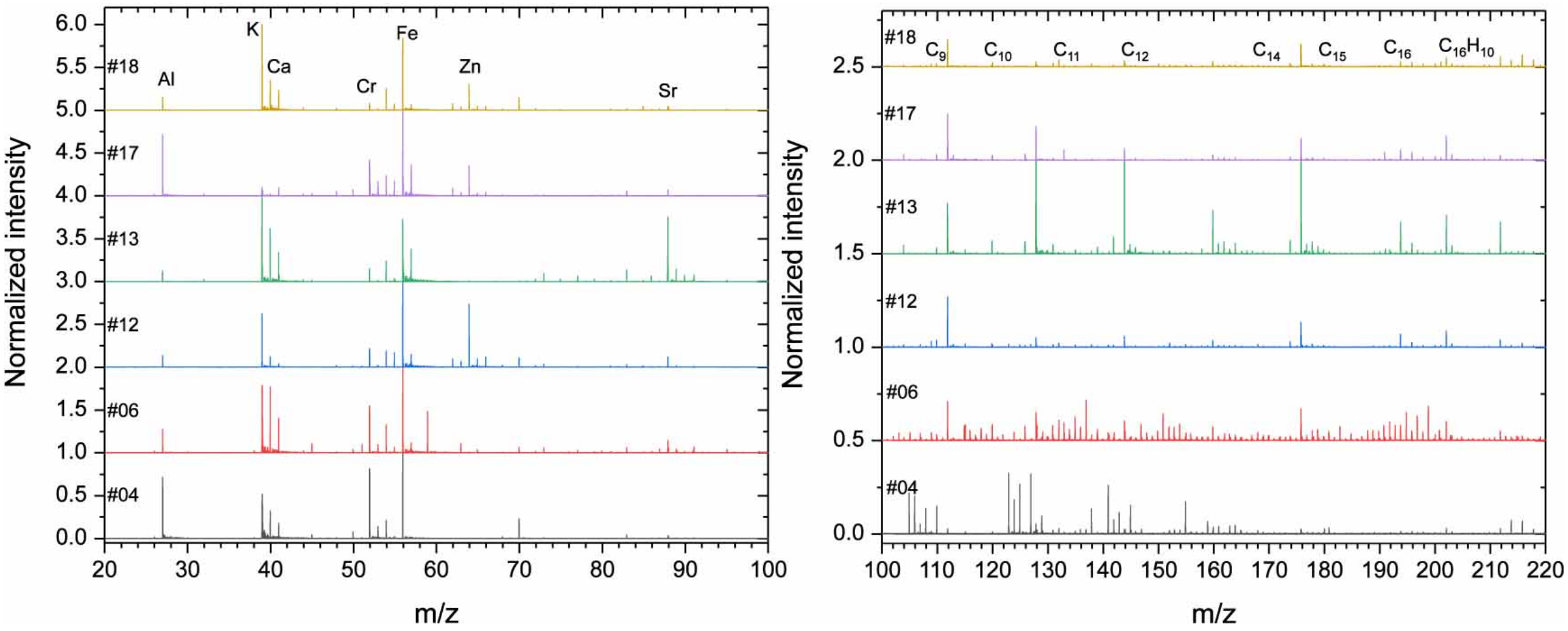

**Fig. 36.** Double-bond equivalent (DBE) = $n_C + 1 - n_H/2$ value as a function of the number of carbon atoms in the molecule. The color code is related to the highest carbon m/z detected.

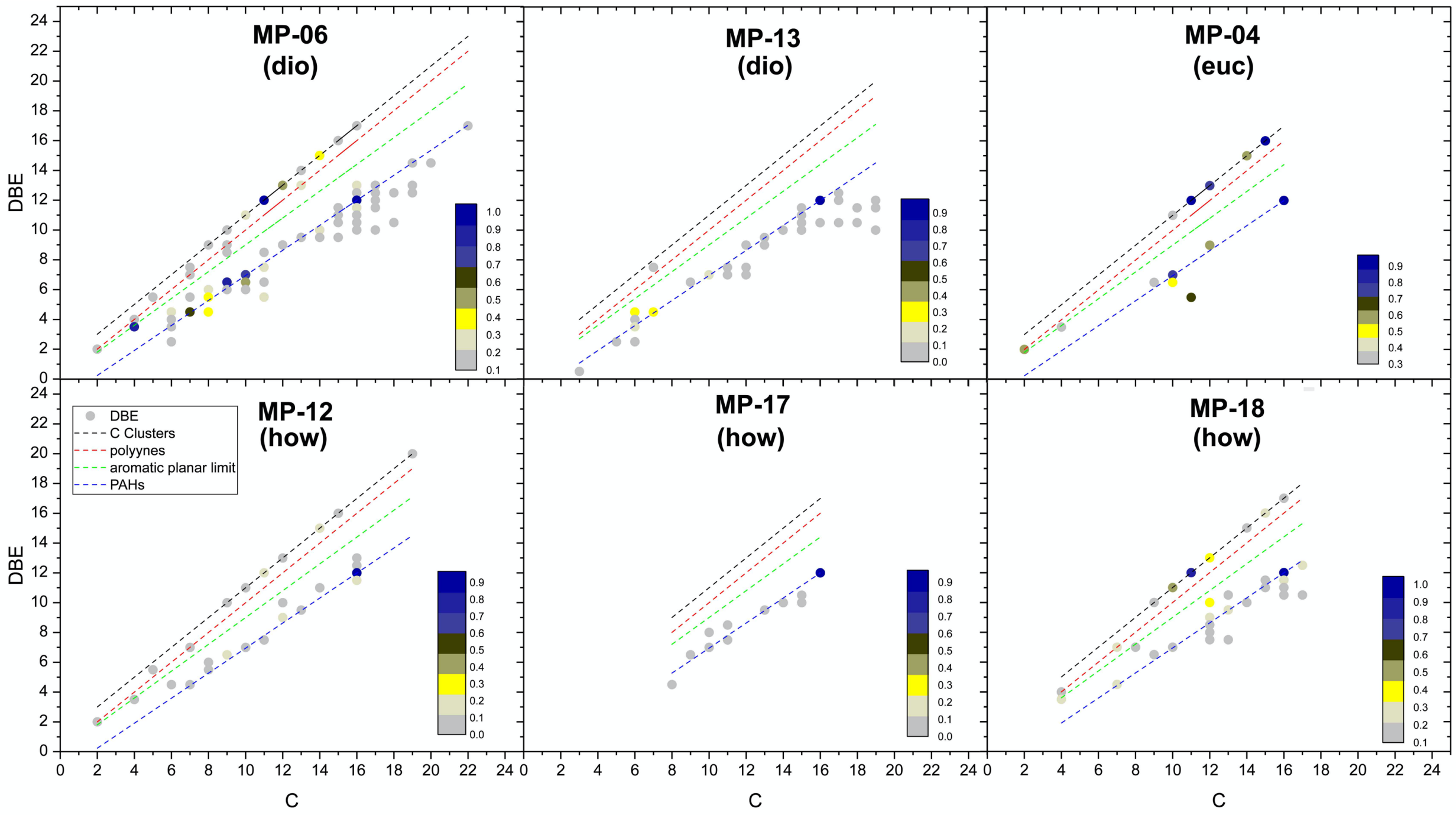

**Fig. 37.** Crater formation age on Vesta. **(a)** From crater counts in the lunar-based chronology scheme (dark points are averages of single terrain counts +) and in the asteroid-based chronology scheme (gray points) for rayed craters of different size, based on (Unsalan *et al.* 2019); **(b)** From cosmic ray exposure age of Motopi Pan (large symbols) and Sariçiçek (small symbols, before (×) and after (•) correction for 2π exposure). The cosmic ray exposure age distribution of other HED meteorites is from (Unsalan *et al.* 2019; Eugster & Michel 1995; Welten *et al.* 1997); **(c)** Inset showing location of likely source craters relative to the Rheasilvia impact basin, based on NASA Dawn image PIA15665.

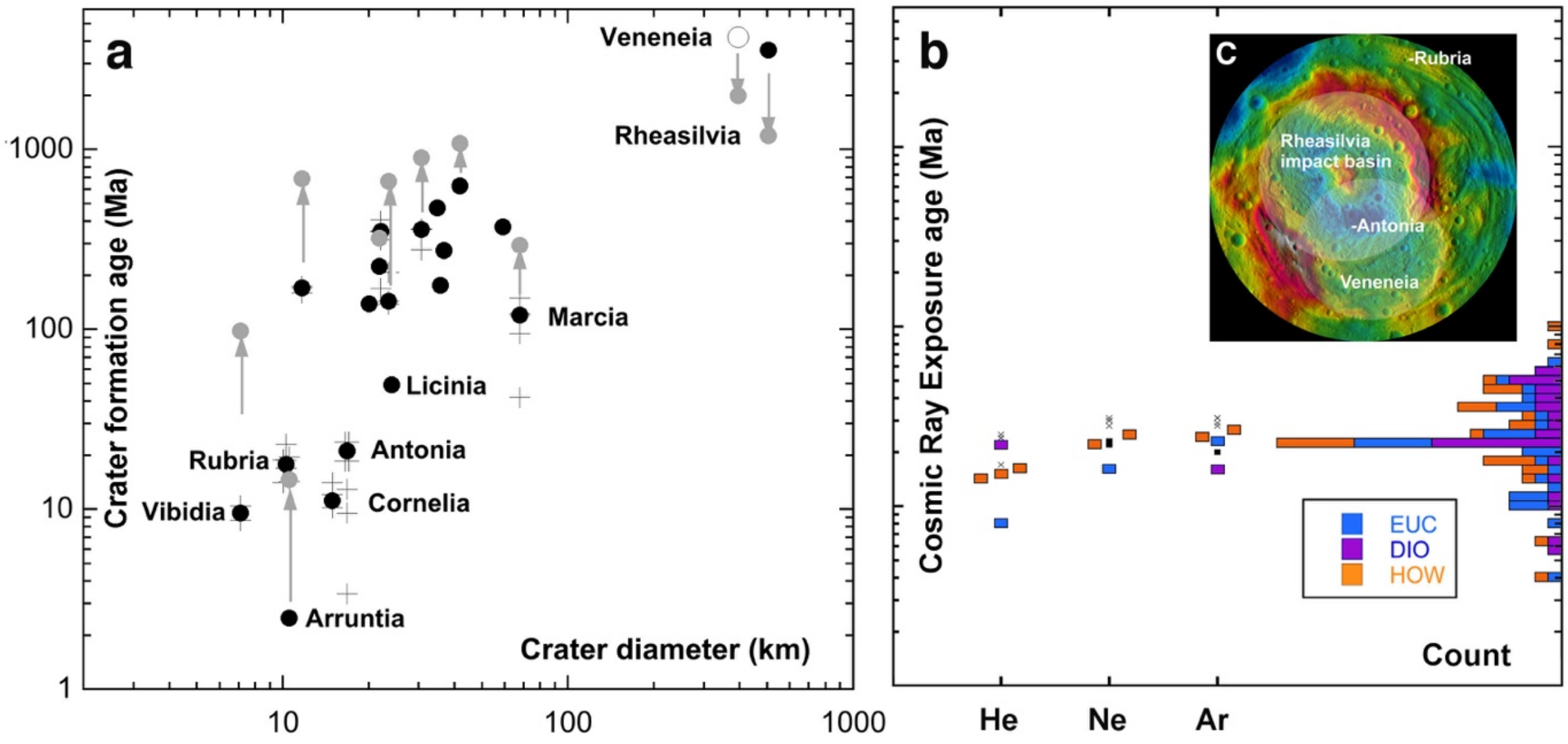